\documentclass[a4paper,twocolumn,11pt,accepted=2023-10-24]{quantumarticle}
\pdfoutput=1
\usepackage[utf8]{inputenc}
\usepackage[english]{babel}
\usepackage[T1]{fontenc}
\usepackage{hyperref}

\usepackage{xcolor}
\definecolor{color1}{rgb}{0,0,0.7}
\definecolor{color2}{rgb}{0.85,0,0}

\usepackage{graphicx}
\usepackage{amsmath}
\usepackage{amssymb}
\usepackage{amsfonts}
\usepackage{amsthm}
\usepackage{bbold}
\usepackage{hyperref}
\usepackage{soul}
\usepackage{textcomp}
\usepackage{bm}
\usepackage{dsfont}
\usepackage{relsize}
\usepackage{braket}
\usepackage{enumitem}   
\usepackage{mathrsfs}
\usepackage{cleveref}
\usepackage{bigints}
\usepackage{float}
\usepackage{framed}
\usepackage[makeroom]{cancel}
\usepackage[most]{tcolorbox}
\usepackage[numbers,sort&compress]{natbib}

\theoremstyle{plain}

\numberwithin{obs}{section}

\newcommand{\eref}[1]{\textcolor{color2}{\hyperref[#1]{eq.$\,$(\ref{#1})}}}
\newcommand{\Eref}[1]{\textcolor{color2}{\hyperref[#1]{Eq.$\,$(\ref{#1})}}}
\newcommand{\fref}[1]{\textcolor{color2}{\hyperref[#1]{Fig.$\,$\bfseries{\ref{#1}}}}}

\newcommand{\aref}[1]{\textcolor{color2}{\hyperref[#1]{App.$\,$\ref{#1}}}}

\newcommand{\<}{\langle}
\renewcommand{\>}{\rangle}

\newcommand{\comments}[1]{}
\newcommand{\ba}{\begin{align}}
\newcommand{\ea}{\end{align}}
\newcommand{\eps}{\varepsilon}

\newcommand{\Tr}[1]{\text{Tr}\left[#1\right]}

\newcommand{\rom}[1]{\uppercase\expandafter{\romannumeral #1\relax}}

\def\Tr{{\rm Tr}}

\makeatletter
\newcommand\footnoteref[1]{\protected@xdef\@thefnmark{\ref{#1}}\@footnotemark}
\makeatother

\begin{document}

\title{Finite-time Landauer principle beyond weak coupling }

\author{Alberto Rolandi}
\email{alberto.rolandi@unige.ch}
\affiliation{D\'{e}partement de Physique Appliqu\'{e}e,  Universit\'{e} de Gen\`{e}ve,  1211 Gen\`{e}ve,  Switzerland}

\author{Martí Perarnau-Llobet}
\email{marti.perarnaullobet@unige.ch}
\affiliation{D\'{e}partement de Physique Appliqu\'{e}e,  Universit\'{e} de Gen\`{e}ve,  1211 Gen\`{e}ve,  Switzerland}

\begin{abstract}
    Landauer's principle gives a fundamental limit to the thermodynamic cost of erasing information. Its saturation requires a reversible isothermal process, and hence infinite time. We develop a finite-time version of Landauer’s principle for  a  bit encoded in the occupation of a single fermionic mode, which can be strongly coupled to a reservoir. By solving the exact non-equilibrium dynamics, we optimize erasure processes (taking both the fermion’s energy and system-bath coupling as control parameters) in the slow driving regime through a geometric approach to thermodynamics. We find analytic expressions for the thermodynamic metric and geodesic equations, which can be solved numerically. Their solution yields optimal processes that allow us to characterize a finite-time correction to Landauer's bound, fully taking into account strong coupling effects. Our result suggests the emergence of the Planckian time, $\tau_{\rm Pl}=\hbar/k_B T$ , as the shortest timescale for information erasure.
\end{abstract}

\maketitle

\vspace{-5pt}
\section{Introduction}
Any logical irreversible operation that will incur a thermodynamic cost in the form of heat dissipated into the environment. Landauer's principle quantifies this relation between information processing and thermodynamics with the bound $Q\geq k_B T \ln 2$ for the erasure of a single bit of information~\cite{landauer1961irreversibility}. Here $Q$ is the dissipated heat, $k_B$ is the Boltzmann constant and $T$ is the absolute temperature at which the process is taking place. In recent years, this principle has been intensively studied within the fields of stochastic and quantum thermodynamics~\cite{Sagawa2009,Esposito2011,Hilt2011,Deffner2013info,Reeb2014,Faist2015,Lorenzo2015,Goold2015,Alhambra2016,Guarnieri2017,Miller2020,Timpanaro2020,Riechers2021,Buffoni2022,taranto2021landauer,browne2014guaranteed}, and  has been approached in several experimental platforms~\cite{berut2012experimental,jun2014high,Brut2015,gavrilov_erasure_2016,Hong2016,Martini2016,Gaudenzi2018,saira_nonequilibrium_2020,Dago2021,dago_dynamics_2022,ciampini2021experimental,scandi2022constant}. 

The unattainability principle suggests that  Landauer's bound \emph{cannot} be saturated 
with finite resources, namely time and energy~\cite{Nernst12,Masanes2017,Freitas2018}.  
In finite time, using tools from optimal transport theory~\cite{Aurell2011a,Aurell2012,vanvu2023} and thermodynamic geometry~\cite{Salamon3,Salamon1,Nulton1985, Andresen,Sivak2012a,Deffner2020a,abiusoGeometricOptimisationQuantum2020,VanVu2021},   
optimal erasure protocols have been  derived both for  classical systems described by overdamped Langevin dynamics~\cite{Zulkowski2014,Proesmans2020,Proesmans2020II,boyd2022shortcuts,Lee2022} and  open quantum systems described by  Lindblad master equations~\cite{Diana2013,scandi19,Zhen2021,VanVu2022,vanvu2023,Zhen2022,Ma2022}.  
Such optimal protocols naturally lead to a finite-time correction to Landauer's bound in different physical set-ups, which has given rise to the term \emph{finite-time Landauer principle}~\cite{Proesmans2020,VanVu2022,Lee2022}. 
For a slowly driven (quantum) two-level system  weakly coupled to a thermal bath, the finite-time bound takes the  simple form~(see \aref{section:WC} and Refs.~\cite{scandi19,Ma2022})
\begin{equation}
    Q \geq k_B T \left(\ln 2 + \frac{ \pi^2 }{4 \Gamma \tau} \right)+ \mathcal{O}\left(\frac{1}{\Gamma^2 \tau^2}\right),
    \label{eq:finitetimeboundweak}
\end{equation}
where  $\tau$ is the total time of the process and  $\Gamma$ is the thermalization rate.
The finite-time correction is positive, in agreement with the second law of thermodynamics, and when $\Gamma \tau \rightarrow \infty$ we recover the standard bound. 
We also note that the optimal protocol saturating the finite bound \eref{eq:finitetimeboundweak} has been recently implemented in a semiconductor quantum dot~\cite{scandi2022constant}.  
More general versions of \eref{eq:finitetimeboundweak} have also been recently developed for Markovian systems driven at any speed~\cite{Zhen2021,VanVu2022,vanvu2023}.


Despite this remarkable progress, previous works on the \emph{finite-time Landauer principle} have focused in Markovian systems which, for quantum systems, can be guaranteed by a sufficiently  weak interaction between system and bath.
In the presence of strong coupling~\cite{Strasberg2016,Jarzynski2017,Miller2018,Nazir2018,Talkner2020}, we expect both new opportunities arising due to faster relaxation rates and non-Markovian dynamics~\cite{Rivas2020,Brenes2020,Pancotti2020,Alipour2020,Ptaszyski2022,Carrega2022,Cavaliere2022,Ivander2022}
, as well as  challenges due to the presence of new sources of irreversibility~\cite{Newman2017,PerarnauLlobet2018,Strasberg2018,Wiedmann_2020,Liu2021,Shirai2021,Koyanagi2022,Liu2022b}. The goal of this work is to take a first step into this exciting regime by deriving the first order to a tight finite-time correction of Landauer's principle for a single fermion that can interact \emph{strongly} with a reservoir, as described by the  resonant-level model~\cite{schaller2014open,Ludovico2014,Esposito2015,Esposito2015b,Bruch2016,Haughian2018,Mitchison2018,Tong2022}. 
Our main result is summarised in what follows:
\begin{tcolorbox}[colback=violet!5!white,colframe=violet!75!black,title=Main result]
Given a two-level system that can be  strongly coupled to a thermal bath, we find that the finite-time version of Landauer's principle can be expressed as 
\begin{align}
    Q \geq k_B T \left(  \ln 2 + a \frac{\tau_{\rm Pl}}{\tau}  \right) + \mathcal{O}\left(\frac{1}{\Gamma^2 \tau^2}\right) 
    \label{MainResult}
\end{align}
where $a\approx 2.57946$,   $\tau_{\rm Pl} = \hbar/k_B T$ is the so-called Planckian time~\cite{Hartnoll22}, and $\Gamma$ is the average thermalization rate (see details below). This extends \eref{eq:finitetimeboundweak} to strong system-bath couplings,  with the transition between the two being  characterized in \fref{fig:main}. The finite-time correction in \eref{MainResult} is of quantum-mechanical nature and independent of the coupling strength, hence prevailing even for   arbitrarily strong system-bath coupling  (roughly speaking, $\Gamma \rightarrow \infty$ in \eref{eq:finitetimeboundweak}).
\end{tcolorbox}
The appearance of the Planckian time $\tau_{\rm Pl} = \hbar/k_B T$ in \eref{MainResult} is particularly interesting.
This timescale  encodes two fundamental constants in nature: Boltzmann's constant $k_B$ and Planck's constant $\hbar$. It arises in several contexts  in many-body physics, including quantum transport and quantum chaos; see Ref.~\cite{Hartnoll22} for a review. In analogy with the ``Planck time'' in quantum gravity, it is 
associated with  the shortest timescale of thermalization~\cite{Hartnoll22,sachdev11,Maldacena2016,Pappalardi22}; that is, the shortest time needed 
to redistribute energy between particles and reach thermal equilibrium. This gives an insightful context to our main result \eref{MainResult}: a fundamental finite-time quantum correction must appear to Landauer's bound due to a minimal time required for thermalization. This also suggests that the form of \eref{MainResult} has a broader range of applicability, with the value $a$ depending on the specific many-body thermalizing dynamics considered.

In order to  obtain \eref{MainResult}, we exploit the framework of thermodynamic geometry~\cite{Salamon3,Salamon1,Nulton1985, Andresen,Sivak2012a,Deffner2020a,abiusoGeometricOptimisationQuantum2020},
which has proven  successful to devise minimally dissipative processes both in classical~\cite{Zulkowski,Sivak2012a,Bonana2014,Rotskoff2017,Li2022,Eglinton2022,Frim2022,chen2022geodesic} and quantum systems~\cite{scandi19,Miller2019,abiuso2020optimal,Brandner2020,Miller2020,Alonso2022Geometric,Mehboudi2022Thermodynamic}.

\section{Framework}
We consider a driven system S that can be put in contact with a thermal bath B, so that the total time-dependent Hamiltonian reads: 
\begin{align}
    H(t)=H_S(t)+H_{\rm int}(t)+H_B.
    \label{eq:GeneralH(t)}
\end{align}
Here, $H_S(t)$, $H_{\rm int}(t)$ are the externally controllable Hamiltonian of S and SB coupling, respectively, whereas $H_B$ is the Hamiltonian of B. 
The state $\rho(t)$ of SB evolves as $\rho(t) =U(t)\rho(0) U^\dagger(t) $ with $U(t)=\mathcal T\exp(-\frac{i}{\hbar}\int_0^t ds~H(s))$. 
The work cost induced by driving $H(t)$, with $t\in [0,\tau]$, reads: 
\begin{align}
    W &= \int_0^\tau \hspace{-1mm} ds{\rm Tr}[\rho(s) \dot{H}(s)] 
    = \rm{Tr} [H(\tau)\rho(\tau) - H(0)\rho(0) ]  
      \label{eq:GeneralW}
\end{align}
Focusing on protocols where  $H_{\rm int}(0)=H_{\rm int}(\tau)=0$, we can naturally identify from the first law of thermodynamics $W=Q+\Delta E_S$ with $\Delta E_S=\Tr[H_S(\tau)\rho(\tau)-H_S(0)\rho(0))]$, the dissipated heat \begin{align}
    Q=\Tr[H_B(\rho(\tau)-\rho(0))]
    \label{eq:GeneralQ}
\end{align}  
as the total energy absorbed by the bath~\cite{Esposito2011}. 

Assuming that the initial state of SB is a thermal state: $\rho(0)= e^{-\beta H(0)}/\mathcal{Z}(0)$ with $\mathcal{Z}(t) \equiv {\rm Tr} [e^{-\beta H(t)}]$, \eref{eq:GeneralW} can be re-expressed as~\cite{Deffner2010}:
\begin{align}
    W=\Delta F + k_B T \Sigma ~,
    \label{eq:WSigma}
\end{align}
where $\Delta F=k_B T \ln [\mathcal{Z}(0)/\mathcal{Z}(\tau)] $ is the change of equilibrium free energy of SB, and the entropy production $\Sigma$ can be expressed as:  $\Sigma=S\!\left(\rho(\tau)\big| \big|\frac{e^{-\beta H(\tau)}}{\mathcal{Z}(\tau)}\right)$. The entropy production  $\Sigma \geq 0$  accounts for the irreversible energetic contribution in finite-time processes, and depends on the particular driving path $H(t)$ linking $H(0)$ to $H(\tau)$. Minimising $\Sigma$ over all finite-time processes leads to thermodynamic protocols that minimize the  work $W$. Furthermore, in an erasure process, $\Delta E_S = 0$ (see details below) therefore these protocols also minimize the dissipated heat $Q$. 

\section{Thermodynamic geometry}

The framework of quantum thermodynamic geometry~\cite{abiusoGeometricOptimisationQuantum2020} allows us to minimize the entropy production $\Sigma$, and therefore the dissipated heat $Q$ for erasure, for protocols that are slow compared to their relaxation time-scale.

\subsection{Strongly coupled systems}

 Let us expand the  Hamiltonian $H(t)$ in \eref{eq:GeneralH(t)} as $H(t)=\sum_j \lambda_j(t) X_j$ where $\{ \lambda_j (t) \}$ are the externally controllable parameters and  $\{ X_j \}$ are the corresponding observables. In order to apply the geometric approach, we need to impose more structure on the possible evolutions $U(t)$ generated by \eref{eq:GeneralH(t)}. We  require two  basic ingredients:  \\
 {\bf Requirement 1: Thermalization.}  In absence of driving, the conjugated observables $X_j$ thermalize. More precisely, for a \emph{frozen} Hamiltonian $H(t)$, we have 
 \begin{align}
 \lim_{s\rightarrow \infty} {\rm Tr}[\tilde{U}_t(s) \rho(0) \tilde{U}_t^\dagger(s)X_j] = \langle X_j(t) \rangle_{\rm eq},
 \end{align}
 where $\tilde U_t(s) \equiv e^{-iH(t)s}$, $\langle X_j(t) \rangle_{\rm eq}=\Tr[ \omega_\beta (t) X_j]$, $\omega_\beta (t)=e^{-\beta H(t)}/\mathcal{Z}(t)$, and  $\beta$ is implicitly defined by the initial energy of the total system.  In the context considered here, namely purely unitary dynamics of SB, this condition is  satisfied both by non-integrable systems satisfying the ETH hypothesis~\cite{Eisert2015,DAlessio2016} and  also for  integrable systems typically appearing in open quantum systems~\cite{Suba2012,Merkli2020,Cresser2021,Trushechkin2022}.
 \\  {\bf Requirement 2: Slow driving}, so that the system  remains close to the instantaneous equilibrium state while being driven. This enables us to keep only leading terms when making a linear-response expansion in the driving speed~\cite{cavina2017slow}, which can be expressed as:
 \begin{align}
      \langle X_j(t) \rangle = \langle X_j(t) \rangle_{\rm eq} + \sum_i m_{ij} \dot{\lambda}_i (t) +...
      \label{Eq:ExpansionOperators}
 \end{align}
The coefficients~$m_{ij}$, which depend on the point~$\{\lambda_i (t)\}$,  can in principle be derived from the exact equations of motion. We should note that it is well known that optimal finite-time protocols feature jumps~\cite{Schmiedl2007o}. However, these jumps disappear near the reversible limit (see \aref{sm:jumps} and~\cite{Esposito10}) and their contribution to the dissipated heat becomes negligible. Therefore this requirement becomes a natural assumption in the context of finding a first order correction to Landauer's bound.

Combining the expansion of \eref{Eq:ExpansionOperators} with \eref{eq:GeneralW} and \eref{eq:WSigma},  we obtain the standard expression for entropy production at leading order in the inverse of the driving speed~\cite{Salamon1,Sivak2012a,abiusoGeometricOptimisationQuantum2020}: 
\begin{align}
    k_B T \Sigma = \sum_{ij} \int_0^\tau dt~\dot{\lambda}_i(t) m_{ij}(t) \dot{\lambda}_j(t)
    \label{eq:GeneralExpansionDiss}
\end{align}
where, in contrast to previous works, the metric $m_{ij}$ depends on the unitary dynamics of SB (this will later be solved for a specific model). 
Because of the second law of thermodynamics, 
it follows that 
$m_{ij}$ can be expressed as a metric, i.e., a symmetric, positive-definite $m\geq 0$ operator that depends smoothly  on the point~$\{\lambda_i(t) \}$.  We can associate a length to a protocol by defining $L = \int_0^\tau dt \sqrt{\sum_{ij} \dot{\lambda}_i(t) m_{ij}(t) \dot{\lambda}_j(t)}$. It is related to the entropy production via a Cauchy-Schwarz inequality~\cite{Salamon1,Sivak2012a,abiusoGeometricOptimisationQuantum2020}:
\begin{equation}
    k_B T\Sigma \geq \frac{1}{\tau}L^2~,
\end{equation}
where equality is satisfied by protocols with constant entropy production rate $\sum_{ij} \dot{\lambda}_i(t) m_{ij}(t) \dot{\lambda}_j(t)$.
Furthermore, to minimize the entropy production of any (slow) protocol we have to find the shortest path between the desired initial and final value of the Hamiltonian's parameters. This corresponds to a geodesic path, with length $\mathcal L$, which naturally defines a minimal entropy production 
\begin{equation}
\label{Sigmamin}
    k_B T\Sigma_{min} = \frac{1}{\tau}\mathcal L^2~.
\end{equation}
We can find $\Sigma_{min}$ by solving the geodesic equation that is derived from the metric  and computing its length~\cite{Salamon1,Sivak2012a,abiusoGeometricOptimisationQuantum2020}. 


\subsection{Resonant-level model}

Having explained the general ideas behind our work, we now focus on finite-time driving processes of a single fermionic mode coupled to a fermionic bath, which can e.g. describe a single-electron quantum dot. The total Hamiltonian reads:
\begin{align}
   H(t)
   = \varepsilon(t) \hat a^\dagger \hat a + \sum_{k=1}^n \omega_k \hat b_k^\dagger \hat b_k+ g(t) \sum_{k=1}^n \lambda_k \hat a^\dagger \hat b_k + \lambda_k^* \hat b_k^\dagger \hat a. 
\end{align}
where $\hat a^\dagger$ is the creation operator of the two-level system and $\hat b_k^\dagger$ is the creation operator of a bath mode with frequency $\omega_k$, following the canonical anti-commutation relations: $\{\hat a^\dagger , \hat a\} = \mathbb{1}$, $\{\hat b_j^\dagger , \hat b_k\} = \delta_{jk}\mathbb{1}$, $\{\hat b_j , \hat b_k\} = \{\hat a , \hat b_k\} = \{\hat a^\dagger , \hat b_k\} = \{\hat a , \hat a\} = 0$; and finally  $\lambda_k$ are the interaction weights which define the spectral density function of the bath $\mathfrak J(\omega) = 2\pi\sum_k|\lambda_k|^2\delta(\omega-\omega_k)$. The energy $\eps$ is the difference between the energy of the two-level system and the chemical potential of the bath\footnote{The chemical $\nu$ potential of the bath is incorporated by subtracting $\nu \hat a^\dagger\hat a$ to the system's Hamiltonian. Since here $H_S = \eps a^\dagger\hat a$ (with $\eps$ the energy of the system), we can simply redefine $\eps$ to be the difference between the system's energy and the chemical potential and set $\nu = 0 $ without loss of generality.}.
We are assuming optimal control over the functions $\eps(t)$ and $g(t)$ so that we can fully optimize the protocol and reach the fundamental limit for this system. While this level of control is, in principle, ambitious experimentally in regards to the coupling, it has been achieved in quantum dots \cite{Rochette2019} where the tunneling rate (i.e. interaction strength) can be modified by several orders of magnitude.
We take the continuum limit and assume that the spectral density of the bath is a Lorentzian
\begin{equation}
    \mathfrak J(\omega) = \frac{\Lambda^2}{\Lambda^2 + \omega^2}~,
\end{equation}
where $\Lambda>0$ is a parameter characterizing its width. Exact and explicit solutions for the resonant-level model are known in the wide-band limit $\Lambda \rightarrow \infty$~\cite{schaller2014open,Ludovico2014,Esposito2015,Esposito2015b,Bruch2016,Haughian2018,Mitchison2018}.   This limit is commonly used to describe quantum systems in contact with fermionic macroscopic baths, e.g. in quantum dots or single-molecule junctions~\cite{Evers2020,Covito2018}. In essence, it neglects the structure of the density of states in the bath and, as a consequence, a  main limitation is that it fails to describe the short-time dynamics~\cite{Covito2018}. Nevertheless,  this problem does not affect this study since we are interested in large times. We should further note that the energy of the system-bath interaction is proportional to $\Lambda$, and therefore is divergent in this limit. We will therefore take $\Lambda$ to be finite but much larger than any other energy scale of the system. For our analysis to be valid we simply require  dynamics much slower than $\Lambda^{-1}$ \cite{schaller2014open}.

The dynamics are solved via a quantum Langevin approach, which is detailed in \aref{sm:solution_rl} (see also~\cite{Pancotti2020,Tong2022}). Taking the initial state to have no correlations between S and B ($\rho(0) = \rho_S(0)\otimes\rho_B(0)$) and the bath to be in a thermal state; we find the probability of occupation of the excited level of the system $p(t) = \<\hat a^\dagger \hat a \>$ and the system-bath interaction energy $v(t) = \sum_k \lambda_k\<\hat a^\dagger \hat b_k\> + h.c.$, 
\begin{align}
    \label{eq:p_exact}
    p(t) =& \left|G(t,0)\right|^2 p(0)  \\
    \nonumber +& \frac{1}{2\pi}\int_{-\infty}^{\infty}\!\!d\omega~ f_\beta(\omega)\left|\int_0^t ds~ g(s) G(t,s) e^{i\frac{\omega}{\hbar}(t-s)}\right|^2  \!\!,\\
    \label{eq:v_exact}
    v(t) =& \frac{1}{\pi}\Im\! \int_{-\infty}^\infty\!\! d\omega~f_\beta(\omega) \int_0^t ds~g(s)G(t,s)e^{i\frac{\omega}{\hbar}(t-s)},
\end{align}
where $f_\beta(\omega) = (1+e^{\beta\omega})^{-1}$ is the Fermi-Dirac distribution and we defined the propagator
\begin{equation}
    G(t,s) = \exp\!\left[-\frac{1}{\hbar}\int_s^t dr~ \mu(r) + i\eps(r)\right]~,
\end{equation}
with $\mu(t):=\frac{1}{2}g(t)^2$. From these expressions we can exactly compute the thermodynamic work \eref{eq:GeneralW}, which  reads: 
\begin{align}
    W = \int_0^\tau dt~ \dot\eps(t)p(t) + \dot \mu(t)v(t)/g(t).
    \label{eq:Wfermion}
\end{align}

From the exact solutions for $p(t)$ and $v(t)$, in \aref{sm:solution_rl} we show that  {\bf Requirement 1} is satisfied, and hence $W=\Delta F$ in the quasistatic limit. For slow but finite-time processes, we perform a slow driving expansion of \eref{eq:p_exact} and \eref{eq:v_exact} (details in \aref{sm:sd}) using that the thermalization rate of the system is $\Gamma := \frac{2}{\hbar\tau}\int_0^\tau dt\, \mu(t)$, so that the expansion can be performed in orders of $1/(\tau\Gamma)$.
We then obtain an expansion for $W$ analogous to \eref{eq:WSigma} where the entropy production $\Sigma$ is described by \eref{eq:GeneralExpansionDiss} with $\vec\lambda(t) = (\eps(t),\mu(t))$ and the thermodynamic metric
\begin{equation}\label{eq:metric}
    m(t) = \frac{\hbar}{\pi}\int_{-\infty}^\infty d\omega~f_\beta(\omega) m_\omega(\eps(t)-\omega,\mu(t))~,
\end{equation}
where
\begin{equation}\label{eq:metric_omega}
    m_\omega(\eps,\mu) = \frac{1}{\left(\mu^2 + \eps^2\right)^3}
	\begin{pmatrix}
		4\eps\mu^2 &
		-\mu(\mu^2-3\eps^2) \\
		\mu(\mu^2-3\eps^2) &
		2\eps(\eps^2-\mu^2)
	\end{pmatrix}~.
\end{equation}
This metric gives a geometrical description of slow thermodynamic protocols performed on the system. By solving the geodesic equations~\cite{Tu17}, we can find the geodesic length $\mathcal L$ and hence the minimal entropy production~\eref{Sigmamin}. 
\vspace{10pt}

\begin{figure*}[ht]
		\centering
        \includegraphics[width=0.8\textwidth]{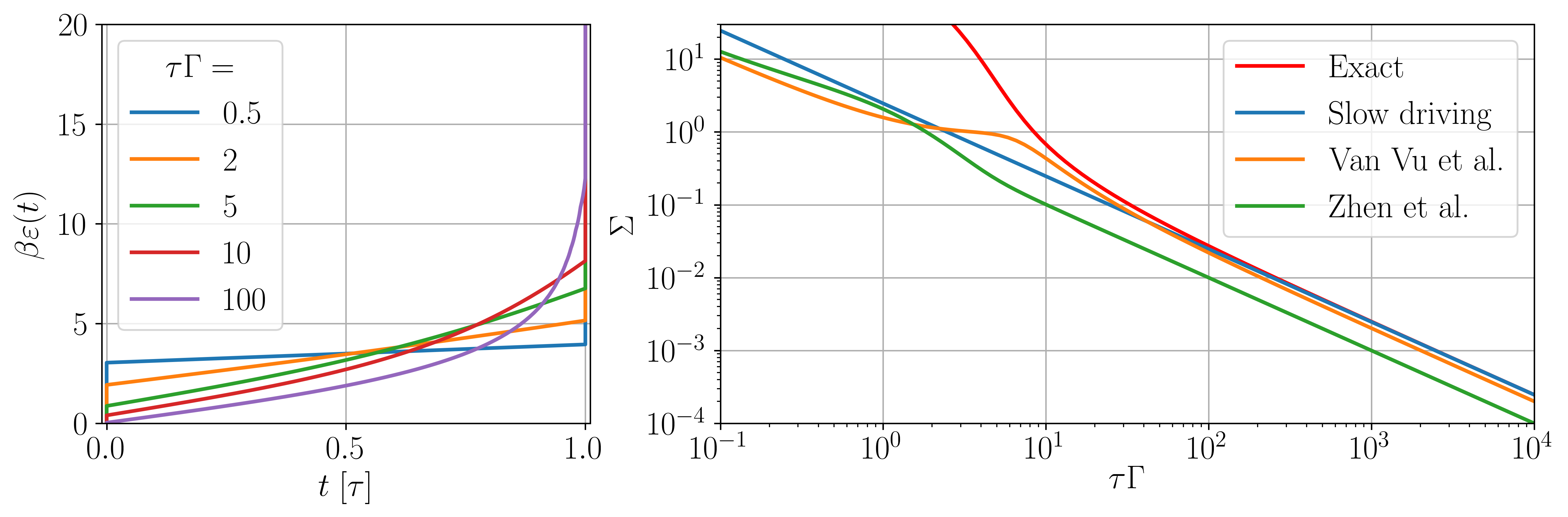}
	    \vspace{-5pt}
	    \caption{\label{fig:WC_comparison}(left) Examples of optimal protocols computed from~\cite{Esposito10} for boundary conditions  $\eps(0) = 0$ and $\beta\eps(\tau) = 20$ at different values of $\tau$. (right) Comparison of the entropy production of the optimal protocols from~\cite{Esposito10} with boundary conditions $\eps(0) = 0$ and $\beta\eps(\tau) = 100$ to the lower bounds given by \eref{eq:finitetimeboundweak}, Van Vu et al.~\cite{VanVu2022} and Zhen et al.~\cite{Zhen2021}.}
	    \vspace{-15pt}
\end{figure*}
\section{Special limits of the metric}

Before attempting to solve the geodesic equations for the  case of  erasure, we now study the high and low temperature limits, as well as the limit of weak coupling, to gain further analytical insights on the form of optimal protocols and the associated entropy production.

\subsection{High temperature limit ($\beta \eps, \beta \mu \ll 1$)}

Since the terms of \eref{eq:metric} quickly decay at high frequencies we can perform the high temperature expansion  $f_\beta(\omega) = \frac{1}{2} - \frac{1}{4}\beta\omega + \mathcal O(\beta^3\omega^3)$ directly in the metric. At leading order, we find:
\begin{equation}\label{eq:metric_HT}
    m_{HT} = \frac{\hbar\beta}{8\mu}\mathbb{1}~.
\end{equation}
This enables an analytical solution of the geodesic equations. 
Given the boundary conditions $\{\eps(0) = \mu(0) = \mu(\tau)=0,\,\eps(\tau)=\eps_*>0 \}$, which  will  later  match  those of an erasure protocol\footnote{\label{fn:init_cond}Usually, the initial condition for erasure would be $\eps(0) = \nu$ for $\nu$ the chemical potential of the bath and $\eps$ the energy of the two-level system (so that the corresponding thermal state is the fully mixed state). But since here we defined $\eps$ to be the difference to the chemical potential we take $\eps(0) = 0$ without loss of generality.}, we find the following geodesic path (cf. \aref{sm:HT})
\begin{align}
    \label{eq:geo_HT_eps}
    \eps(t) &= \eps_*\left(t/\tau - \frac{\sin(2\pi t/\tau)}{2\pi}\right)~, \\
    \label{eq:geo_HT_mu}
    \mu(t) &= \frac{\eps_*}{\pi}\sin(\pi t/\tau)^2~.
\end{align}
In the regime $\beta \eps(t) \ll 1$, we observe that minimising entropy production requires a maximal coupling strength~$\eps(\tau)/\pi$. The  entropy production of the geodesic protocol is 
\begin{equation}\label{eq:work_HT}
    k_B T\Sigma_{min} = \frac{\pi\hbar\beta\eps_*}{2\tau} + \mathcal O(\beta^3 \eps_*^3)~,
\end{equation}
which linearly scales with the final energy $\beta\eps_*$.

\subsection{Zero temperature limit ($\beta \eps$ or  $\beta \mu \rightarrow \infty$)}
In the limit of $T=0$\footnote{The zero temperature limit is achieved whenever the energy gaps of the system are too large for thermal fluctuations to occur between the energy levels. Bringing either $\beta\eps$ or $\beta\mu$ to infinity achieves this effect. It is the opposite in the infinite temperature limit, where the thermal fluctuations need to overcome any energy gap, therefore in that limit, both $\beta\eps$ and $\beta\mu$ need to be brought to zero.} we have $f_\beta(\omega)\rightarrow f_\infty(\omega) = \Theta(-\omega)$, where $\Theta$ is the Heaviside step function. Therefore the metric becomes (cf. \aref{sm:LT})
\begin{equation}\label{eq:metric_T0}
    m_{T=0} = \frac{\hbar}{\pi}\frac{1}{\left(\mu^2 + \eps^2\right)^2}
	\begin{pmatrix}
		\mu^2 &
		-\eps\mu \\
		-\eps\mu &
		\eps^2
	\end{pmatrix}~,
\end{equation}
which coincides with the metric of an angle distance in the $(\eps,\mu)$ space -hence the metric is singular.
If we re-parameterize $(\eps,\mu)$ as $(r \cos{\phi},r \sin{\phi})$ we find $k_B T\Sigma = \frac{1}{\pi}\int_0^\tau dt \hspace{1mm} \dot\phi(t)^2$. Therefore any protocol that keeps $\dot{\phi}(t)$ constant is a geodesic, leading to the minimal entropy production: 
\begin{equation}\label{eq:work_LT}
    k_B T\Sigma_{min}\bigg|_{T=0} = \frac{\hbar (\Delta \phi)^2}{\pi\tau},
\end{equation}
with $\phi=\arctan (\mu/\eps)$.
Note that there are multiple (infinitely many) geodesics for any pair of boundary points. 
This fact prevents us from continuing the expansion to further orders in temperature. Nevertheless, this limit provides analytical insights on optimal protocols with $\beta \eps$ or  $\beta \mu \gg 1$. In particular, we note that  there is no need for a diverging coupling even when $\eps(\tau) \rightarrow \infty$ as, once $\mu$ has become large, \eref{eq:work_LT} shows  that it is optimal to reduce the coupling while increasing the energy. Furthermore, \eref{eq:work_LT} shows that at zero temperature, while the reversible cost of the operation goes to zero, the dissipation remains strictly positive. This result is complementary to the findings of Ref.~\cite{Timpanaro2020} which demonstrate a finite-size correction to Landauer's bound that does not disappear in the zero-temperature regime.

\begin{figure*}[ht]
		\centering
        \includegraphics[width=\textwidth]{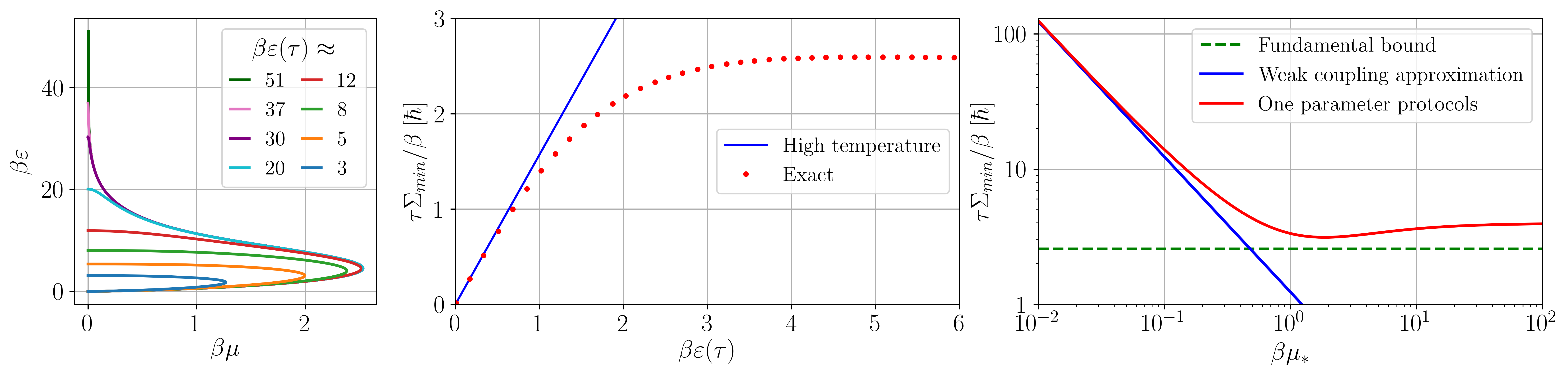}
	    \vspace{-15pt}
	    \caption{(left) A series of optimal protocols depicted in the $(\mu,\eps)$ space. They all start with zero energy and coupling and end with finite energy and zero coupling. In the limit of large $\beta\eps(\tau)$ they can be considered as erasure protocols. (middle) The entropy production of the optimal erasure protocol as a function of the final energy, compared to the high temperature regime cost \eref{eq:work_HT}. (right)  
	    Comparison of the entropy production for a geodesic protocol in which one parameter is varied at a time  (with $\mu$ being increased until $\mu^*$)  and the weak coupling approximation \eref{eq:finitetimeboundweak}; the minimal possible entropy production, $\tau\Sigma_{min}/\beta = 2.57946 \pm 1\cdot 10^{-5}\;[\hbar]$, obtained when both parameters are changed simultaneously is also shown. 
	    }
	    \vspace{-17pt}
		\label{fig:main}
\end{figure*}
\subsection{Weak coupling limit}
Lastly, we take take the weak coupling limit to compare to previous erasure results that are obtained via Lindbladian dynamics, a common assumption in previous works on optimal thermodynamic control in the quantum regime~\cite{Diana2013,scandi19,Zhen2021,VanVu2022,vanvu2023,Zhen2022,Ma2022}.
In this limit the coupling is taken to be small and constant, therefore the metric becomes a scalar (cf. \aref{section:WC}):
\begin{equation}\label{eq:metric_wc}
    m_{weak}(\eps) = \frac{\beta\hbar}{\Gamma}f_\beta(\eps)(1-f_\beta(\eps))~,
\end{equation}
which matches with the results of~\cite{scandi19,Ma2022} which were also obtained with thermodynamic geometry. Indeed, this metric can be obtained from the rate equation
\begin{equation}\label{eq:rate_eq}
    \frac{dp(t)}{dt} = -\Gamma\left(p(t) - \frac{1}{1+e^{\beta\eps(t)}} \right)~,
\end{equation}
which can also be obtained by taking the weak coupling limit in the Heisenberg equations that define \eref{eq:p_exact}. In this regime protocols that minimize dissipated heat at arbitrary speed were found by~\cite{Esposito10}. Therefore we will compare the results one obtains in slow driving and the results of~\cite{Zhen2021,VanVu2022} to the exact minimization of~\cite{Esposito10}.\\ 

We are interested in erasure processes, where $\eps(t)$ is driven from\footnoteref{fn:init_cond} $\eps(0)=0$ to $\eps(\tau)=\eps_*$ with $\eps_* \gg k_B T$ in a time $\tau$. Optimal finite-time protocols are those which minimize the work cost $W = \int_0^\tau dt~ \dot\eps(t)p(t)$, and hence the heat dissipated to the environment $Q= W - \Delta E$. 
The results of~\cite{Esposito10} provide an exact solution to this problem, which is shown in \fref{fig:WC_comparison}. As it is well-known in finite time stochastic thermodynamics~\cite{Schmiedl2007o}, jumps appear in the optimal solution. However, as we approach the quasistatic limit where $\tau \Gamma \gg 1$, the jumps progressively disappear. In \aref{sm:jumps} we prove why the jumps should also disappear in the long times limit at strong coupling. 
As detailed in \aref{section:WC}, and also discussed in previous references~\cite{scandi19}, the optimal driving solution in this limit has the simple analytical form 
\begin{equation} 
\varepsilon(t) = 2\beta^{-1}\ln\tan\!\left[\frac{\pi}{4}(t/\tau+1)\right]~,
\end{equation}
leading to the work cost 
\begin{equation}\label{eq:work_wc}
W = k_B T\left( \ln 2 + \frac{\pi^2}{4\tau \Gamma} \right)~,
\end{equation} 
from where we can directly recover \eref{eq:finitetimeboundweak} through the first law of thermodynamics (note that $\Delta E_S \approx 0$). 
In \fref{fig:WC_comparison} we notice that the exact solution of~\cite{Esposito10} agrees well with this analytical form in the slow driving limit. For completeness, we also show recent results of~\cite{Zhen2021,VanVu2022}. These results apply more generally to any Markovian master equation (here we apply them to the particular case of \eref{eq:rate_eq}), and one can see that they provide a bound to the exact numerical (and approximate analytical) solutions. 

\section{Optimized erasure}
We now focus on erasure outside of any approximation, where we will optimize the driving over both the energy and coupling. In what follows, we focus on minimising $\Sigma$ in an erasure process, which imposes specific boundary conditions to the geodesic equations. We assume that we have no prior knowledge of the system, therefore its initial state is $\rho_S(0) = \mathbb{1}/2$. This translates in taking $\eps(0) = 0$ so that it coincides with the thermal state of $H_S$. For the qubit to be erased we want its final state to be $\rho_S(\tau) \approx \ket{0}\!\!\bra{0}$ (i.e. $p(\tau)\approx0$). Since the driving is done slowly, $p(t)$ is always close to its thermal expectation value. Therefore by choosing $\beta\eps(\tau) \rightarrow \infty$ we ensure $p(\tau) \approx 0$\footnote{Strictly speaking, in order to ensure consistency with the slow driving limit,   $\beta\eps(\tau)$ has to remain finite (so that the speed $\dot{\lambda}$ remain finite). However, the final population $p(\tau)$ is exponentially small with  $\beta \eps(\tau)$, leading to exponentially small corrections.  Our  results are valid up to such corrections, and for sufficiently large~$\tau$ to ensure the validity of the approximation. }. For the coupling, the boundary conditions are $\mu(0) = \mu(\tau) = 0$, because we want to think of this as an ``erasure machine'' that the qubit is ``brought to'' at the start and ``retrieved from'' at the end. Given this family of protocols, we recognise from \eref{eq:WSigma} that $W = k_B T (\ln 2 + \Sigma) $, and similarly $Q=k_B T (\ln 2 + \Sigma)$, thus justifying the minimisation of $\Sigma$ as given in \eref{eq:GeneralExpansionDiss}.
\emph{After} the qubit has been decoupled (i.e. at $t>\tau$), we bring the Hamiltonian of the system back to its starting value ($\eps = 0$) to close the cycle. Since  $p(\tau)\approx 0$, this step requires no work
, and it can be done arbitrarily quickly.

The geodesic equations we obtain for this process are not solvable analytically. The integral of \eref{eq:metric} can be solved to give us an expression of the metric in terms of polygamma functions (cf. \aref{sm:sd}) but it does not simplify the geodesic equations into an analytically solvable form. We therefore turn to numerical tools to obtain the optimal protocol and compute the the dissipated work. Though, in our case, we want to impose the aforementioned boundary conditions; this is known as a Boundary Value Problem (BVP), which is famously hard to solve numerically~\cite{Fox87}. Though we can use the fact that the high temperature limit is accurate at the start of an erasure protocol, therefore the initial conditions of the optimal protocol for erasure will match the initial conditions of~\eref{eq:geo_HT_eps} and \eref{eq:geo_HT_mu}. This allows us to turn the BVP into an Initial Value Problem (cf. \aref{sm:num}) which is much simpler to solve.

In \fref{fig:main} we show  optimal erasure protocols in the $(\mu,\eps)$ space for different final values of $\beta\eps$. We can notice that the predictions of the high and low temperature limit are verified: at the start of the protocols the coupling is increased but once we reach a certain value there is no more need to increase it, regardless of the final value of $\beta\eps$ we try to reach. Interestingly, the maximal value reached by $\beta\mu$ is larger than $1$. 
This shows that reaching the strong coupling is needed to achieve optimal erasure, which is one of the main insights of our work.
In the same figure we also show the value of $\tau \Sigma_{min}/\beta$ for the optimal protocol as a function of the final energy. We can see that for small values of $\beta\eps(\tau)$ \eref{eq:work_HT} gives an accurate description of the work cost, but as we reach higher values it saturates around $\tau\Sigma_{min}/\beta = 2.57946 \pm 1\cdot 10^{-5}\;[\hbar]$. This provides a  finite-time correction to Landauer's principle in this set-up, thus leading to a generalisation of \eref{eq:finitetimeboundweak}:
\begin{equation}\label{eq:finitetimebound_gen}
    Q \geq k_B T \left(\ln 2 + a\frac{\tau_{\rm Pl}}{\tau}\right) + \mathcal{O}\left(\frac{1}{\Gamma^2 \tau^2}\right).
\end{equation}
with $a\approx 2.57946$ and $\tau_{\rm Pl} = \beta\hbar$. This is one of the main results of this work  and can be seen as a generalization of \eref{eq:finitetimeboundweak}. As opposed to the results of~\cite{Zhen2021} and~\cite{VanVu2022}, \eref{eq:finitetimebound_gen} is only valid for large protocol times; yet, it has the advantage of taking into account strong coupling effects (including any possible variation of the coupling strength), having a much simpler form for the correction (which is independent of any chosen relaxation timescale), and we provide an explicit protocol to  achieve it.  By turning around \eref{eq:finitetimebound_gen} one can highlight a quantum speed limit for erasure of a qubit, furthermore this speed limit is of the order of the Planckian time $\tau_{\rm Pl} = \hbar/k_B T$ which is conjectured to be the fastest relaxation timescale for thermalization~\cite{Hartnoll22}. In particular, one can see that \eref{eq:finitetimebound_gen} bounds the speed of erasure by the order of $\tau_{\rm Pl}$ regardless of how large is the coupling strength used in the protocol.

Interestingly, we now argue that the form of the correction \eref{eq:finitetimebound_gen} is in fact general of any Landauer erasure protocol with control on S and the SB coupling.  Indeed, first note that the geodesic length $\mathcal L$ is dimensionless and  can only depend on $\beta$ and the boundary conditions as we optimize over $\mu$ and $\eps$. In an erasure process, the boundary conditions read:  $\eps(0)=0$, $\eps(\tau) \rightarrow \infty$, and $\mu_i(0)=\mu_i(\tau)=0$ where $i$ runs over all the possible control parameters on SB.  But this implies that $\mathcal L$ is independent of them and hence of $\beta$. Therefore $k_BT\Sigma_{min}$ will take the form of a constant, independent of any parameter of the system and bath, divided by $\tau$. This is a crucial difference from~\eref{eq:finitetimeboundweak}. 

This simple argument based on dimensional analysis thus shows that  \eref{eq:finitetimebound_gen} is rather general, with the value of $a$ depending on the specific implementation (e.g. the ohmicity of the bath). It is important to highlight that the bound \eref{eq:finitetimebound_gen}  implies that, even when having access to arbitrary strong SB interactions  (naively taking $\Gamma \rightarrow \infty$ in \eref{eq:finitetimeboundweak}), infinite time is still required for perfect erasure due to the quantum-mechanical correction derived here. 

Finally, we analyze a scenario where the coupling is kept constant while $\eps(t)$ is driven, which is  motivated both by experimental set-ups and for a comparison with the weakly interacting case. 
Therefore, we restrict to \emph{one-parameter protocols} consisting of the three following steps: 1.~while keeping $\eps$ at $0$ we turn on the coupling to some value $\mu_*$; 2.~while keeping the coupling fixed we bring $\eps$ from $0$ to some value $\eps_* \gg k_B T$; 3.~while keeping $\eps$ constant we turn off the coupling. 
Each step contributes positively to the entropy production, and its minimisation is discussed in~\aref{sm:OP}. In \fref{fig:main}, we show $\Sigma_{\rm min}$ for different values of  $\mu_*$, ranging from the weak  to the super-strong coupling regime. It can be appreciated how~\eref{eq:finitetimeboundweak} breaks down, and also how such one-parameter protocols become close to the fundamental limit~\eref{eq:finitetimebound_gen} for $\beta \mu_*>1$.

\section{Conclusions and outlook}

Deriving finite-time corrections to the seminal Landauer bound is a challenging endeavour in stochastic and quantum thermodynamics~\cite{Zulkowski2014,Proesmans2020,Proesmans2020II,boyd2022shortcuts,Lee2022,Diana2013,scandi19,Zhen2021,VanVu2022,vanvu2023,Zhen2022,Ma2022}. Previous works have focused on markovian systems only, which in the quantum regime is obtained through the weak coupling limit~($\beta g^2 \rightarrow 0$). However, should a general finite-time correction exist, it will require the presence of strong coupling at some point during the process as the dissipation generated in finite time is proportional to $g^{-2}$ when $g$ is small\footnote{This can be seen by expanding the finite time dissipation around $g^2 = 0$ for long times: $k_B T\Sigma \propto 1/g^2\tau$, which follows by noticing that the relaxation time-scale is of the order of $g^{-2}$.}. Motivated by this observation, we have developed new insights into the form of optimal protocols for erasure beyond the weak coupling limit.

We have focused on a bit encoded in the occupation of a single fermionic mode, which can be strongly coupled to a reservoir. We have derived analytically the thermodynamic metric, which governs the dissipation rate in the slow driving regime, and showed that it takes a simple form in the high and low temperature limits. From the general form of the metric we obtained the optimal erasure protocol, which requires increasing the coupling strength to $g^2 \sim k_B T$, which corresponds to a relaxation timescale of the order of the Planckian time $\tau_{\rm Pl}$. The corresponding dissipation yields a finite-time correction to Landauer's bound for this setup, which is substantially lower than similar results in the weak coupling regime. Furthermore, by using the obtained bound as a quantum speed limit, this result adds further evidence to the conjecture \cite{Hartnoll22} that $\tau_{\rm Pl}$ is fastest relaxation timescale many-body systems can achieve.

While our results were derived in a fermionic model there are some general insights that follow from our work. First there is a fundamental quantum correction that prevails, see \eref{eq:finitetimebound_gen}, which can  be compared with \eref{eq:finitetimeboundweak} derived in the weak coupling regime. While the specific value of $a$ in \eref{eq:finitetimebound_gen} will depend on the specific setup, it will never approach $0$ (even for diverging system-bath coupling) due to the inherent cost of changing the interaction strength. Furthermore, to obtain these results we adapted the framework of thermodynamic geometry to system-bath unitary dynamics in which the coupling can be arbitrarily large or small. This is in contrast to recent claims of failure of this approach in closed quantum systems~\cite{soriani2022failure}. Finally, as was argued before, our results make evident the need of strong coupling in a general finite-time correction to Landauer's principle.


This work opens exciting directions for the future. On the one hand, the level of experimental control required to implement such protocols is in principle possible in quantum dots~\cite{Ciorga2000Addition,Elzerman2003Few,Koski2014,Koski2014Exp}, where   the energy-level $\eps(t)$ and coupling $g(t)$ can be independently controlled, even by several orders of magnitude~\cite{Rochette2019}.  On the other hand, it would be interesting to characterise the dependence of $a$ in the nature of the bath and the SB coupling (e.g. its spectral density),  more generally  to derive similar quantum-mechanical finite-time corrections that are independent of the specific implementation, and to gain further insights in the connection between Landauer erasure and the Planckian time.

\emph{Acknowledgments.} We warmly thank Mark T. Mitchinson, Ludovico Machet, and Gabriel Landi for insightful discussions. We also sincerely  thank Luca V. Delacretaz for introducing us to the concept of Planckian time. This work was supported by the the Swiss National Science Foundation through an Ambizione Grant No. PZ00P2-186067.




\bibliographystyle{apsrev4-1}
\bibliography{mybib.bib,bibPaolo.bib}

\begin{thebibliography}{121}%
\makeatletter
\providecommand \@ifxundefined [1]{%
 \@ifx{#1\undefined}
}%
\providecommand \@ifnum [1]{%
 \ifnum #1\expandafter \@firstoftwo
 \else \expandafter \@secondoftwo
 \fi
}%
\providecommand \@ifx [1]{%
 \ifx #1\expandafter \@firstoftwo
 \else \expandafter \@secondoftwo
 \fi
}%
\providecommand \natexlab [1]{#1}%
\providecommand \enquote  [1]{``#1''}%
\providecommand \bibnamefont  [1]{#1}%
\providecommand \bibfnamefont [1]{#1}%
\providecommand \citenamefont [1]{#1}%
\providecommand \href@noop [0]{\@secondoftwo}%
\providecommand \href [0]{\begingroup \@sanitize@url \@href}%
\providecommand \@href[1]{\@@startlink{#1}\@@href}%
\providecommand \@@href[1]{\endgroup#1\@@endlink}%
\providecommand \@sanitize@url [0]{\catcode `\\12\catcode `\$12\catcode
  `\&12\catcode `\#12\catcode `\^12\catcode `\_12\catcode `\%12\relax}%
\providecommand \@@startlink[1]{}%
\providecommand \@@endlink[0]{}%
\providecommand \url  [0]{\begingroup\@sanitize@url \@url }%
\providecommand \@url [1]{\endgroup\@href {#1}{\urlprefix }}%
\providecommand \urlprefix  [0]{URL }%
\providecommand \Eprint [0]{\href }%
\providecommand \doibase [0]{http://dx.doi.org/}%
\providecommand \selectlanguage [0]{\@gobble}%
\providecommand \bibinfo  [0]{\@secondoftwo}%
\providecommand \bibfield  [0]{\@secondoftwo}%
\providecommand \translation [1]{[#1]}%
\providecommand \BibitemOpen [0]{}%
\providecommand \bibitemStop [0]{}%
\providecommand \bibitemNoStop [0]{.\EOS\space}%
\providecommand \EOS [0]{\spacefactor3000\relax}%
\providecommand \BibitemShut  [1]{\csname bibitem#1\endcsname}%
\let\auto@bib@innerbib\@empty
\bibitem [{\citenamefont {Landauer}(1961)}]{landauer1961irreversibility}%
  \BibitemOpen
  \bibfield  {author} {\bibinfo {author} {\bibfnamefont {R.}~\bibnamefont
  {Landauer}},\ }\href {\doibase https://doi.org/10.1147/rd.53.0183} {\bibfield
   {journal} {\bibinfo  {journal} {IBM Journal of Research and Development}\
  }\textbf {\bibinfo {volume} {5}},\ \bibinfo {pages} {183} (\bibinfo {year}
  {1961})}\BibitemShut {NoStop}%
\bibitem [{\citenamefont {Sagawa}\ and\ \citenamefont
  {Ueda}(2009)}]{Sagawa2009}%
  \BibitemOpen
  \bibfield  {author} {\bibinfo {author} {\bibfnamefont {T.}~\bibnamefont
  {Sagawa}}\ and\ \bibinfo {author} {\bibfnamefont {M.}~\bibnamefont {Ueda}},\
  }\href {\doibase 10.1103/PhysRevLett.102.250602} {\bibfield  {journal}
  {\bibinfo  {journal} {Phys. Rev. Lett.}\ }\textbf {\bibinfo {volume} {102}},\
  \bibinfo {pages} {250602} (\bibinfo {year} {2009})}\BibitemShut {NoStop}%
\bibitem [{\citenamefont {Esposito}\ and\ \citenamefont {den
  Broeck}(2011)}]{Esposito2011}%
  \BibitemOpen
  \bibfield  {author} {\bibinfo {author} {\bibfnamefont {M.}~\bibnamefont
  {Esposito}}\ and\ \bibinfo {author} {\bibfnamefont {C.~V.}\ \bibnamefont {den
  Broeck}},\ }\href {\doibase 10.1209/0295-5075/95/40004} {\bibfield  {journal}
  {\bibinfo  {journal} {{EPL} (Europhysics Letters)}\ }\textbf {\bibinfo
  {volume} {95}},\ \bibinfo {pages} {40004} (\bibinfo {year}
  {2011})}\BibitemShut {NoStop}%
\bibitem [{\citenamefont {Hilt}\ \emph {et~al.}(2011)\citenamefont {Hilt},
  \citenamefont {Shabbir}, \citenamefont {Anders},\ and\ \citenamefont
  {Lutz}}]{Hilt2011}%
  \BibitemOpen
  \bibfield  {author} {\bibinfo {author} {\bibfnamefont {S.}~\bibnamefont
  {Hilt}}, \bibinfo {author} {\bibfnamefont {S.}~\bibnamefont {Shabbir}},
  \bibinfo {author} {\bibfnamefont {J.}~\bibnamefont {Anders}}, \ and\ \bibinfo
  {author} {\bibfnamefont {E.}~\bibnamefont {Lutz}},\ }\href {\doibase
  10.1103/PhysRevE.83.030102} {\bibfield  {journal} {\bibinfo  {journal} {Phys.
  Rev. E}\ }\textbf {\bibinfo {volume} {83}},\ \bibinfo {pages} {030102}
  (\bibinfo {year} {2011})}\BibitemShut {NoStop}%
\bibitem [{\citenamefont {Deffner}\ and\ \citenamefont
  {Jarzynski}(2013)}]{Deffner2013info}%
  \BibitemOpen
  \bibfield  {author} {\bibinfo {author} {\bibfnamefont {S.}~\bibnamefont
  {Deffner}}\ and\ \bibinfo {author} {\bibfnamefont {C.}~\bibnamefont
  {Jarzynski}},\ }\href {\doibase 10.1103/PhysRevX.3.041003} {\bibfield
  {journal} {\bibinfo  {journal} {Phys. Rev. X}\ }\textbf {\bibinfo {volume}
  {3}},\ \bibinfo {pages} {041003} (\bibinfo {year} {2013})}\BibitemShut
  {NoStop}%
\bibitem [{\citenamefont {Reeb}\ and\ \citenamefont {Wolf}(2014)}]{Reeb2014}%
  \BibitemOpen
  \bibfield  {author} {\bibinfo {author} {\bibfnamefont {D.}~\bibnamefont
  {Reeb}}\ and\ \bibinfo {author} {\bibfnamefont {M.~M.}\ \bibnamefont
  {Wolf}},\ }\href {\doibase 10.1088/1367-2630/16/10/103011} {\bibfield
  {journal} {\bibinfo  {journal} {New Journal of Physics}\ }\textbf {\bibinfo
  {volume} {16}},\ \bibinfo {pages} {103011} (\bibinfo {year}
  {2014})}\BibitemShut {NoStop}%
\bibitem [{\citenamefont {Faist}\ \emph {et~al.}(2015)\citenamefont {Faist},
  \citenamefont {Dupuis}, \citenamefont {Oppenheim},\ and\ \citenamefont
  {Renner}}]{Faist2015}%
  \BibitemOpen
  \bibfield  {author} {\bibinfo {author} {\bibfnamefont {P.}~\bibnamefont
  {Faist}}, \bibinfo {author} {\bibfnamefont {F.}~\bibnamefont {Dupuis}},
  \bibinfo {author} {\bibfnamefont {J.}~\bibnamefont {Oppenheim}}, \ and\
  \bibinfo {author} {\bibfnamefont {R.}~\bibnamefont {Renner}},\ }\href
  {\doibase 10.1038/ncomms8669} {\bibfield  {journal} {\bibinfo  {journal}
  {Nature Communications}\ }\textbf {\bibinfo {volume} {6}} (\bibinfo {year}
  {2015}),\ 10.1038/ncomms8669}\BibitemShut {NoStop}%
\bibitem [{\citenamefont {Lorenzo}\ \emph {et~al.}(2015)\citenamefont
  {Lorenzo}, \citenamefont {McCloskey}, \citenamefont {Ciccarello},
  \citenamefont {Paternostro},\ and\ \citenamefont {Palma}}]{Lorenzo2015}%
  \BibitemOpen
  \bibfield  {author} {\bibinfo {author} {\bibfnamefont {S.}~\bibnamefont
  {Lorenzo}}, \bibinfo {author} {\bibfnamefont {R.}~\bibnamefont {McCloskey}},
  \bibinfo {author} {\bibfnamefont {F.}~\bibnamefont {Ciccarello}}, \bibinfo
  {author} {\bibfnamefont {M.}~\bibnamefont {Paternostro}}, \ and\ \bibinfo
  {author} {\bibfnamefont {G.~M.}\ \bibnamefont {Palma}},\ }\href {\doibase
  10.1103/PhysRevLett.115.120403} {\bibfield  {journal} {\bibinfo  {journal}
  {Phys. Rev. Lett.}\ }\textbf {\bibinfo {volume} {115}},\ \bibinfo {pages}
  {120403} (\bibinfo {year} {2015})}\BibitemShut {NoStop}%
\bibitem [{\citenamefont {Goold}\ \emph {et~al.}(2015)\citenamefont {Goold},
  \citenamefont {Paternostro},\ and\ \citenamefont {Modi}}]{Goold2015}%
  \BibitemOpen
  \bibfield  {author} {\bibinfo {author} {\bibfnamefont {J.}~\bibnamefont
  {Goold}}, \bibinfo {author} {\bibfnamefont {M.}~\bibnamefont {Paternostro}},
  \ and\ \bibinfo {author} {\bibfnamefont {K.}~\bibnamefont {Modi}},\ }\href
  {\doibase 10.1103/PhysRevLett.114.060602} {\bibfield  {journal} {\bibinfo
  {journal} {Phys. Rev. Lett.}\ }\textbf {\bibinfo {volume} {114}},\ \bibinfo
  {pages} {060602} (\bibinfo {year} {2015})}\BibitemShut {NoStop}%
\bibitem [{\citenamefont {Alhambra}\ \emph {et~al.}(2016)\citenamefont
  {Alhambra}, \citenamefont {Masanes}, \citenamefont {Oppenheim},\ and\
  \citenamefont {Perry}}]{Alhambra2016}%
  \BibitemOpen
  \bibfield  {author} {\bibinfo {author} {\bibfnamefont {A.~M.}\ \bibnamefont
  {Alhambra}}, \bibinfo {author} {\bibfnamefont {L.}~\bibnamefont {Masanes}},
  \bibinfo {author} {\bibfnamefont {J.}~\bibnamefont {Oppenheim}}, \ and\
  \bibinfo {author} {\bibfnamefont {C.}~\bibnamefont {Perry}},\ }\href
  {\doibase 10.1103/PhysRevX.6.041017} {\bibfield  {journal} {\bibinfo
  {journal} {Phys. Rev. X}\ }\textbf {\bibinfo {volume} {6}},\ \bibinfo {pages}
  {041017} (\bibinfo {year} {2016})}\BibitemShut {NoStop}%
\bibitem [{\citenamefont {Guarnieri}\ \emph {et~al.}(2017)\citenamefont
  {Guarnieri}, \citenamefont {Campbell}, \citenamefont {Goold}, \citenamefont
  {Pigeon}, \citenamefont {Vacchini},\ and\ \citenamefont
  {Paternostro}}]{Guarnieri2017}%
  \BibitemOpen
  \bibfield  {author} {\bibinfo {author} {\bibfnamefont {G.}~\bibnamefont
  {Guarnieri}}, \bibinfo {author} {\bibfnamefont {S.}~\bibnamefont {Campbell}},
  \bibinfo {author} {\bibfnamefont {J.}~\bibnamefont {Goold}}, \bibinfo
  {author} {\bibfnamefont {S.}~\bibnamefont {Pigeon}}, \bibinfo {author}
  {\bibfnamefont {B.}~\bibnamefont {Vacchini}}, \ and\ \bibinfo {author}
  {\bibfnamefont {M.}~\bibnamefont {Paternostro}},\ }\href {\doibase
  10.1088/1367-2630/aa8cf1} {\bibfield  {journal} {\bibinfo  {journal} {New
  Journal of Physics}\ }\textbf {\bibinfo {volume} {19}},\ \bibinfo {pages}
  {103038} (\bibinfo {year} {2017})}\BibitemShut {NoStop}%
\bibitem [{\citenamefont {Miller}\ \emph {et~al.}(2020)\citenamefont {Miller},
  \citenamefont {Guarnieri}, \citenamefont {Mitchison},\ and\ \citenamefont
  {Goold}}]{Miller2020}%
  \BibitemOpen
  \bibfield  {author} {\bibinfo {author} {\bibfnamefont {H.~J.}\ \bibnamefont
  {Miller}}, \bibinfo {author} {\bibfnamefont {G.}~\bibnamefont {Guarnieri}},
  \bibinfo {author} {\bibfnamefont {M.~T.}\ \bibnamefont {Mitchison}}, \ and\
  \bibinfo {author} {\bibfnamefont {J.}~\bibnamefont {Goold}},\ }\href
  {\doibase 10.1103/physrevlett.125.160602} {\bibfield  {journal} {\bibinfo
  {journal} {Physical Review Letters}\ }\textbf {\bibinfo {volume} {125}}
  (\bibinfo {year} {2020}),\ 10.1103/physrevlett.125.160602}\BibitemShut
  {NoStop}%
\bibitem [{\citenamefont {Timpanaro}\ \emph {et~al.}(2020)\citenamefont
  {Timpanaro}, \citenamefont {Santos},\ and\ \citenamefont
  {Landi}}]{Timpanaro2020}%
  \BibitemOpen
  \bibfield  {author} {\bibinfo {author} {\bibfnamefont {A.~M.}\ \bibnamefont
  {Timpanaro}}, \bibinfo {author} {\bibfnamefont {J.~P.}\ \bibnamefont
  {Santos}}, \ and\ \bibinfo {author} {\bibfnamefont {G.~T.}\ \bibnamefont
  {Landi}},\ }\href {\doibase 10.1103/PhysRevLett.124.240601} {\bibfield
  {journal} {\bibinfo  {journal} {Phys. Rev. Lett.}\ }\textbf {\bibinfo
  {volume} {124}},\ \bibinfo {pages} {240601} (\bibinfo {year}
  {2020})}\BibitemShut {NoStop}%
\bibitem [{\citenamefont {Riechers}\ and\ \citenamefont
  {Gu}(2021)}]{Riechers2021}%
  \BibitemOpen
  \bibfield  {author} {\bibinfo {author} {\bibfnamefont {P.~M.}\ \bibnamefont
  {Riechers}}\ and\ \bibinfo {author} {\bibfnamefont {M.}~\bibnamefont {Gu}},\
  }\href {\doibase 10.1103/PhysRevA.104.012214} {\bibfield  {journal} {\bibinfo
   {journal} {Phys. Rev. A}\ }\textbf {\bibinfo {volume} {104}},\ \bibinfo
  {pages} {012214} (\bibinfo {year} {2021})}\BibitemShut {NoStop}%
\bibitem [{\citenamefont {Buffoni}\ and\ \citenamefont
  {Campisi}(2022)}]{Buffoni2022}%
  \BibitemOpen
  \bibfield  {author} {\bibinfo {author} {\bibfnamefont {L.}~\bibnamefont
  {Buffoni}}\ and\ \bibinfo {author} {\bibfnamefont {M.}~\bibnamefont
  {Campisi}},\ }\href {\doibase 10.1007/s10955-022-02877-8} {\bibfield
  {journal} {\bibinfo  {journal} {Journal of Statistical Physics}\ }\textbf
  {\bibinfo {volume} {186}} (\bibinfo {year} {2022}),\
  10.1007/s10955-022-02877-8}\BibitemShut {NoStop}%
\bibitem [{\citenamefont {Taranto}\ \emph {et~al.}(2023)\citenamefont
  {Taranto}, \citenamefont {Bakhshinezhad}, \citenamefont {Bluhm},
  \citenamefont {Silva}, \citenamefont {Friis}, \citenamefont {Lock},
  \citenamefont {Vitagliano}, \citenamefont {Binder}, \citenamefont {Debarba},
  \citenamefont {Schwarzhans}, \citenamefont {Clivaz},\ and\ \citenamefont
  {Huber}}]{taranto2021landauer}%
  \BibitemOpen
  \bibfield  {author} {\bibinfo {author} {\bibfnamefont {P.}~\bibnamefont
  {Taranto}}, \bibinfo {author} {\bibfnamefont {F.}~\bibnamefont
  {Bakhshinezhad}}, \bibinfo {author} {\bibfnamefont {A.}~\bibnamefont
  {Bluhm}}, \bibinfo {author} {\bibfnamefont {R.}~\bibnamefont {Silva}},
  \bibinfo {author} {\bibfnamefont {N.}~\bibnamefont {Friis}}, \bibinfo
  {author} {\bibfnamefont {M.~P.}\ \bibnamefont {Lock}}, \bibinfo {author}
  {\bibfnamefont {G.}~\bibnamefont {Vitagliano}}, \bibinfo {author}
  {\bibfnamefont {F.~C.}\ \bibnamefont {Binder}}, \bibinfo {author}
  {\bibfnamefont {T.}~\bibnamefont {Debarba}}, \bibinfo {author} {\bibfnamefont
  {E.}~\bibnamefont {Schwarzhans}}, \bibinfo {author} {\bibfnamefont
  {F.}~\bibnamefont {Clivaz}}, \ and\ \bibinfo {author} {\bibfnamefont
  {M.}~\bibnamefont {Huber}},\ }\href {\doibase 10.1103/PRXQuantum.4.010332}
  {\bibfield  {journal} {\bibinfo  {journal} {PRX Quantum}\ }\textbf {\bibinfo
  {volume} {4}},\ \bibinfo {pages} {010332} (\bibinfo {year}
  {2023})}\BibitemShut {NoStop}%
\bibitem [{\citenamefont {Browne}\ \emph {et~al.}(2014)\citenamefont {Browne},
  \citenamefont {Garner}, \citenamefont {Dahlsten},\ and\ \citenamefont
  {Vedral}}]{browne2014guaranteed}%
  \BibitemOpen
  \bibfield  {author} {\bibinfo {author} {\bibfnamefont {C.}~\bibnamefont
  {Browne}}, \bibinfo {author} {\bibfnamefont {A.~J.~P.}\ \bibnamefont
  {Garner}}, \bibinfo {author} {\bibfnamefont {O.~C.~O.}\ \bibnamefont
  {Dahlsten}}, \ and\ \bibinfo {author} {\bibfnamefont {V.}~\bibnamefont
  {Vedral}},\ }\href {\doibase 10.1103/PhysRevLett.113.100603} {\bibfield
  {journal} {\bibinfo  {journal} {Phys. Rev. Lett.}\ }\textbf {\bibinfo
  {volume} {113}},\ \bibinfo {pages} {100603} (\bibinfo {year}
  {2014})}\BibitemShut {NoStop}%
\bibitem [{\citenamefont {B{\'e}rut}\ \emph {et~al.}(2012)\citenamefont
  {B{\'e}rut}, \citenamefont {Arakelyan}, \citenamefont {Petrosyan},
  \citenamefont {Ciliberto}, \citenamefont {Dillenschneider},\ and\
  \citenamefont {Lutz}}]{berut2012experimental}%
  \BibitemOpen
  \bibfield  {author} {\bibinfo {author} {\bibfnamefont {A.}~\bibnamefont
  {B{\'e}rut}}, \bibinfo {author} {\bibfnamefont {A.}~\bibnamefont
  {Arakelyan}}, \bibinfo {author} {\bibfnamefont {A.}~\bibnamefont
  {Petrosyan}}, \bibinfo {author} {\bibfnamefont {S.}~\bibnamefont
  {Ciliberto}}, \bibinfo {author} {\bibfnamefont {R.}~\bibnamefont
  {Dillenschneider}}, \ and\ \bibinfo {author} {\bibfnamefont {E.}~\bibnamefont
  {Lutz}},\ }\href {\doibase https://doi.org/10.1088/1742-5468/2015/06/P06015}
  {\bibfield  {journal} {\bibinfo  {journal} {Nature}\ }\textbf {\bibinfo
  {volume} {483}},\ \bibinfo {pages} {187} (\bibinfo {year}
  {2012})}\BibitemShut {NoStop}%
\bibitem [{\citenamefont {Jun}\ \emph {et~al.}(2014)\citenamefont {Jun},
  \citenamefont {Gavrilov},\ and\ \citenamefont {Bechhoefer}}]{jun2014high}%
  \BibitemOpen
  \bibfield  {author} {\bibinfo {author} {\bibfnamefont {Y.}~\bibnamefont
  {Jun}}, \bibinfo {author} {\bibfnamefont {M.}~\bibnamefont {Gavrilov}}, \
  and\ \bibinfo {author} {\bibfnamefont {J.}~\bibnamefont {Bechhoefer}},\
  }\href {\doibase https://doi.org/10.1103/PhysRevLett.113.190601} {\bibfield
  {journal} {\bibinfo  {journal} {Phys. Rev. Lett.}\ }\textbf {\bibinfo
  {volume} {113}},\ \bibinfo {pages} {190601} (\bibinfo {year}
  {2014})}\BibitemShut {NoStop}%
\bibitem [{\citenamefont {B{\'{e}}rut}\ \emph {et~al.}(2015)\citenamefont
  {B{\'{e}}rut}, \citenamefont {Petrosyan},\ and\ \citenamefont
  {Ciliberto}}]{Brut2015}%
  \BibitemOpen
  \bibfield  {author} {\bibinfo {author} {\bibfnamefont {A.}~\bibnamefont
  {B{\'{e}}rut}}, \bibinfo {author} {\bibfnamefont {A.}~\bibnamefont
  {Petrosyan}}, \ and\ \bibinfo {author} {\bibfnamefont {S.}~\bibnamefont
  {Ciliberto}},\ }\href {\doibase 10.1088/1742-5468/2015/06/p06015} {\bibfield
  {journal} {\bibinfo  {journal} {J. Stat. Mech.}\ }\textbf {\bibinfo {volume}
  {2015}},\ \bibinfo {pages} {P06015} (\bibinfo {year} {2015})}\BibitemShut
  {NoStop}%
\bibitem [{\citenamefont {Gavrilov}\ and\ \citenamefont
  {Bechhoefer}(2016)}]{gavrilov_erasure_2016}%
  \BibitemOpen
  \bibfield  {author} {\bibinfo {author} {\bibfnamefont {M.}~\bibnamefont
  {Gavrilov}}\ and\ \bibinfo {author} {\bibfnamefont {J.}~\bibnamefont
  {Bechhoefer}},\ }\href {\doibase 10.1103/PhysRevLett.117.200601} {\bibfield
  {journal} {\bibinfo  {journal} {Phys. Rev. Lett.}\ }\textbf {\bibinfo
  {volume} {117}},\ \bibinfo {pages} {200601} (\bibinfo {year}
  {2016})}\BibitemShut {NoStop}%
\bibitem [{\citenamefont {Hong}\ \emph {et~al.}(2016)\citenamefont {Hong},
  \citenamefont {Lambson}, \citenamefont {Dhuey},\ and\ \citenamefont
  {Bokor}}]{Hong2016}%
  \BibitemOpen
  \bibfield  {author} {\bibinfo {author} {\bibfnamefont {J.}~\bibnamefont
  {Hong}}, \bibinfo {author} {\bibfnamefont {B.}~\bibnamefont {Lambson}},
  \bibinfo {author} {\bibfnamefont {S.}~\bibnamefont {Dhuey}}, \ and\ \bibinfo
  {author} {\bibfnamefont {J.}~\bibnamefont {Bokor}},\ }\href {\doibase
  10.1126/sciadv.1501492} {\bibfield  {journal} {\bibinfo  {journal} {Sci.
  Adv.}\ }\textbf {\bibinfo {volume} {2}} (\bibinfo {year} {2016}),\
  10.1126/sciadv.1501492}\BibitemShut {NoStop}%
\bibitem [{\citenamefont {Martini}\ \emph {et~al.}(2016)\citenamefont
  {Martini}, \citenamefont {Pancaldi}, \citenamefont {Madami}, \citenamefont
  {Vavassori}, \citenamefont {Gubbiotti}, \citenamefont {Tacchi}, \citenamefont
  {Hartmann}, \citenamefont {Emmerling}, \citenamefont {H\"{o}fling},
  \citenamefont {Worschech},\ and\ \citenamefont {Carlotti}}]{Martini2016}%
  \BibitemOpen
  \bibfield  {author} {\bibinfo {author} {\bibfnamefont {L.}~\bibnamefont
  {Martini}}, \bibinfo {author} {\bibfnamefont {M.}~\bibnamefont {Pancaldi}},
  \bibinfo {author} {\bibfnamefont {M.}~\bibnamefont {Madami}}, \bibinfo
  {author} {\bibfnamefont {P.}~\bibnamefont {Vavassori}}, \bibinfo {author}
  {\bibfnamefont {G.}~\bibnamefont {Gubbiotti}}, \bibinfo {author}
  {\bibfnamefont {S.}~\bibnamefont {Tacchi}}, \bibinfo {author} {\bibfnamefont
  {F.}~\bibnamefont {Hartmann}}, \bibinfo {author} {\bibfnamefont
  {M.}~\bibnamefont {Emmerling}}, \bibinfo {author} {\bibfnamefont
  {S.}~\bibnamefont {H\"{o}fling}}, \bibinfo {author} {\bibfnamefont
  {L.}~\bibnamefont {Worschech}}, \ and\ \bibinfo {author} {\bibfnamefont
  {G.}~\bibnamefont {Carlotti}},\ }\href {\doibase
  10.1016/j.nanoen.2015.10.028} {\bibfield  {journal} {\bibinfo  {journal}
  {Nano Energy}\ }\textbf {\bibinfo {volume} {19}},\ \bibinfo {pages} {108}
  (\bibinfo {year} {2016})}\BibitemShut {NoStop}%
\bibitem [{\citenamefont {Gaudenzi}\ \emph {et~al.}(2018)\citenamefont
  {Gaudenzi}, \citenamefont {Burzur{\'{\i}}}, \citenamefont {Maegawa},
  \citenamefont {van~der Zant},\ and\ \citenamefont {Luis}}]{Gaudenzi2018}%
  \BibitemOpen
  \bibfield  {author} {\bibinfo {author} {\bibfnamefont {R.}~\bibnamefont
  {Gaudenzi}}, \bibinfo {author} {\bibfnamefont {E.}~\bibnamefont
  {Burzur{\'{\i}}}}, \bibinfo {author} {\bibfnamefont {S.}~\bibnamefont
  {Maegawa}}, \bibinfo {author} {\bibfnamefont {H.~S.~J.}\ \bibnamefont
  {van~der Zant}}, \ and\ \bibinfo {author} {\bibfnamefont {F.}~\bibnamefont
  {Luis}},\ }\href {\doibase 10.1038/s41567-018-0070-7} {\bibfield  {journal}
  {\bibinfo  {journal} {Nat. Phys.}\ }\textbf {\bibinfo {volume} {14}},\
  \bibinfo {pages} {565} (\bibinfo {year} {2018})}\BibitemShut {NoStop}%
\bibitem [{\citenamefont {Saira}\ \emph {et~al.}(2020)\citenamefont {Saira},
  \citenamefont {Matheny}, \citenamefont {Katti}, \citenamefont {Fon},
  \citenamefont {Wimsatt}, \citenamefont {Crutchfield}, \citenamefont {Han},\
  and\ \citenamefont {Roukes}}]{saira_nonequilibrium_2020}%
  \BibitemOpen
  \bibfield  {author} {\bibinfo {author} {\bibfnamefont {O.}~\bibnamefont
  {Saira}}, \bibinfo {author} {\bibfnamefont {M.~H.}\ \bibnamefont {Matheny}},
  \bibinfo {author} {\bibfnamefont {R.}~\bibnamefont {Katti}}, \bibinfo
  {author} {\bibfnamefont {W.}~\bibnamefont {Fon}}, \bibinfo {author}
  {\bibfnamefont {G.}~\bibnamefont {Wimsatt}}, \bibinfo {author} {\bibfnamefont
  {J.~P.}\ \bibnamefont {Crutchfield}}, \bibinfo {author} {\bibfnamefont
  {S.}~\bibnamefont {Han}}, \ and\ \bibinfo {author} {\bibfnamefont {M.~L.}\
  \bibnamefont {Roukes}},\ }\href {\doibase 10.1103/PhysRevResearch.2.013249}
  {\bibfield  {journal} {\bibinfo  {journal} {Phys. Rev. Res.}\ }\textbf
  {\bibinfo {volume} {2}},\ \bibinfo {pages} {013249} (\bibinfo {year}
  {2020})}\BibitemShut {NoStop}%
\bibitem [{\citenamefont {Dago}\ \emph {et~al.}(2021)\citenamefont {Dago},
  \citenamefont {Pereda}, \citenamefont {Barros}, \citenamefont {Ciliberto},\
  and\ \citenamefont {Bellon}}]{Dago2021}%
  \BibitemOpen
  \bibfield  {author} {\bibinfo {author} {\bibfnamefont {S.}~\bibnamefont
  {Dago}}, \bibinfo {author} {\bibfnamefont {J.}~\bibnamefont {Pereda}},
  \bibinfo {author} {\bibfnamefont {N.}~\bibnamefont {Barros}}, \bibinfo
  {author} {\bibfnamefont {S.}~\bibnamefont {Ciliberto}}, \ and\ \bibinfo
  {author} {\bibfnamefont {L.}~\bibnamefont {Bellon}},\ }\href {\doibase
  10.1103/PhysRevLett.126.170601} {\bibfield  {journal} {\bibinfo  {journal}
  {Phys. Rev. Lett.}\ }\textbf {\bibinfo {volume} {126}},\ \bibinfo {pages}
  {170601} (\bibinfo {year} {2021})}\BibitemShut {NoStop}%
\bibitem [{\citenamefont {Dago}\ and\ \citenamefont
  {Bellon}(2022)}]{dago_dynamics_2022}%
  \BibitemOpen
  \bibfield  {author} {\bibinfo {author} {\bibfnamefont {S.}~\bibnamefont
  {Dago}}\ and\ \bibinfo {author} {\bibfnamefont {L.}~\bibnamefont {Bellon}},\
  }\href {\doibase 10.1103/PhysRevLett.128.070604} {\bibfield  {journal}
  {\bibinfo  {journal} {Phys. Rev. Lett.}\ }\textbf {\bibinfo {volume} {128}},\
  \bibinfo {pages} {070604} (\bibinfo {year} {2022})}\BibitemShut {NoStop}%
\bibitem [{\citenamefont {Ciampini}\ \emph {et~al.}(2021)\citenamefont
  {Ciampini}, \citenamefont {Wenzl}, \citenamefont {Konopik}, \citenamefont
  {Thalhammer}, \citenamefont {Aspelmeyer}, \citenamefont {Lutz},\ and\
  \citenamefont {Kiesel}}]{ciampini2021experimental}%
  \BibitemOpen
  \bibfield  {author} {\bibinfo {author} {\bibfnamefont {M.~A.}\ \bibnamefont
  {Ciampini}}, \bibinfo {author} {\bibfnamefont {T.}~\bibnamefont {Wenzl}},
  \bibinfo {author} {\bibfnamefont {M.}~\bibnamefont {Konopik}}, \bibinfo
  {author} {\bibfnamefont {G.}~\bibnamefont {Thalhammer}}, \bibinfo {author}
  {\bibfnamefont {M.}~\bibnamefont {Aspelmeyer}}, \bibinfo {author}
  {\bibfnamefont {E.}~\bibnamefont {Lutz}}, \ and\ \bibinfo {author}
  {\bibfnamefont {N.}~\bibnamefont {Kiesel}},\ }\href
  {https://doi.org/10.48550/arXiv.2107.04429} {\bibfield  {journal} {\bibinfo
  {journal} {arXiv:2107.04429}\ } (\bibinfo {year} {2021})}\BibitemShut
  {NoStop}%
\bibitem [{\citenamefont {Scandi}\ \emph {et~al.}(2022)\citenamefont {Scandi},
  \citenamefont {Barker}, \citenamefont {Lehmann}, \citenamefont {Dick},
  \citenamefont {Maisi},\ and\ \citenamefont
  {Perarnau-Llobet}}]{scandi2022constant}%
  \BibitemOpen
  \bibfield  {author} {\bibinfo {author} {\bibfnamefont {M.}~\bibnamefont
  {Scandi}}, \bibinfo {author} {\bibfnamefont {D.}~\bibnamefont {Barker}},
  \bibinfo {author} {\bibfnamefont {S.}~\bibnamefont {Lehmann}}, \bibinfo
  {author} {\bibfnamefont {K.~A.}\ \bibnamefont {Dick}}, \bibinfo {author}
  {\bibfnamefont {V.~F.}\ \bibnamefont {Maisi}}, \ and\ \bibinfo {author}
  {\bibfnamefont {M.}~\bibnamefont {Perarnau-Llobet}},\ }\href {\doibase
  10.1103/PhysRevLett.129.270601} {\bibfield  {journal} {\bibinfo  {journal}
  {Phys. Rev. Lett.}\ }\textbf {\bibinfo {volume} {129}},\ \bibinfo {pages}
  {270601} (\bibinfo {year} {2022})}\BibitemShut {NoStop}%
\bibitem [{\citenamefont {Nernst}(1912)}]{Nernst12}%
  \BibitemOpen
  \bibfield  {author} {\bibinfo {author} {\bibnamefont {Nernst}},\ }\href
  {\doibase missing} {\bibfield  {journal} {\bibinfo  {journal} {W. Sitzber.
  Kgl. Preuss. Akad. Wiss. Physik-Math. Kl. 134}\ } (\bibinfo {year} {1912}),\
  missing}\BibitemShut {NoStop}%
\bibitem [{\citenamefont {Masanes}\ and\ \citenamefont
  {Oppenheim}(2017)}]{Masanes2017}%
  \BibitemOpen
  \bibfield  {author} {\bibinfo {author} {\bibfnamefont {L.}~\bibnamefont
  {Masanes}}\ and\ \bibinfo {author} {\bibfnamefont {J.}~\bibnamefont
  {Oppenheim}},\ }\href {\doibase 10.1038/ncomms14538} {\bibfield  {journal}
  {\bibinfo  {journal} {Nature Communications}\ }\textbf {\bibinfo {volume}
  {8}} (\bibinfo {year} {2017}),\ 10.1038/ncomms14538}\BibitemShut {NoStop}%
\bibitem [{\citenamefont {Freitas}\ \emph {et~al.}(2018)\citenamefont
  {Freitas}, \citenamefont {Gallego}, \citenamefont {Masanes},\ and\
  \citenamefont {Paz}}]{Freitas2018}%
  \BibitemOpen
  \bibfield  {author} {\bibinfo {author} {\bibfnamefont {N.}~\bibnamefont
  {Freitas}}, \bibinfo {author} {\bibfnamefont {R.}~\bibnamefont {Gallego}},
  \bibinfo {author} {\bibfnamefont {L.}~\bibnamefont {Masanes}}, \ and\
  \bibinfo {author} {\bibfnamefont {J.~P.}\ \bibnamefont {Paz}},\ }in\ \href
  {\doibase 10.1007/978-3-319-99046-0_25} {\emph {\bibinfo {booktitle}
  {Fundamental Theories of Physics}}}\ (\bibinfo  {publisher} {Springer
  International Publishing},\ \bibinfo {year} {2018})\ pp.\ \bibinfo {pages}
  {597--622}\BibitemShut {NoStop}%
\bibitem [{\citenamefont {Aurell}\ \emph {et~al.}(2011)\citenamefont {Aurell},
  \citenamefont {Mej\'{\i}a-Monasterio},\ and\ \citenamefont
  {Muratore-Ginanneschi}}]{Aurell2011a}%
  \BibitemOpen
  \bibfield  {author} {\bibinfo {author} {\bibfnamefont {E.}~\bibnamefont
  {Aurell}}, \bibinfo {author} {\bibfnamefont {C.}~\bibnamefont
  {Mej\'{\i}a-Monasterio}}, \ and\ \bibinfo {author} {\bibfnamefont
  {P.}~\bibnamefont {Muratore-Ginanneschi}},\ }\href {\doibase
  10.1103/PhysRevLett.106.250601} {\bibfield  {journal} {\bibinfo  {journal}
  {Phys. Rev. Lett.}\ }\textbf {\bibinfo {volume} {106}},\ \bibinfo {pages}
  {250601} (\bibinfo {year} {2011})}\BibitemShut {NoStop}%
\bibitem [{\citenamefont {Aurell}\ \emph {et~al.}(2012)\citenamefont {Aurell},
  \citenamefont {Gaw\c{e}dzki}, \citenamefont {Mej{\'{\i}}a-Monasterio},
  \citenamefont {Mohayaee},\ and\ \citenamefont
  {Muratore-Ginanneschi}}]{Aurell2012}%
  \BibitemOpen
  \bibfield  {author} {\bibinfo {author} {\bibfnamefont {E.}~\bibnamefont
  {Aurell}}, \bibinfo {author} {\bibfnamefont {K.}~\bibnamefont
  {Gaw\c{e}dzki}}, \bibinfo {author} {\bibfnamefont {C.}~\bibnamefont
  {Mej{\'{\i}}a-Monasterio}}, \bibinfo {author} {\bibfnamefont
  {R.}~\bibnamefont {Mohayaee}}, \ and\ \bibinfo {author} {\bibfnamefont
  {P.}~\bibnamefont {Muratore-Ginanneschi}},\ }\href {\doibase
  10.1007/s10955-012-0478-x} {\bibfield  {journal} {\bibinfo  {journal}
  {Journal of Statistical Physics}\ }\textbf {\bibinfo {volume} {147}},\
  \bibinfo {pages} {487} (\bibinfo {year} {2012})}\BibitemShut {NoStop}%
\bibitem [{\citenamefont {Van~Vu}\ and\ \citenamefont
  {Saito}(2023)}]{vanvu2023}%
  \BibitemOpen
  \bibfield  {author} {\bibinfo {author} {\bibfnamefont {T.}~\bibnamefont
  {Van~Vu}}\ and\ \bibinfo {author} {\bibfnamefont {K.}~\bibnamefont {Saito}},\
  }\href {\doibase 10.1103/PhysRevX.13.011013} {\bibfield  {journal} {\bibinfo
  {journal} {Phys. Rev. X}\ }\textbf {\bibinfo {volume} {13}},\ \bibinfo
  {pages} {011013} (\bibinfo {year} {2023})}\BibitemShut {NoStop}%
\bibitem [{\citenamefont {Salamon}\ \emph {et~al.}(1980)\citenamefont
  {Salamon}, \citenamefont {Andresen}, \citenamefont {Gait},\ and\
  \citenamefont {Berry}}]{Salamon3}%
  \BibitemOpen
  \bibfield  {author} {\bibinfo {author} {\bibfnamefont {P.}~\bibnamefont
  {Salamon}}, \bibinfo {author} {\bibfnamefont {B.}~\bibnamefont {Andresen}},
  \bibinfo {author} {\bibfnamefont {P.~D.}\ \bibnamefont {Gait}}, \ and\
  \bibinfo {author} {\bibfnamefont {R.~S.}\ \bibnamefont {Berry}},\ }\href
  {\doibase 10.1063/1.440217} {\bibfield  {journal} {\bibinfo  {journal} {J.
  Chem. Phys.}\ }\textbf {\bibinfo {volume} {73}},\ \bibinfo {pages} {1001}
  (\bibinfo {year} {1980})}\BibitemShut {NoStop}%
\bibitem [{\citenamefont {Salamon}\ and\ \citenamefont
  {Berry}(1983)}]{Salamon1}%
  \BibitemOpen
  \bibfield  {author} {\bibinfo {author} {\bibfnamefont {P.}~\bibnamefont
  {Salamon}}\ and\ \bibinfo {author} {\bibfnamefont {R.~S.}\ \bibnamefont
  {Berry}},\ }\href {\doibase 10.1103/PhysRevLett.51.1127} {\bibfield
  {journal} {\bibinfo  {journal} {Phys. Rev. Lett.}\ }\textbf {\bibinfo
  {volume} {51}},\ \bibinfo {pages} {1127} (\bibinfo {year}
  {1983})}\BibitemShut {NoStop}%
\bibitem [{\citenamefont {Nulton}\ \emph {et~al.}(1985)\citenamefont {Nulton},
  \citenamefont {Salamon}, \citenamefont {Andresen},\ and\ \citenamefont
  {Qi}}]{Nulton1985}%
  \BibitemOpen
  \bibfield  {author} {\bibinfo {author} {\bibfnamefont {J.}~\bibnamefont
  {Nulton}}, \bibinfo {author} {\bibfnamefont {P.}~\bibnamefont {Salamon}},
  \bibinfo {author} {\bibfnamefont {B.}~\bibnamefont {Andresen}}, \ and\
  \bibinfo {author} {\bibfnamefont {A.}~\bibnamefont {Qi}},\ }\href {\doibase
  10.1063/1.449774} {\bibfield  {journal} {\bibinfo  {journal} {J. Chem.
  Phys.}\ }\textbf {\bibinfo {volume} {83}},\ \bibinfo {pages} {334} (\bibinfo
  {year} {1985})}\BibitemShut {NoStop}%
\bibitem [{\citenamefont {Andresen}\ \emph {et~al.}(1988)\citenamefont
  {Andresen}, \citenamefont {Berry}, \citenamefont {Gilmore}, \citenamefont
  {Ihrig},\ and\ \citenamefont {Salamon}}]{Andresen}%
  \BibitemOpen
  \bibfield  {author} {\bibinfo {author} {\bibfnamefont {B.}~\bibnamefont
  {Andresen}}, \bibinfo {author} {\bibfnamefont {R.~S.}\ \bibnamefont {Berry}},
  \bibinfo {author} {\bibfnamefont {R.}~\bibnamefont {Gilmore}}, \bibinfo
  {author} {\bibfnamefont {E.}~\bibnamefont {Ihrig}}, \ and\ \bibinfo {author}
  {\bibfnamefont {P.}~\bibnamefont {Salamon}},\ }\href {\doibase
  10.1103/PhysRevA.37.845} {\bibfield  {journal} {\bibinfo  {journal} {Phys.
  Rev. A}\ }\textbf {\bibinfo {volume} {37}},\ \bibinfo {pages} {845} (\bibinfo
  {year} {1988})}\BibitemShut {NoStop}%
\bibitem [{\citenamefont {Sivak}\ and\ \citenamefont
  {Crooks}(2012)}]{Sivak2012a}%
  \BibitemOpen
  \bibfield  {author} {\bibinfo {author} {\bibfnamefont {D.~A.}\ \bibnamefont
  {Sivak}}\ and\ \bibinfo {author} {\bibfnamefont {G.~E.}\ \bibnamefont
  {Crooks}},\ }\href {\doibase 10.1103/PhysRevLett.108.190602} {\bibfield
  {journal} {\bibinfo  {journal} {Phys. Rev. Lett.}\ }\textbf {\bibinfo
  {volume} {108}},\ \bibinfo {pages} {190602} (\bibinfo {year}
  {2012})}\BibitemShut {NoStop}%
\bibitem [{\citenamefont {Deffner}\ and\ \citenamefont
  {Bonan{\c{c}}a}(2020)}]{Deffner2020a}%
  \BibitemOpen
  \bibfield  {author} {\bibinfo {author} {\bibfnamefont {S.}~\bibnamefont
  {Deffner}}\ and\ \bibinfo {author} {\bibfnamefont {M.~V.~S.}\ \bibnamefont
  {Bonan{\c{c}}a}},\ }\href {\doibase 10.1209/0295-5075/131/20001} {\bibfield
  {journal} {\bibinfo  {journal} {{EPL} (Europhysics Letters)}\ }\textbf
  {\bibinfo {volume} {131}},\ \bibinfo {pages} {20001} (\bibinfo {year}
  {2020})}\BibitemShut {NoStop}%
\bibitem [{\citenamefont {Abiuso}\ \emph {et~al.}(2020)\citenamefont {Abiuso},
  \citenamefont {Miller}, \citenamefont {{Perarnau-Llobet}},\ and\
  \citenamefont {Scandi}}]{abiusoGeometricOptimisationQuantum2020}%
  \BibitemOpen
  \bibfield  {author} {\bibinfo {author} {\bibfnamefont {P.}~\bibnamefont
  {Abiuso}}, \bibinfo {author} {\bibfnamefont {H.~J.~D.}\ \bibnamefont
  {Miller}}, \bibinfo {author} {\bibfnamefont {M.}~\bibnamefont
  {{Perarnau-Llobet}}}, \ and\ \bibinfo {author} {\bibfnamefont
  {M.}~\bibnamefont {Scandi}},\ }\href {\doibase 10.3390/e22101076} {\bibfield
  {journal} {\bibinfo  {journal} {Entropy}\ }\textbf {\bibinfo {volume} {22}},\
  \bibinfo {pages} {1076} (\bibinfo {year} {2020})}\BibitemShut {NoStop}%
\bibitem [{\citenamefont {Vu}\ and\ \citenamefont
  {Hasegawa}(2021)}]{VanVu2021}%
  \BibitemOpen
  \bibfield  {author} {\bibinfo {author} {\bibfnamefont {T.~V.}\ \bibnamefont
  {Vu}}\ and\ \bibinfo {author} {\bibfnamefont {Y.}~\bibnamefont {Hasegawa}},\
  }\href {\doibase 10.1103/physrevlett.126.010601} {\bibfield  {journal}
  {\bibinfo  {journal} {Physical Review Letters}\ }\textbf {\bibinfo {volume}
  {126}} (\bibinfo {year} {2021}),\ 10.1103/physrevlett.126.010601}\BibitemShut
  {NoStop}%
\bibitem [{\citenamefont {Zulkowski}\ and\ \citenamefont
  {DeWeese}(2014)}]{Zulkowski2014}%
  \BibitemOpen
  \bibfield  {author} {\bibinfo {author} {\bibfnamefont {P.~R.}\ \bibnamefont
  {Zulkowski}}\ and\ \bibinfo {author} {\bibfnamefont {M.~R.}\ \bibnamefont
  {DeWeese}},\ }\href {\doibase 10.1103/PhysRevE.89.052140} {\bibfield
  {journal} {\bibinfo  {journal} {Phys. Rev. E}\ }\textbf {\bibinfo {volume}
  {89}},\ \bibinfo {pages} {052140} (\bibinfo {year} {2014})}\BibitemShut
  {NoStop}%
\bibitem [{\citenamefont {Proesmans}\ \emph
  {et~al.}(2020{\natexlab{a}})\citenamefont {Proesmans}, \citenamefont
  {Ehrich},\ and\ \citenamefont {Bechhoefer}}]{Proesmans2020}%
  \BibitemOpen
  \bibfield  {author} {\bibinfo {author} {\bibfnamefont {K.}~\bibnamefont
  {Proesmans}}, \bibinfo {author} {\bibfnamefont {J.}~\bibnamefont {Ehrich}}, \
  and\ \bibinfo {author} {\bibfnamefont {J.}~\bibnamefont {Bechhoefer}},\
  }\href {\doibase 10.1103/PhysRevLett.125.100602} {\bibfield  {journal}
  {\bibinfo  {journal} {Phys. Rev. Lett.}\ }\textbf {\bibinfo {volume} {125}},\
  \bibinfo {pages} {100602} (\bibinfo {year} {2020}{\natexlab{a}})}\BibitemShut
  {NoStop}%
\bibitem [{\citenamefont {Proesmans}\ \emph
  {et~al.}(2020{\natexlab{b}})\citenamefont {Proesmans}, \citenamefont
  {Ehrich},\ and\ \citenamefont {Bechhoefer}}]{Proesmans2020II}%
  \BibitemOpen
  \bibfield  {author} {\bibinfo {author} {\bibfnamefont {K.}~\bibnamefont
  {Proesmans}}, \bibinfo {author} {\bibfnamefont {J.}~\bibnamefont {Ehrich}}, \
  and\ \bibinfo {author} {\bibfnamefont {J.}~\bibnamefont {Bechhoefer}},\
  }\href {\doibase 10.1103/PhysRevE.102.032105} {\bibfield  {journal} {\bibinfo
   {journal} {Phys. Rev. E}\ }\textbf {\bibinfo {volume} {102}},\ \bibinfo
  {pages} {032105} (\bibinfo {year} {2020}{\natexlab{b}})}\BibitemShut
  {NoStop}%
\bibitem [{\citenamefont {Boyd}\ \emph {et~al.}(2022)\citenamefont {Boyd},
  \citenamefont {Patra}, \citenamefont {Jarzynski},\ and\ \citenamefont
  {Crutchfield}}]{boyd2022shortcuts}%
  \BibitemOpen
  \bibfield  {author} {\bibinfo {author} {\bibfnamefont {A.~B.}\ \bibnamefont
  {Boyd}}, \bibinfo {author} {\bibfnamefont {A.}~\bibnamefont {Patra}},
  \bibinfo {author} {\bibfnamefont {C.}~\bibnamefont {Jarzynski}}, \ and\
  \bibinfo {author} {\bibfnamefont {J.~P.}\ \bibnamefont {Crutchfield}},\
  }\href {\doibase 10.1007/s10955-022-02871-0} {\bibfield  {journal} {\bibinfo
  {journal} {Journal of Statistical Physics}\ }\textbf {\bibinfo {volume}
  {187}},\ \bibinfo {pages} {1} (\bibinfo {year} {2022})}\BibitemShut {NoStop}%
\bibitem [{\citenamefont {Lee}\ \emph {et~al.}(2022)\citenamefont {Lee},
  \citenamefont {Lee}, \citenamefont {Kwon},\ and\ \citenamefont
  {Park}}]{Lee2022}%
  \BibitemOpen
  \bibfield  {author} {\bibinfo {author} {\bibfnamefont {J.~S.}\ \bibnamefont
  {Lee}}, \bibinfo {author} {\bibfnamefont {S.}~\bibnamefont {Lee}}, \bibinfo
  {author} {\bibfnamefont {H.}~\bibnamefont {Kwon}}, \ and\ \bibinfo {author}
  {\bibfnamefont {H.}~\bibnamefont {Park}},\ }\href {\doibase
  10.1103/PhysRevLett.129.120603} {\bibfield  {journal} {\bibinfo  {journal}
  {Phys. Rev. Lett.}\ }\textbf {\bibinfo {volume} {129}},\ \bibinfo {pages}
  {120603} (\bibinfo {year} {2022})}\BibitemShut {NoStop}%
\bibitem [{\citenamefont {Diana}\ \emph {et~al.}(2013)\citenamefont {Diana},
  \citenamefont {Bagci},\ and\ \citenamefont {Esposito}}]{Diana2013}%
  \BibitemOpen
  \bibfield  {author} {\bibinfo {author} {\bibfnamefont {G.}~\bibnamefont
  {Diana}}, \bibinfo {author} {\bibfnamefont {G.~B.}\ \bibnamefont {Bagci}}, \
  and\ \bibinfo {author} {\bibfnamefont {M.}~\bibnamefont {Esposito}},\ }\href
  {\doibase 10.1103/PhysRevE.87.012111} {\bibfield  {journal} {\bibinfo
  {journal} {Phys. Rev. E}\ }\textbf {\bibinfo {volume} {87}},\ \bibinfo
  {pages} {012111} (\bibinfo {year} {2013})}\BibitemShut {NoStop}%
\bibitem [{\citenamefont {Scandi}\ and\ \citenamefont
  {{Perarnau-Llobet}}(2019)}]{scandi19}%
  \BibitemOpen
  \bibfield  {author} {\bibinfo {author} {\bibfnamefont {M.}~\bibnamefont
  {Scandi}}\ and\ \bibinfo {author} {\bibfnamefont {M.}~\bibnamefont
  {{Perarnau-Llobet}}},\ }\href {\doibase 10.22331/q-2019-10-24-197} {\bibfield
   {journal} {\bibinfo  {journal} {Quantum}\ }\textbf {\bibinfo {volume} {3}},\
  \bibinfo {pages} {197} (\bibinfo {year} {2019})}\BibitemShut {NoStop}%
\bibitem [{\citenamefont {Zhen}\ \emph {et~al.}(2021)\citenamefont {Zhen},
  \citenamefont {Egloff}, \citenamefont {Modi},\ and\ \citenamefont
  {Dahlsten}}]{Zhen2021}%
  \BibitemOpen
  \bibfield  {author} {\bibinfo {author} {\bibfnamefont {Y.-Z.}\ \bibnamefont
  {Zhen}}, \bibinfo {author} {\bibfnamefont {D.}~\bibnamefont {Egloff}},
  \bibinfo {author} {\bibfnamefont {K.}~\bibnamefont {Modi}}, \ and\ \bibinfo
  {author} {\bibfnamefont {O.}~\bibnamefont {Dahlsten}},\ }\href {\doibase
  10.1103/PhysRevLett.127.190602} {\bibfield  {journal} {\bibinfo  {journal}
  {Phys. Rev. Lett.}\ }\textbf {\bibinfo {volume} {127}},\ \bibinfo {pages}
  {190602} (\bibinfo {year} {2021})}\BibitemShut {NoStop}%
\bibitem [{\citenamefont {Van~Vu}\ and\ \citenamefont
  {Saito}(2022)}]{VanVu2022}%
  \BibitemOpen
  \bibfield  {author} {\bibinfo {author} {\bibfnamefont {T.}~\bibnamefont
  {Van~Vu}}\ and\ \bibinfo {author} {\bibfnamefont {K.}~\bibnamefont {Saito}},\
  }\href {\doibase 10.1103/PhysRevLett.128.010602} {\bibfield  {journal}
  {\bibinfo  {journal} {Phys. Rev. Lett.}\ }\textbf {\bibinfo {volume} {128}},\
  \bibinfo {pages} {010602} (\bibinfo {year} {2022})}\BibitemShut {NoStop}%
\bibitem [{\citenamefont {Zhen}\ \emph {et~al.}(2022)\citenamefont {Zhen},
  \citenamefont {Egloff}, \citenamefont {Modi},\ and\ \citenamefont
  {Dahlsten}}]{Zhen2022}%
  \BibitemOpen
  \bibfield  {author} {\bibinfo {author} {\bibfnamefont {Y.-Z.}\ \bibnamefont
  {Zhen}}, \bibinfo {author} {\bibfnamefont {D.}~\bibnamefont {Egloff}},
  \bibinfo {author} {\bibfnamefont {K.}~\bibnamefont {Modi}}, \ and\ \bibinfo
  {author} {\bibfnamefont {O.}~\bibnamefont {Dahlsten}},\ }\href {\doibase
  10.1103/PhysRevE.105.044147} {\bibfield  {journal} {\bibinfo  {journal}
  {Phys. Rev. E}\ }\textbf {\bibinfo {volume} {105}},\ \bibinfo {pages}
  {044147} (\bibinfo {year} {2022})}\BibitemShut {NoStop}%
\bibitem [{\citenamefont {Ma}\ \emph {et~al.}(2022)\citenamefont {Ma},
  \citenamefont {Chen}, \citenamefont {Sun},\ and\ \citenamefont
  {Dong}}]{Ma2022}%
  \BibitemOpen
  \bibfield  {author} {\bibinfo {author} {\bibfnamefont {Y.-H.}\ \bibnamefont
  {Ma}}, \bibinfo {author} {\bibfnamefont {J.-F.}\ \bibnamefont {Chen}},
  \bibinfo {author} {\bibfnamefont {C.~P.}\ \bibnamefont {Sun}}, \ and\
  \bibinfo {author} {\bibfnamefont {H.}~\bibnamefont {Dong}},\ }\href {\doibase
  10.1103/PhysRevE.106.034112} {\bibfield  {journal} {\bibinfo  {journal}
  {Phys. Rev. E}\ }\textbf {\bibinfo {volume} {106}},\ \bibinfo {pages}
  {034112} (\bibinfo {year} {2022})}\BibitemShut {NoStop}%
\bibitem [{\citenamefont {Strasberg}\ \emph {et~al.}(2016)\citenamefont
  {Strasberg}, \citenamefont {Schaller}, \citenamefont {Lambert},\ and\
  \citenamefont {Brandes}}]{Strasberg2016}%
  \BibitemOpen
  \bibfield  {author} {\bibinfo {author} {\bibfnamefont {P.}~\bibnamefont
  {Strasberg}}, \bibinfo {author} {\bibfnamefont {G.}~\bibnamefont {Schaller}},
  \bibinfo {author} {\bibfnamefont {N.}~\bibnamefont {Lambert}}, \ and\
  \bibinfo {author} {\bibfnamefont {T.}~\bibnamefont {Brandes}},\ }\href
  {\doibase 10.1088/1367-2630/18/7/073007} {\bibfield  {journal} {\bibinfo
  {journal} {New Journal of Physics}\ }\textbf {\bibinfo {volume} {18}},\
  \bibinfo {pages} {073007} (\bibinfo {year} {2016})}\BibitemShut {NoStop}%
\bibitem [{\citenamefont {Jarzynski}(2017)}]{Jarzynski2017}%
  \BibitemOpen
  \bibfield  {author} {\bibinfo {author} {\bibfnamefont {C.}~\bibnamefont
  {Jarzynski}},\ }\href {\doibase 10.1103/PhysRevX.7.011008} {\bibfield
  {journal} {\bibinfo  {journal} {Phys. Rev. X}\ }\textbf {\bibinfo {volume}
  {7}},\ \bibinfo {pages} {011008} (\bibinfo {year} {2017})}\BibitemShut
  {NoStop}%
\bibitem [{\citenamefont {Miller}(2018)}]{Miller2018}%
  \BibitemOpen
  \bibfield  {author} {\bibinfo {author} {\bibfnamefont {H.~J.~D.}\
  \bibnamefont {Miller}},\ }in\ \href {\doibase 10.1007/978-3-319-99046-0_22}
  {\emph {\bibinfo {booktitle} {Fundamental Theories of Physics}}}\ (\bibinfo
  {publisher} {Springer International Publishing},\ \bibinfo {year} {2018})\
  pp.\ \bibinfo {pages} {531--549}\BibitemShut {NoStop}%
\bibitem [{\citenamefont {Nazir}\ and\ \citenamefont
  {Schaller}(2018)}]{Nazir2018}%
  \BibitemOpen
  \bibfield  {author} {\bibinfo {author} {\bibfnamefont {A.}~\bibnamefont
  {Nazir}}\ and\ \bibinfo {author} {\bibfnamefont {G.}~\bibnamefont
  {Schaller}},\ }in\ \href {\doibase 10.1007/978-3-319-99046-0_23} {\emph
  {\bibinfo {booktitle} {Fundamental Theories of Physics}}}\ (\bibinfo
  {publisher} {Springer International Publishing},\ \bibinfo {year} {2018})\
  pp.\ \bibinfo {pages} {551--577}\BibitemShut {NoStop}%
\bibitem [{\citenamefont {Talkner}\ and\ \citenamefont
  {H\"anggi}(2020)}]{Talkner2020}%
  \BibitemOpen
  \bibfield  {author} {\bibinfo {author} {\bibfnamefont {P.}~\bibnamefont
  {Talkner}}\ and\ \bibinfo {author} {\bibfnamefont {P.}~\bibnamefont
  {H\"anggi}},\ }\href {\doibase 10.1103/RevModPhys.92.041002} {\bibfield
  {journal} {\bibinfo  {journal} {Rev. Mod. Phys.}\ }\textbf {\bibinfo {volume}
  {92}},\ \bibinfo {pages} {041002} (\bibinfo {year} {2020})}\BibitemShut
  {NoStop}%
\bibitem [{\citenamefont {Rivas}(2020)}]{Rivas2020}%
  \BibitemOpen
  \bibfield  {author} {\bibinfo {author} {\bibfnamefont {A.}~\bibnamefont
  {Rivas}},\ }\href {\doibase 10.1103/PhysRevLett.124.160601} {\bibfield
  {journal} {\bibinfo  {journal} {Phys. Rev. Lett.}\ }\textbf {\bibinfo
  {volume} {124}},\ \bibinfo {pages} {160601} (\bibinfo {year}
  {2020})}\BibitemShut {NoStop}%
\bibitem [{\citenamefont {Brenes}\ \emph {et~al.}(2020)\citenamefont {Brenes},
  \citenamefont {Mendoza-Arenas}, \citenamefont {Purkayastha}, \citenamefont
  {Mitchison}, \citenamefont {Clark},\ and\ \citenamefont
  {Goold}}]{Brenes2020}%
  \BibitemOpen
  \bibfield  {author} {\bibinfo {author} {\bibfnamefont {M.}~\bibnamefont
  {Brenes}}, \bibinfo {author} {\bibfnamefont {J.~J.}\ \bibnamefont
  {Mendoza-Arenas}}, \bibinfo {author} {\bibfnamefont {A.}~\bibnamefont
  {Purkayastha}}, \bibinfo {author} {\bibfnamefont {M.~T.}\ \bibnamefont
  {Mitchison}}, \bibinfo {author} {\bibfnamefont {S.~R.}\ \bibnamefont
  {Clark}}, \ and\ \bibinfo {author} {\bibfnamefont {J.}~\bibnamefont
  {Goold}},\ }\href {\doibase 10.1103/PhysRevX.10.031040} {\bibfield  {journal}
  {\bibinfo  {journal} {Phys. Rev. X}\ }\textbf {\bibinfo {volume} {10}},\
  \bibinfo {pages} {031040} (\bibinfo {year} {2020})}\BibitemShut {NoStop}%
\bibitem [{\citenamefont {Pancotti}\ \emph {et~al.}(2020)\citenamefont
  {Pancotti}, \citenamefont {Scandi}, \citenamefont {Mitchison},\ and\
  \citenamefont {Perarnau-Llobet}}]{Pancotti2020}%
  \BibitemOpen
  \bibfield  {author} {\bibinfo {author} {\bibfnamefont {N.}~\bibnamefont
  {Pancotti}}, \bibinfo {author} {\bibfnamefont {M.}~\bibnamefont {Scandi}},
  \bibinfo {author} {\bibfnamefont {M.~T.}\ \bibnamefont {Mitchison}}, \ and\
  \bibinfo {author} {\bibfnamefont {M.}~\bibnamefont {Perarnau-Llobet}},\
  }\href {\doibase 10.1103/PhysRevX.10.031015} {\bibfield  {journal} {\bibinfo
  {journal} {Phys. Rev. X}\ }\textbf {\bibinfo {volume} {10}},\ \bibinfo
  {pages} {031015} (\bibinfo {year} {2020})}\BibitemShut {NoStop}%
\bibitem [{\citenamefont {Alipour}\ \emph {et~al.}(2020)\citenamefont
  {Alipour}, \citenamefont {Chenu}, \citenamefont {Rezakhani},\ and\
  \citenamefont {del Campo}}]{Alipour2020}%
  \BibitemOpen
  \bibfield  {author} {\bibinfo {author} {\bibfnamefont {S.}~\bibnamefont
  {Alipour}}, \bibinfo {author} {\bibfnamefont {A.}~\bibnamefont {Chenu}},
  \bibinfo {author} {\bibfnamefont {A.~T.}\ \bibnamefont {Rezakhani}}, \ and\
  \bibinfo {author} {\bibfnamefont {A.}~\bibnamefont {del Campo}},\ }\href
  {\doibase 10.22331/q-2020-09-28-336} {\bibfield  {journal} {\bibinfo
  {journal} {Quantum}\ }\textbf {\bibinfo {volume} {4}},\ \bibinfo {pages}
  {336} (\bibinfo {year} {2020})}\BibitemShut {NoStop}%
\bibitem [{\citenamefont {Ptaszy{\'{n}}ski}(2022)}]{Ptaszyski2022}%
  \BibitemOpen
  \bibfield  {author} {\bibinfo {author} {\bibfnamefont {K.}~\bibnamefont
  {Ptaszy{\'{n}}ski}},\ }\href {\doibase 10.1103/physreve.106.014114}
  {\bibfield  {journal} {\bibinfo  {journal} {Physical Review E}\ }\textbf
  {\bibinfo {volume} {106}} (\bibinfo {year} {2022}),\
  10.1103/physreve.106.014114}\BibitemShut {NoStop}%
\bibitem [{\citenamefont {Carrega}\ \emph {et~al.}(2022)\citenamefont
  {Carrega}, \citenamefont {Cangemi}, \citenamefont {De~Filippis},
  \citenamefont {Cataudella}, \citenamefont {Benenti},\ and\ \citenamefont
  {Sassetti}}]{Carrega2022}%
  \BibitemOpen
  \bibfield  {author} {\bibinfo {author} {\bibfnamefont {M.}~\bibnamefont
  {Carrega}}, \bibinfo {author} {\bibfnamefont {L.~M.}\ \bibnamefont
  {Cangemi}}, \bibinfo {author} {\bibfnamefont {G.}~\bibnamefont
  {De~Filippis}}, \bibinfo {author} {\bibfnamefont {V.}~\bibnamefont
  {Cataudella}}, \bibinfo {author} {\bibfnamefont {G.}~\bibnamefont {Benenti}},
  \ and\ \bibinfo {author} {\bibfnamefont {M.}~\bibnamefont {Sassetti}},\
  }\href {\doibase 10.1103/PRXQuantum.3.010323} {\bibfield  {journal} {\bibinfo
   {journal} {PRX Quantum}\ }\textbf {\bibinfo {volume} {3}},\ \bibinfo {pages}
  {010323} (\bibinfo {year} {2022})}\BibitemShut {NoStop}%
\bibitem [{\citenamefont {Cavaliere}\ \emph {et~al.}(2022)\citenamefont
  {Cavaliere}, \citenamefont {Carrega}, \citenamefont {Filippis}, \citenamefont
  {Cataudella}, \citenamefont {Benenti},\ and\ \citenamefont
  {Sassetti}}]{Cavaliere2022}%
  \BibitemOpen
  \bibfield  {author} {\bibinfo {author} {\bibfnamefont {F.}~\bibnamefont
  {Cavaliere}}, \bibinfo {author} {\bibfnamefont {M.}~\bibnamefont {Carrega}},
  \bibinfo {author} {\bibfnamefont {G.~D.}\ \bibnamefont {Filippis}}, \bibinfo
  {author} {\bibfnamefont {V.}~\bibnamefont {Cataudella}}, \bibinfo {author}
  {\bibfnamefont {G.}~\bibnamefont {Benenti}}, \ and\ \bibinfo {author}
  {\bibfnamefont {M.}~\bibnamefont {Sassetti}},\ }\href {\doibase
  10.1103/physrevresearch.4.033233} {\bibfield  {journal} {\bibinfo  {journal}
  {Physical Review Research}\ }\textbf {\bibinfo {volume} {4}} (\bibinfo {year}
  {2022}),\ 10.1103/physrevresearch.4.033233}\BibitemShut {NoStop}%
\bibitem [{\citenamefont {Ivander}\ \emph {et~al.}(2022)\citenamefont
  {Ivander}, \citenamefont {Anto-Sztrikacs},\ and\ \citenamefont
  {Segal}}]{Ivander2022}%
  \BibitemOpen
  \bibfield  {author} {\bibinfo {author} {\bibfnamefont {F.}~\bibnamefont
  {Ivander}}, \bibinfo {author} {\bibfnamefont {N.}~\bibnamefont
  {Anto-Sztrikacs}}, \ and\ \bibinfo {author} {\bibfnamefont {D.}~\bibnamefont
  {Segal}},\ }\href {\doibase 10.1103/PhysRevE.105.034112} {\bibfield
  {journal} {\bibinfo  {journal} {Phys. Rev. E}\ }\textbf {\bibinfo {volume}
  {105}},\ \bibinfo {pages} {034112} (\bibinfo {year} {2022})}\BibitemShut
  {NoStop}%
\bibitem [{\citenamefont {Newman}\ \emph {et~al.}(2017)\citenamefont {Newman},
  \citenamefont {Mintert},\ and\ \citenamefont {Nazir}}]{Newman2017}%
  \BibitemOpen
  \bibfield  {author} {\bibinfo {author} {\bibfnamefont {D.}~\bibnamefont
  {Newman}}, \bibinfo {author} {\bibfnamefont {F.}~\bibnamefont {Mintert}}, \
  and\ \bibinfo {author} {\bibfnamefont {A.}~\bibnamefont {Nazir}},\ }\href
  {\doibase 10.1103/PhysRevE.95.032139} {\bibfield  {journal} {\bibinfo
  {journal} {Phys. Rev. E}\ }\textbf {\bibinfo {volume} {95}},\ \bibinfo
  {pages} {032139} (\bibinfo {year} {2017})}\BibitemShut {NoStop}%
\bibitem [{\citenamefont {Perarnau-Llobet}\ \emph {et~al.}(2018)\citenamefont
  {Perarnau-Llobet}, \citenamefont {Wilming}, \citenamefont {Riera},
  \citenamefont {Gallego},\ and\ \citenamefont {Eisert}}]{PerarnauLlobet2018}%
  \BibitemOpen
  \bibfield  {author} {\bibinfo {author} {\bibfnamefont {M.}~\bibnamefont
  {Perarnau-Llobet}}, \bibinfo {author} {\bibfnamefont {H.}~\bibnamefont
  {Wilming}}, \bibinfo {author} {\bibfnamefont {A.}~\bibnamefont {Riera}},
  \bibinfo {author} {\bibfnamefont {R.}~\bibnamefont {Gallego}}, \ and\
  \bibinfo {author} {\bibfnamefont {J.}~\bibnamefont {Eisert}},\ }\href
  {\doibase 10.1103/physrevlett.120.120602} {\bibfield  {journal} {\bibinfo
  {journal} {Physical Review Letters}\ }\textbf {\bibinfo {volume} {120}}
  (\bibinfo {year} {2018}),\ 10.1103/physrevlett.120.120602}\BibitemShut
  {NoStop}%
\bibitem [{\citenamefont {Strasberg}\ \emph {et~al.}(2018)\citenamefont
  {Strasberg}, \citenamefont {Schaller}, \citenamefont {Schmidt},\ and\
  \citenamefont {Esposito}}]{Strasberg2018}%
  \BibitemOpen
  \bibfield  {author} {\bibinfo {author} {\bibfnamefont {P.}~\bibnamefont
  {Strasberg}}, \bibinfo {author} {\bibfnamefont {G.}~\bibnamefont {Schaller}},
  \bibinfo {author} {\bibfnamefont {T.~L.}\ \bibnamefont {Schmidt}}, \ and\
  \bibinfo {author} {\bibfnamefont {M.}~\bibnamefont {Esposito}},\ }\href
  {\doibase 10.1103/physrevb.97.205405} {\bibfield  {journal} {\bibinfo
  {journal} {Physical Review B}\ }\textbf {\bibinfo {volume} {97}} (\bibinfo
  {year} {2018}),\ 10.1103/physrevb.97.205405}\BibitemShut {NoStop}%
\bibitem [{\citenamefont {Wiedmann}\ \emph {et~al.}(2020)\citenamefont
  {Wiedmann}, \citenamefont {Stockburger},\ and\ \citenamefont
  {Ankerhold}}]{Wiedmann_2020}%
  \BibitemOpen
  \bibfield  {author} {\bibinfo {author} {\bibfnamefont {M.}~\bibnamefont
  {Wiedmann}}, \bibinfo {author} {\bibfnamefont {J.~T.}\ \bibnamefont
  {Stockburger}}, \ and\ \bibinfo {author} {\bibfnamefont {J.}~\bibnamefont
  {Ankerhold}},\ }\href {\doibase 10.1088/1367-2630/ab725a} {\bibfield
  {journal} {\bibinfo  {journal} {New Journal of Physics}\ }\textbf {\bibinfo
  {volume} {22}},\ \bibinfo {pages} {033007} (\bibinfo {year}
  {2020})}\BibitemShut {NoStop}%
\bibitem [{\citenamefont {Liu}\ \emph {et~al.}(2021)\citenamefont {Liu},
  \citenamefont {Jung},\ and\ \citenamefont {Segal}}]{Liu2021}%
  \BibitemOpen
  \bibfield  {author} {\bibinfo {author} {\bibfnamefont {J.}~\bibnamefont
  {Liu}}, \bibinfo {author} {\bibfnamefont {K.~A.}\ \bibnamefont {Jung}}, \
  and\ \bibinfo {author} {\bibfnamefont {D.}~\bibnamefont {Segal}},\ }\href
  {\doibase 10.1103/PhysRevLett.127.200602} {\bibfield  {journal} {\bibinfo
  {journal} {Phys. Rev. Lett.}\ }\textbf {\bibinfo {volume} {127}},\ \bibinfo
  {pages} {200602} (\bibinfo {year} {2021})}\BibitemShut {NoStop}%
\bibitem [{\citenamefont {Shirai}\ \emph {et~al.}(2021)\citenamefont {Shirai},
  \citenamefont {Hashimoto}, \citenamefont {Tezuka}, \citenamefont {Uchiyama},\
  and\ \citenamefont {Hatano}}]{Shirai2021}%
  \BibitemOpen
  \bibfield  {author} {\bibinfo {author} {\bibfnamefont {Y.}~\bibnamefont
  {Shirai}}, \bibinfo {author} {\bibfnamefont {K.}~\bibnamefont {Hashimoto}},
  \bibinfo {author} {\bibfnamefont {R.}~\bibnamefont {Tezuka}}, \bibinfo
  {author} {\bibfnamefont {C.}~\bibnamefont {Uchiyama}}, \ and\ \bibinfo
  {author} {\bibfnamefont {N.}~\bibnamefont {Hatano}},\ }\href {\doibase
  10.1103/PhysRevResearch.3.023078} {\bibfield  {journal} {\bibinfo  {journal}
  {Phys. Rev. Research}\ }\textbf {\bibinfo {volume} {3}},\ \bibinfo {pages}
  {023078} (\bibinfo {year} {2021})}\BibitemShut {NoStop}%
\bibitem [{\citenamefont {Koyanagi}\ and\ \citenamefont
  {Tanimura}(2022)}]{Koyanagi2022}%
  \BibitemOpen
  \bibfield  {author} {\bibinfo {author} {\bibfnamefont {S.}~\bibnamefont
  {Koyanagi}}\ and\ \bibinfo {author} {\bibfnamefont {Y.}~\bibnamefont
  {Tanimura}},\ }\href {\doibase 10.1063/5.0107305} {\bibfield  {journal}
  {\bibinfo  {journal} {The Journal of Chemical Physics}\ }\textbf {\bibinfo
  {volume} {157}},\ \bibinfo {pages} {084110} (\bibinfo {year}
  {2022})}\BibitemShut {NoStop}%
\bibitem [{\citenamefont {Liu}\ and\ \citenamefont {Jung}(2022)}]{Liu2022b}%
  \BibitemOpen
  \bibfield  {author} {\bibinfo {author} {\bibfnamefont {J.}~\bibnamefont
  {Liu}}\ and\ \bibinfo {author} {\bibfnamefont {K.~A.}\ \bibnamefont {Jung}},\
  }\href {\doibase 10.1103/PhysRevE.106.L022105} {\bibfield  {journal}
  {\bibinfo  {journal} {Phys. Rev. E}\ }\textbf {\bibinfo {volume} {106}},\
  \bibinfo {pages} {L022105} (\bibinfo {year} {2022})}\BibitemShut {NoStop}%
\bibitem [{\citenamefont {Schaller}(2014)}]{schaller2014open}%
  \BibitemOpen
  \bibfield  {author} {\bibinfo {author} {\bibfnamefont {G.}~\bibnamefont
  {Schaller}},\ }\href {\doibase 10.1007/978-3-319-03877-3} {\emph {\bibinfo
  {title} {Open quantum systems far from equilibrium}}},\ Vol.\ \bibinfo
  {volume} {881}\ (\bibinfo  {publisher} {Springer},\ \bibinfo {year}
  {2014})\BibitemShut {NoStop}%
\bibitem [{\citenamefont {Ludovico}\ \emph {et~al.}(2014)\citenamefont
  {Ludovico}, \citenamefont {Lim}, \citenamefont {Moskalets}, \citenamefont
  {Arrachea},\ and\ \citenamefont {S\'anchez}}]{Ludovico2014}%
  \BibitemOpen
  \bibfield  {author} {\bibinfo {author} {\bibfnamefont {M.~F.}\ \bibnamefont
  {Ludovico}}, \bibinfo {author} {\bibfnamefont {J.~S.}\ \bibnamefont {Lim}},
  \bibinfo {author} {\bibfnamefont {M.}~\bibnamefont {Moskalets}}, \bibinfo
  {author} {\bibfnamefont {L.}~\bibnamefont {Arrachea}}, \ and\ \bibinfo
  {author} {\bibfnamefont {D.}~\bibnamefont {S\'anchez}},\ }\href {\doibase
  10.1103/PhysRevB.89.161306} {\bibfield  {journal} {\bibinfo  {journal} {Phys.
  Rev. B}\ }\textbf {\bibinfo {volume} {89}},\ \bibinfo {pages} {161306}
  (\bibinfo {year} {2014})}\BibitemShut {NoStop}%
\bibitem [{\citenamefont {Esposito}\ \emph
  {et~al.}(2015{\natexlab{a}})\citenamefont {Esposito}, \citenamefont {Ochoa},\
  and\ \citenamefont {Galperin}}]{Esposito2015}%
  \BibitemOpen
  \bibfield  {author} {\bibinfo {author} {\bibfnamefont {M.}~\bibnamefont
  {Esposito}}, \bibinfo {author} {\bibfnamefont {M.~A.}\ \bibnamefont {Ochoa}},
  \ and\ \bibinfo {author} {\bibfnamefont {M.}~\bibnamefont {Galperin}},\
  }\href {\doibase 10.1103/PhysRevLett.114.080602} {\bibfield  {journal}
  {\bibinfo  {journal} {Phys. Rev. Lett.}\ }\textbf {\bibinfo {volume} {114}},\
  \bibinfo {pages} {080602} (\bibinfo {year} {2015}{\natexlab{a}})}\BibitemShut
  {NoStop}%
\bibitem [{\citenamefont {Esposito}\ \emph
  {et~al.}(2015{\natexlab{b}})\citenamefont {Esposito}, \citenamefont {Ochoa},\
  and\ \citenamefont {Galperin}}]{Esposito2015b}%
  \BibitemOpen
  \bibfield  {author} {\bibinfo {author} {\bibfnamefont {M.}~\bibnamefont
  {Esposito}}, \bibinfo {author} {\bibfnamefont {M.~A.}\ \bibnamefont {Ochoa}},
  \ and\ \bibinfo {author} {\bibfnamefont {M.}~\bibnamefont {Galperin}},\
  }\href {\doibase 10.1103/PhysRevB.92.235440} {\bibfield  {journal} {\bibinfo
  {journal} {Phys. Rev. B}\ }\textbf {\bibinfo {volume} {92}},\ \bibinfo
  {pages} {235440} (\bibinfo {year} {2015}{\natexlab{b}})}\BibitemShut
  {NoStop}%
\bibitem [{\citenamefont {Bruch}\ \emph {et~al.}(2016)\citenamefont {Bruch},
  \citenamefont {Thomas}, \citenamefont {Viola~Kusminskiy}, \citenamefont {von
  Oppen},\ and\ \citenamefont {Nitzan}}]{Bruch2016}%
  \BibitemOpen
  \bibfield  {author} {\bibinfo {author} {\bibfnamefont {A.}~\bibnamefont
  {Bruch}}, \bibinfo {author} {\bibfnamefont {M.}~\bibnamefont {Thomas}},
  \bibinfo {author} {\bibfnamefont {S.}~\bibnamefont {Viola~Kusminskiy}},
  \bibinfo {author} {\bibfnamefont {F.}~\bibnamefont {von Oppen}}, \ and\
  \bibinfo {author} {\bibfnamefont {A.}~\bibnamefont {Nitzan}},\ }\href
  {\doibase 10.1103/PhysRevB.93.115318} {\bibfield  {journal} {\bibinfo
  {journal} {Phys. Rev. B}\ }\textbf {\bibinfo {volume} {93}},\ \bibinfo
  {pages} {115318} (\bibinfo {year} {2016})}\BibitemShut {NoStop}%
\bibitem [{\citenamefont {Haughian}\ \emph {et~al.}(2018)\citenamefont
  {Haughian}, \citenamefont {Esposito},\ and\ \citenamefont
  {Schmidt}}]{Haughian2018}%
  \BibitemOpen
  \bibfield  {author} {\bibinfo {author} {\bibfnamefont {P.}~\bibnamefont
  {Haughian}}, \bibinfo {author} {\bibfnamefont {M.}~\bibnamefont {Esposito}},
  \ and\ \bibinfo {author} {\bibfnamefont {T.~L.}\ \bibnamefont {Schmidt}},\
  }\href {\doibase 10.1103/PhysRevB.97.085435} {\bibfield  {journal} {\bibinfo
  {journal} {Phys. Rev. B}\ }\textbf {\bibinfo {volume} {97}},\ \bibinfo
  {pages} {085435} (\bibinfo {year} {2018})}\BibitemShut {NoStop}%
\bibitem [{\citenamefont {Mitchison}\ and\ \citenamefont
  {Plenio}(2018)}]{Mitchison2018}%
  \BibitemOpen
  \bibfield  {author} {\bibinfo {author} {\bibfnamefont {M.~T.}\ \bibnamefont
  {Mitchison}}\ and\ \bibinfo {author} {\bibfnamefont {M.~B.}\ \bibnamefont
  {Plenio}},\ }\href {\doibase 10.1088/1367-2630/aa9f70} {\bibfield  {journal}
  {\bibinfo  {journal} {New Journal of Physics}\ }\textbf {\bibinfo {volume}
  {20}},\ \bibinfo {pages} {033005} (\bibinfo {year} {2018})}\BibitemShut
  {NoStop}%
\bibitem [{\citenamefont {Tong}\ and\ \citenamefont {Dou}(2022)}]{Tong2022}%
  \BibitemOpen
  \bibfield  {author} {\bibinfo {author} {\bibfnamefont {K.}~\bibnamefont
  {Tong}}\ and\ \bibinfo {author} {\bibfnamefont {W.}~\bibnamefont {Dou}},\
  }\href {\doibase 10.1088/1361-648x/ac99c8} {\bibfield  {journal} {\bibinfo
  {journal} {Journal of Physics: Condensed Matter}\ }\textbf {\bibinfo {volume}
  {34}},\ \bibinfo {pages} {495703} (\bibinfo {year} {2022})}\BibitemShut
  {NoStop}%
\bibitem [{\citenamefont {Hartnoll}\ and\ \citenamefont
  {Mackenzie}(2022)}]{Hartnoll22}%
  \BibitemOpen
  \bibfield  {author} {\bibinfo {author} {\bibfnamefont {S.~A.}\ \bibnamefont
  {Hartnoll}}\ and\ \bibinfo {author} {\bibfnamefont {A.~P.}\ \bibnamefont
  {Mackenzie}},\ }\href {\doibase 10.1103/RevModPhys.94.041002} {\bibfield
  {journal} {\bibinfo  {journal} {Rev. Mod. Phys.}\ }\textbf {\bibinfo {volume}
  {94}},\ \bibinfo {pages} {041002} (\bibinfo {year} {2022})}\BibitemShut
  {NoStop}%
\bibitem [{\citenamefont {Sachdev}(2011)}]{sachdev11}%
  \BibitemOpen
  \bibfield  {author} {\bibinfo {author} {\bibfnamefont {S.}~\bibnamefont
  {Sachdev}},\ }\href {\doibase 10.1017/CBO9780511973765} {\emph {\bibinfo
  {title} {Quantum Phase Transitions}}},\ \bibinfo {edition} {2nd}\ ed.\
  (\bibinfo  {publisher} {Cambridge University Press},\ \bibinfo {year}
  {2011})\BibitemShut {NoStop}%
\bibitem [{\citenamefont {Maldacena}\ \emph {et~al.}(2016)\citenamefont
  {Maldacena}, \citenamefont {Shenker},\ and\ \citenamefont
  {Stanford}}]{Maldacena2016}%
  \BibitemOpen
  \bibfield  {author} {\bibinfo {author} {\bibfnamefont {J.}~\bibnamefont
  {Maldacena}}, \bibinfo {author} {\bibfnamefont {S.~H.}\ \bibnamefont
  {Shenker}}, \ and\ \bibinfo {author} {\bibfnamefont {D.}~\bibnamefont
  {Stanford}},\ }\href {\doibase 10.1007/jhep08(2016)106} {\bibfield  {journal}
  {\bibinfo  {journal} {Journal of High Energy Physics}\ }\textbf {\bibinfo
  {volume} {2016}} (\bibinfo {year} {2016}),\
  10.1007/jhep08(2016)106}\BibitemShut {NoStop}%
\bibitem [{\citenamefont {Pappalardi}\ and\ \citenamefont
  {Kurchan}(2022)}]{Pappalardi22}%
  \BibitemOpen
  \bibfield  {author} {\bibinfo {author} {\bibfnamefont {S.}~\bibnamefont
  {Pappalardi}}\ and\ \bibinfo {author} {\bibfnamefont {J.}~\bibnamefont
  {Kurchan}},\ }\href {\doibase 10.21468/SciPostPhys.13.1.006} {\bibfield
  {journal} {\bibinfo  {journal} {SciPost Phys.}\ }\textbf {\bibinfo {volume}
  {13}},\ \bibinfo {pages} {006} (\bibinfo {year} {2022})}\BibitemShut
  {NoStop}%
\bibitem [{\citenamefont {Zulkowski}\ \emph {et~al.}(2012)\citenamefont
  {Zulkowski}, \citenamefont {Sivak}, \citenamefont {Crooks},\ and\
  \citenamefont {DeWeese}}]{Zulkowski}%
  \BibitemOpen
  \bibfield  {author} {\bibinfo {author} {\bibfnamefont {P.~R.}\ \bibnamefont
  {Zulkowski}}, \bibinfo {author} {\bibfnamefont {D.~A.}\ \bibnamefont
  {Sivak}}, \bibinfo {author} {\bibfnamefont {G.~E.}\ \bibnamefont {Crooks}}, \
  and\ \bibinfo {author} {\bibfnamefont {M.~R.}\ \bibnamefont {DeWeese}},\
  }\href {\doibase 10.1103/PhysRevE.86.041148} {\bibfield  {journal} {\bibinfo
  {journal} {Phys. Rev. E}\ }\textbf {\bibinfo {volume} {86}},\ \bibinfo
  {pages} {041148} (\bibinfo {year} {2012})}\BibitemShut {NoStop}%
\bibitem [{\citenamefont {Bonan{\c{c}}a}\ and\ \citenamefont
  {Deffner}(2014)}]{Bonana2014}%
  \BibitemOpen
  \bibfield  {author} {\bibinfo {author} {\bibfnamefont {M.~V.~S.}\
  \bibnamefont {Bonan{\c{c}}a}}\ and\ \bibinfo {author} {\bibfnamefont
  {S.}~\bibnamefont {Deffner}},\ }\href {\doibase 10.1063/1.4885277} {\bibfield
   {journal} {\bibinfo  {journal} {J. Chem. Phys.}\ }\textbf {\bibinfo {volume}
  {140}},\ \bibinfo {pages} {244119} (\bibinfo {year} {2014})}\BibitemShut
  {NoStop}%
\bibitem [{\citenamefont {Rotskoff}\ \emph {et~al.}(2017)\citenamefont
  {Rotskoff}, \citenamefont {Crooks},\ and\ \citenamefont
  {Vanden-Eijnden}}]{Rotskoff2017}%
  \BibitemOpen
  \bibfield  {author} {\bibinfo {author} {\bibfnamefont {G.~M.}\ \bibnamefont
  {Rotskoff}}, \bibinfo {author} {\bibfnamefont {G.~E.}\ \bibnamefont
  {Crooks}}, \ and\ \bibinfo {author} {\bibfnamefont {E.}~\bibnamefont
  {Vanden-Eijnden}},\ }\href {\doibase 10.1103/PhysRevE.95.012148} {\bibfield
  {journal} {\bibinfo  {journal} {Phys. Rev. E}\ }\textbf {\bibinfo {volume}
  {95}},\ \bibinfo {pages} {012148} (\bibinfo {year} {2017})}\BibitemShut
  {NoStop}%
\bibitem [{\citenamefont {Li}\ \emph {et~al.}(2022)\citenamefont {Li},
  \citenamefont {Chen}, \citenamefont {Sun},\ and\ \citenamefont
  {Dong}}]{Li2022}%
  \BibitemOpen
  \bibfield  {author} {\bibinfo {author} {\bibfnamefont {G.}~\bibnamefont
  {Li}}, \bibinfo {author} {\bibfnamefont {J.-F.}\ \bibnamefont {Chen}},
  \bibinfo {author} {\bibfnamefont {C.~P.}\ \bibnamefont {Sun}}, \ and\
  \bibinfo {author} {\bibfnamefont {H.}~\bibnamefont {Dong}},\ }\href {\doibase
  10.1103/PhysRevLett.128.230603} {\bibfield  {journal} {\bibinfo  {journal}
  {Phys. Rev. Lett.}\ }\textbf {\bibinfo {volume} {128}},\ \bibinfo {pages}
  {230603} (\bibinfo {year} {2022})}\BibitemShut {NoStop}%
\bibitem [{\citenamefont {Eglinton}\ and\ \citenamefont
  {Brandner}(2022)}]{Eglinton2022}%
  \BibitemOpen
  \bibfield  {author} {\bibinfo {author} {\bibfnamefont {J.}~\bibnamefont
  {Eglinton}}\ and\ \bibinfo {author} {\bibfnamefont {K.}~\bibnamefont
  {Brandner}},\ }\href {\doibase 10.1103/PhysRevE.105.L052102} {\bibfield
  {journal} {\bibinfo  {journal} {Phys. Rev. E}\ }\textbf {\bibinfo {volume}
  {105}},\ \bibinfo {pages} {L052102} (\bibinfo {year} {2022})}\BibitemShut
  {NoStop}%
\bibitem [{\citenamefont {Frim}\ and\ \citenamefont
  {DeWeese}(2022)}]{Frim2022}%
  \BibitemOpen
  \bibfield  {author} {\bibinfo {author} {\bibfnamefont {A.~G.}\ \bibnamefont
  {Frim}}\ and\ \bibinfo {author} {\bibfnamefont {M.~R.}\ \bibnamefont
  {DeWeese}},\ }\href {\doibase 10.1103/PhysRevLett.128.230601} {\bibfield
  {journal} {\bibinfo  {journal} {Phys. Rev. Lett.}\ }\textbf {\bibinfo
  {volume} {128}},\ \bibinfo {pages} {230601} (\bibinfo {year}
  {2022})}\BibitemShut {NoStop}%
\bibitem [{\citenamefont {Chen}\ \emph {et~al.}(2022)\citenamefont {Chen},
  \citenamefont {Zhai}, \citenamefont {Sun},\ and\ \citenamefont
  {Dong}}]{chen2022geodesic}%
  \BibitemOpen
  \bibfield  {author} {\bibinfo {author} {\bibfnamefont {J.-F.}\ \bibnamefont
  {Chen}}, \bibinfo {author} {\bibfnamefont {R.-X.}\ \bibnamefont {Zhai}},
  \bibinfo {author} {\bibfnamefont {C.}~\bibnamefont {Sun}}, \ and\ \bibinfo
  {author} {\bibfnamefont {H.}~\bibnamefont {Dong}},\ }\href {\doibase
  10.48550/arXiv.2209.07269} {\bibfield  {journal} {\bibinfo  {journal} {arXiv
  preprint arXiv:2209.07269}\ } (\bibinfo {year} {2022}),\
  10.48550/arXiv.2209.07269}\BibitemShut {NoStop}%
\bibitem [{\citenamefont {Miller}\ \emph {et~al.}(2019)\citenamefont {Miller},
  \citenamefont {Scandi}, \citenamefont {Anders},\ and\ \citenamefont
  {Perarnau-Llobet}}]{Miller2019}%
  \BibitemOpen
  \bibfield  {author} {\bibinfo {author} {\bibfnamefont {H.~J.~D.}\
  \bibnamefont {Miller}}, \bibinfo {author} {\bibfnamefont {M.}~\bibnamefont
  {Scandi}}, \bibinfo {author} {\bibfnamefont {J.}~\bibnamefont {Anders}}, \
  and\ \bibinfo {author} {\bibfnamefont {M.}~\bibnamefont {Perarnau-Llobet}},\
  }\href {\doibase 10.1103/PhysRevLett.123.230603} {\bibfield  {journal}
  {\bibinfo  {journal} {Phys. Rev. Lett.}\ }\textbf {\bibinfo {volume} {123}},\
  \bibinfo {pages} {230603} (\bibinfo {year} {2019})}\BibitemShut {NoStop}%
\bibitem [{\citenamefont {Abiuso}\ and\ \citenamefont
  {Perarnau-Llobet}(2020)}]{abiuso2020optimal}%
  \BibitemOpen
  \bibfield  {author} {\bibinfo {author} {\bibfnamefont {P.}~\bibnamefont
  {Abiuso}}\ and\ \bibinfo {author} {\bibfnamefont {M.}~\bibnamefont
  {Perarnau-Llobet}},\ }\href {\doibase
  https://doi.org/10.1103/PhysRevLett.124.110606} {\bibfield  {journal}
  {\bibinfo  {journal} {Phys. Rev. Lett.}\ }\textbf {\bibinfo {volume} {124}},\
  \bibinfo {pages} {110606} (\bibinfo {year} {2020})}\BibitemShut {NoStop}%
\bibitem [{\citenamefont {Brandner}\ and\ \citenamefont
  {Saito}(2020)}]{Brandner2020}%
  \BibitemOpen
  \bibfield  {author} {\bibinfo {author} {\bibfnamefont {K.}~\bibnamefont
  {Brandner}}\ and\ \bibinfo {author} {\bibfnamefont {K.}~\bibnamefont
  {Saito}},\ }\href {\doibase 10.1103/PhysRevLett.124.040602} {\bibfield
  {journal} {\bibinfo  {journal} {Phys. Rev. Lett.}\ }\textbf {\bibinfo
  {volume} {124}},\ \bibinfo {pages} {040602} (\bibinfo {year}
  {2020})}\BibitemShut {NoStop}%
\bibitem [{\citenamefont {Terr\'en~Alonso}\ \emph {et~al.}(2022)\citenamefont
  {Terr\'en~Alonso}, \citenamefont {Abiuso}, \citenamefont {Perarnau-Llobet},\
  and\ \citenamefont {Arrachea}}]{Alonso2022Geometric}%
  \BibitemOpen
  \bibfield  {author} {\bibinfo {author} {\bibfnamefont {P.}~\bibnamefont
  {Terr\'en~Alonso}}, \bibinfo {author} {\bibfnamefont {P.}~\bibnamefont
  {Abiuso}}, \bibinfo {author} {\bibfnamefont {M.}~\bibnamefont
  {Perarnau-Llobet}}, \ and\ \bibinfo {author} {\bibfnamefont {L.}~\bibnamefont
  {Arrachea}},\ }\href {\doibase 10.1103/PRXQuantum.3.010326} {\bibfield
  {journal} {\bibinfo  {journal} {PRX Quantum}\ }\textbf {\bibinfo {volume}
  {3}},\ \bibinfo {pages} {010326} (\bibinfo {year} {2022})}\BibitemShut
  {NoStop}%
\bibitem [{\citenamefont {Mehboudi}\ and\ \citenamefont
  {Miller}(2022)}]{Mehboudi2022Thermodynamic}%
  \BibitemOpen
  \bibfield  {author} {\bibinfo {author} {\bibfnamefont {M.}~\bibnamefont
  {Mehboudi}}\ and\ \bibinfo {author} {\bibfnamefont {H.~J.~D.}\ \bibnamefont
  {Miller}},\ }\href {\doibase 10.1103/PhysRevA.105.062434} {\bibfield
  {journal} {\bibinfo  {journal} {Phys. Rev. A}\ }\textbf {\bibinfo {volume}
  {105}},\ \bibinfo {pages} {062434} (\bibinfo {year} {2022})}\BibitemShut
  {NoStop}%
\bibitem [{\citenamefont {Deffner}\ and\ \citenamefont
  {Lutz}(2010)}]{Deffner2010}%
  \BibitemOpen
  \bibfield  {author} {\bibinfo {author} {\bibfnamefont {S.}~\bibnamefont
  {Deffner}}\ and\ \bibinfo {author} {\bibfnamefont {E.}~\bibnamefont {Lutz}},\
  }\href {\doibase 10.1103/PhysRevLett.105.170402} {\bibfield  {journal}
  {\bibinfo  {journal} {Phys. Rev. Lett.}\ }\textbf {\bibinfo {volume} {105}},\
  \bibinfo {pages} {170402} (\bibinfo {year} {2010})}\BibitemShut {NoStop}%
\bibitem [{\citenamefont {Eisert}\ \emph {et~al.}(2015)\citenamefont {Eisert},
  \citenamefont {Friesdorf},\ and\ \citenamefont {Gogolin}}]{Eisert2015}%
  \BibitemOpen
  \bibfield  {author} {\bibinfo {author} {\bibfnamefont {J.}~\bibnamefont
  {Eisert}}, \bibinfo {author} {\bibfnamefont {M.}~\bibnamefont {Friesdorf}}, \
  and\ \bibinfo {author} {\bibfnamefont {C.}~\bibnamefont {Gogolin}},\ }\href
  {\doibase 10.1038/nphys3215} {\bibfield  {journal} {\bibinfo  {journal}
  {Nature Physics}\ }\textbf {\bibinfo {volume} {11}},\ \bibinfo {pages} {124}
  (\bibinfo {year} {2015})}\BibitemShut {NoStop}%
\bibitem [{\citenamefont {D{\textquotesingle}Alessio}\ \emph
  {et~al.}(2016)\citenamefont {D{\textquotesingle}Alessio}, \citenamefont
  {Kafri}, \citenamefont {Polkovnikov},\ and\ \citenamefont
  {Rigol}}]{DAlessio2016}%
  \BibitemOpen
  \bibfield  {author} {\bibinfo {author} {\bibfnamefont {L.}~\bibnamefont
  {D{\textquotesingle}Alessio}}, \bibinfo {author} {\bibfnamefont
  {Y.}~\bibnamefont {Kafri}}, \bibinfo {author} {\bibfnamefont
  {A.}~\bibnamefont {Polkovnikov}}, \ and\ \bibinfo {author} {\bibfnamefont
  {M.}~\bibnamefont {Rigol}},\ }\href {\doibase 10.1080/00018732.2016.1198134}
  {\bibfield  {journal} {\bibinfo  {journal} {Advances in Physics}\ }\textbf
  {\bibinfo {volume} {65}},\ \bibinfo {pages} {239} (\bibinfo {year}
  {2016})}\BibitemShut {NoStop}%
\bibitem [{\citenamefont {Suba{\c{s}}{\i}}\ \emph {et~al.}(2012)\citenamefont
  {Suba{\c{s}}{\i}}, \citenamefont {Fleming}, \citenamefont {Taylor},\ and\
  \citenamefont {Hu}}]{Suba2012}%
  \BibitemOpen
  \bibfield  {author} {\bibinfo {author} {\bibfnamefont {Y.}~\bibnamefont
  {Suba{\c{s}}{\i}}}, \bibinfo {author} {\bibfnamefont {C.~H.}\ \bibnamefont
  {Fleming}}, \bibinfo {author} {\bibfnamefont {J.~M.}\ \bibnamefont {Taylor}},
  \ and\ \bibinfo {author} {\bibfnamefont {B.~L.}\ \bibnamefont {Hu}},\ }\href
  {\doibase 10.1103/physreve.86.061132} {\bibfield  {journal} {\bibinfo
  {journal} {Physical Review E}\ }\textbf {\bibinfo {volume} {86}} (\bibinfo
  {year} {2012}),\ 10.1103/physreve.86.061132}\BibitemShut {NoStop}%
\bibitem [{\citenamefont {Merkli}(2020)}]{Merkli2020}%
  \BibitemOpen
  \bibfield  {author} {\bibinfo {author} {\bibfnamefont {M.}~\bibnamefont
  {Merkli}},\ }\href {\doibase 10.1016/j.aop.2019.167996} {\bibfield  {journal}
  {\bibinfo  {journal} {Annals of Physics}\ }\textbf {\bibinfo {volume}
  {412}},\ \bibinfo {pages} {167996} (\bibinfo {year} {2020})}\BibitemShut
  {NoStop}%
\bibitem [{\citenamefont {Cresser}\ and\ \citenamefont
  {Anders}(2021)}]{Cresser2021}%
  \BibitemOpen
  \bibfield  {author} {\bibinfo {author} {\bibfnamefont {J.~D.}\ \bibnamefont
  {Cresser}}\ and\ \bibinfo {author} {\bibfnamefont {J.}~\bibnamefont
  {Anders}},\ }\href {\doibase 10.1103/PhysRevLett.127.250601} {\bibfield
  {journal} {\bibinfo  {journal} {Phys. Rev. Lett.}\ }\textbf {\bibinfo
  {volume} {127}},\ \bibinfo {pages} {250601} (\bibinfo {year}
  {2021})}\BibitemShut {NoStop}%
\bibitem [{\citenamefont {Trushechkin}\ \emph {et~al.}(2022)\citenamefont
  {Trushechkin}, \citenamefont {Merkli}, \citenamefont {Cresser},\ and\
  \citenamefont {Anders}}]{Trushechkin2022}%
  \BibitemOpen
  \bibfield  {author} {\bibinfo {author} {\bibfnamefont {A.~S.}\ \bibnamefont
  {Trushechkin}}, \bibinfo {author} {\bibfnamefont {M.}~\bibnamefont {Merkli}},
  \bibinfo {author} {\bibfnamefont {J.~D.}\ \bibnamefont {Cresser}}, \ and\
  \bibinfo {author} {\bibfnamefont {J.}~\bibnamefont {Anders}},\ }\href
  {\doibase 10.1116/5.0073853} {\bibfield  {journal} {\bibinfo  {journal}
  {{AVS} Quantum Science}\ }\textbf {\bibinfo {volume} {4}},\ \bibinfo {pages}
  {012301} (\bibinfo {year} {2022})}\BibitemShut {NoStop}%
\bibitem [{\citenamefont {Cavina}\ \emph {et~al.}(2017)\citenamefont {Cavina},
  \citenamefont {Mari},\ and\ \citenamefont {Giovannetti}}]{cavina2017slow}%
  \BibitemOpen
  \bibfield  {author} {\bibinfo {author} {\bibfnamefont {V.}~\bibnamefont
  {Cavina}}, \bibinfo {author} {\bibfnamefont {A.}~\bibnamefont {Mari}}, \ and\
  \bibinfo {author} {\bibfnamefont {V.}~\bibnamefont {Giovannetti}},\ }\href
  {\doibase 10.1103/PhysRevLett.119.050601} {\bibfield  {journal} {\bibinfo
  {journal} {Phys. Rev. Lett.}\ }\textbf {\bibinfo {volume} {119}},\ \bibinfo
  {pages} {050601} (\bibinfo {year} {2017})}\BibitemShut {NoStop}%
\bibitem [{\citenamefont {Schmiedl}\ and\ \citenamefont
  {Seifert}(2007)}]{Schmiedl2007o}%
  \BibitemOpen
  \bibfield  {author} {\bibinfo {author} {\bibfnamefont {T.}~\bibnamefont
  {Schmiedl}}\ and\ \bibinfo {author} {\bibfnamefont {U.}~\bibnamefont
  {Seifert}},\ }\href {\doibase 10.1103/PhysRevLett.98.108301} {\bibfield
  {journal} {\bibinfo  {journal} {Phys. Rev. Lett.}\ }\textbf {\bibinfo
  {volume} {98}},\ \bibinfo {pages} {108301} (\bibinfo {year}
  {2007})}\BibitemShut {NoStop}%
\bibitem [{\citenamefont {Esposito}\ \emph {et~al.}(2010)\citenamefont
  {Esposito}, \citenamefont {Kawai}, \citenamefont {Lindenberg},\ and\
  \citenamefont {{Van Den Broeck}}}]{Esposito10}%
  \BibitemOpen
  \bibfield  {author} {\bibinfo {author} {\bibfnamefont {M.}~\bibnamefont
  {Esposito}}, \bibinfo {author} {\bibfnamefont {R.}~\bibnamefont {Kawai}},
  \bibinfo {author} {\bibfnamefont {K.}~\bibnamefont {Lindenberg}}, \ and\
  \bibinfo {author} {\bibfnamefont {C.}~\bibnamefont {{Van Den Broeck}}},\
  }\href {\doibase 10.1209/0295-5075/89/20003} {\bibfield  {journal} {\bibinfo
  {journal} {EPL}\ }\textbf {\bibinfo {volume} {89}},\ \bibinfo {pages} {20003}
  (\bibinfo {year} {2010})}\BibitemShut {NoStop}%
\bibitem [{\citenamefont {Rochette}\ \emph {et~al.}(2019)\citenamefont
  {Rochette}, \citenamefont {Rudolph}, \citenamefont {Roy}, \citenamefont
  {Curry}, \citenamefont {Eyck}, \citenamefont {Manginell}, \citenamefont
  {Wendt}, \citenamefont {Pluym}, \citenamefont {Carr}, \citenamefont {Ward},
  \citenamefont {Lilly}, \citenamefont {Carroll},\ and\ \citenamefont
  {Pioro-Ladri{\`{e}}re}}]{Rochette2019}%
  \BibitemOpen
  \bibfield  {author} {\bibinfo {author} {\bibfnamefont {S.}~\bibnamefont
  {Rochette}}, \bibinfo {author} {\bibfnamefont {M.}~\bibnamefont {Rudolph}},
  \bibinfo {author} {\bibfnamefont {A.-M.}\ \bibnamefont {Roy}}, \bibinfo
  {author} {\bibfnamefont {M.~J.}\ \bibnamefont {Curry}}, \bibinfo {author}
  {\bibfnamefont {G.~A.~T.}\ \bibnamefont {Eyck}}, \bibinfo {author}
  {\bibfnamefont {R.~P.}\ \bibnamefont {Manginell}}, \bibinfo {author}
  {\bibfnamefont {J.~R.}\ \bibnamefont {Wendt}}, \bibinfo {author}
  {\bibfnamefont {T.}~\bibnamefont {Pluym}}, \bibinfo {author} {\bibfnamefont
  {S.~M.}\ \bibnamefont {Carr}}, \bibinfo {author} {\bibfnamefont {D.~R.}\
  \bibnamefont {Ward}}, \bibinfo {author} {\bibfnamefont {M.~P.}\ \bibnamefont
  {Lilly}}, \bibinfo {author} {\bibfnamefont {M.~S.}\ \bibnamefont {Carroll}},
  \ and\ \bibinfo {author} {\bibfnamefont {M.}~\bibnamefont
  {Pioro-Ladri{\`{e}}re}},\ }\href {\doibase 10.1063/1.5091111} {\bibfield
  {journal} {\bibinfo  {journal} {Applied Physics Letters}\ }\textbf {\bibinfo
  {volume} {114}},\ \bibinfo {pages} {083101} (\bibinfo {year}
  {2019})}\BibitemShut {NoStop}%
\bibitem [{\citenamefont {Evers}\ \emph {et~al.}(2020)\citenamefont {Evers},
  \citenamefont {Koryt\'ar}, \citenamefont {Tewari},\ and\ \citenamefont {van
  Ruitenbeek}}]{Evers2020}%
  \BibitemOpen
  \bibfield  {author} {\bibinfo {author} {\bibfnamefont {F.}~\bibnamefont
  {Evers}}, \bibinfo {author} {\bibfnamefont {R.}~\bibnamefont {Koryt\'ar}},
  \bibinfo {author} {\bibfnamefont {S.}~\bibnamefont {Tewari}}, \ and\ \bibinfo
  {author} {\bibfnamefont {J.~M.}\ \bibnamefont {van Ruitenbeek}},\ }\href
  {\doibase 10.1103/RevModPhys.92.035001} {\bibfield  {journal} {\bibinfo
  {journal} {Rev. Mod. Phys.}\ }\textbf {\bibinfo {volume} {92}},\ \bibinfo
  {pages} {035001} (\bibinfo {year} {2020})}\BibitemShut {NoStop}%
\bibitem [{\citenamefont {Covito}\ \emph {et~al.}(2018)\citenamefont {Covito},
  \citenamefont {Eich}, \citenamefont {Tuovinen}, \citenamefont {Sentef},\ and\
  \citenamefont {Rubio}}]{Covito2018}%
  \BibitemOpen
  \bibfield  {author} {\bibinfo {author} {\bibfnamefont {F.}~\bibnamefont
  {Covito}}, \bibinfo {author} {\bibfnamefont {F.~G.}\ \bibnamefont {Eich}},
  \bibinfo {author} {\bibfnamefont {R.}~\bibnamefont {Tuovinen}}, \bibinfo
  {author} {\bibfnamefont {M.~A.}\ \bibnamefont {Sentef}}, \ and\ \bibinfo
  {author} {\bibfnamefont {A.}~\bibnamefont {Rubio}},\ }\href {\doibase
  10.1021/acs.jctc.8b00077} {\bibfield  {journal} {\bibinfo  {journal} {Journal
  of Chemical Theory and Computation}\ }\textbf {\bibinfo {volume} {14}},\
  \bibinfo {pages} {2495} (\bibinfo {year} {2018})}\BibitemShut {NoStop}%
\bibitem [{\citenamefont {Tu}(2017)}]{Tu17}%
  \BibitemOpen
  \bibfield  {author} {\bibinfo {author} {\bibfnamefont {L.~W.}\ \bibnamefont
  {Tu}},\ }\href {\doibase 10.1007/978-3-319-55084-8} {\emph {\bibinfo {title}
  {Differential Geometry}}}\ (\bibinfo  {publisher} {Springer International
  Publishing},\ \bibinfo {year} {2017})\BibitemShut {NoStop}%
\bibitem [{\citenamefont {Fox}\ and\ \citenamefont {Mayers}(1987)}]{Fox87}%
  \BibitemOpen
  \bibfield  {author} {\bibinfo {author} {\bibfnamefont {L.}~\bibnamefont
  {Fox}}\ and\ \bibinfo {author} {\bibfnamefont {D.~F.}\ \bibnamefont
  {Mayers}},\ }\href {\doibase 10.1007/978-94-009-3129-9} {\emph {\bibinfo
  {title} {Numerical Solution of Ordinary Differential Equations}}}\ (\bibinfo
  {publisher} {Springer Netherlands},\ \bibinfo {year} {1987})\BibitemShut
  {NoStop}%
\bibitem [{\citenamefont {Soriani}\ \emph {et~al.}(2022)\citenamefont
  {Soriani}, \citenamefont {Miranda},\ and\ \citenamefont
  {Bonança}}]{soriani2022failure}%
  \BibitemOpen
  \bibfield  {author} {\bibinfo {author} {\bibfnamefont {A.}~\bibnamefont
  {Soriani}}, \bibinfo {author} {\bibfnamefont {E.}~\bibnamefont {Miranda}}, \
  and\ \bibinfo {author} {\bibfnamefont {M.~V.~S.}\ \bibnamefont {Bonança}},\
  }\href {\doibase 10.1088/1367-2630/aca177} {\bibfield  {journal} {\bibinfo
  {journal} {New Journal of Physics}\ }\textbf {\bibinfo {volume} {24}},\
  \bibinfo {pages} {113037} (\bibinfo {year} {2022})}\BibitemShut {NoStop}%
\bibitem [{\citenamefont {Ciorga}\ \emph {et~al.}(2000)\citenamefont {Ciorga},
  \citenamefont {Sachrajda}, \citenamefont {Hawrylak}, \citenamefont {Gould},
  \citenamefont {Zawadzki}, \citenamefont {Jullian}, \citenamefont {Feng},\
  and\ \citenamefont {Wasilewski}}]{Ciorga2000Addition}%
  \BibitemOpen
  \bibfield  {author} {\bibinfo {author} {\bibfnamefont {M.}~\bibnamefont
  {Ciorga}}, \bibinfo {author} {\bibfnamefont {A.~S.}\ \bibnamefont
  {Sachrajda}}, \bibinfo {author} {\bibfnamefont {P.}~\bibnamefont {Hawrylak}},
  \bibinfo {author} {\bibfnamefont {C.}~\bibnamefont {Gould}}, \bibinfo
  {author} {\bibfnamefont {P.}~\bibnamefont {Zawadzki}}, \bibinfo {author}
  {\bibfnamefont {S.}~\bibnamefont {Jullian}}, \bibinfo {author} {\bibfnamefont
  {Y.}~\bibnamefont {Feng}}, \ and\ \bibinfo {author} {\bibfnamefont
  {Z.}~\bibnamefont {Wasilewski}},\ }\href {\doibase
  10.1103/PhysRevB.61.R16315} {\bibfield  {journal} {\bibinfo  {journal} {Phys.
  Rev. B}\ }\textbf {\bibinfo {volume} {61}},\ \bibinfo {pages} {R16315}
  (\bibinfo {year} {2000})}\BibitemShut {NoStop}%
\bibitem [{\citenamefont {Elzerman}\ \emph {et~al.}(2003)\citenamefont
  {Elzerman}, \citenamefont {Hanson}, \citenamefont {Greidanus}, \citenamefont
  {Willems~van Beveren}, \citenamefont {De~Franceschi}, \citenamefont
  {Vandersypen}, \citenamefont {Tarucha},\ and\ \citenamefont
  {Kouwenhoven}}]{Elzerman2003Few}%
  \BibitemOpen
  \bibfield  {author} {\bibinfo {author} {\bibfnamefont {J.~M.}\ \bibnamefont
  {Elzerman}}, \bibinfo {author} {\bibfnamefont {R.}~\bibnamefont {Hanson}},
  \bibinfo {author} {\bibfnamefont {J.~S.}\ \bibnamefont {Greidanus}}, \bibinfo
  {author} {\bibfnamefont {L.~H.}\ \bibnamefont {Willems~van Beveren}},
  \bibinfo {author} {\bibfnamefont {S.}~\bibnamefont {De~Franceschi}}, \bibinfo
  {author} {\bibfnamefont {L.~M.~K.}\ \bibnamefont {Vandersypen}}, \bibinfo
  {author} {\bibfnamefont {S.}~\bibnamefont {Tarucha}}, \ and\ \bibinfo
  {author} {\bibfnamefont {L.~P.}\ \bibnamefont {Kouwenhoven}},\ }\href
  {\doibase 10.1103/PhysRevB.67.161308} {\bibfield  {journal} {\bibinfo
  {journal} {Phys. Rev. B}\ }\textbf {\bibinfo {volume} {67}},\ \bibinfo
  {pages} {161308} (\bibinfo {year} {2003})}\BibitemShut {NoStop}%
\bibitem [{\citenamefont {Koski}\ \emph
  {et~al.}(2014{\natexlab{a}})\citenamefont {Koski}, \citenamefont {Maisi},
  \citenamefont {Pekola},\ and\ \citenamefont {Averin}}]{Koski2014}%
  \BibitemOpen
  \bibfield  {author} {\bibinfo {author} {\bibfnamefont {J.~V.}\ \bibnamefont
  {Koski}}, \bibinfo {author} {\bibfnamefont {V.~F.}\ \bibnamefont {Maisi}},
  \bibinfo {author} {\bibfnamefont {J.~P.}\ \bibnamefont {Pekola}}, \ and\
  \bibinfo {author} {\bibfnamefont {D.~V.}\ \bibnamefont {Averin}},\ }\href
  {\doibase 10.1073/pnas.1406966111} {\bibfield  {journal} {\bibinfo  {journal}
  {Proceedings of the National Academy of Sciences}\ }\textbf {\bibinfo
  {volume} {111}},\ \bibinfo {pages} {13786} (\bibinfo {year}
  {2014}{\natexlab{a}})}\BibitemShut {NoStop}%
\bibitem [{\citenamefont {Koski}\ \emph
  {et~al.}(2014{\natexlab{b}})\citenamefont {Koski}, \citenamefont {Maisi},
  \citenamefont {Sagawa},\ and\ \citenamefont {Pekola}}]{Koski2014Exp}%
  \BibitemOpen
  \bibfield  {author} {\bibinfo {author} {\bibfnamefont {J.~V.}\ \bibnamefont
  {Koski}}, \bibinfo {author} {\bibfnamefont {V.~F.}\ \bibnamefont {Maisi}},
  \bibinfo {author} {\bibfnamefont {T.}~\bibnamefont {Sagawa}}, \ and\ \bibinfo
  {author} {\bibfnamefont {J.~P.}\ \bibnamefont {Pekola}},\ }\href {\doibase
  10.1103/PhysRevLett.113.030601} {\bibfield  {journal} {\bibinfo  {journal}
  {Phys. Rev. Lett.}\ }\textbf {\bibinfo {volume} {113}},\ \bibinfo {pages}
  {030601} (\bibinfo {year} {2014}{\natexlab{b}})}\BibitemShut {NoStop}%
\bibitem [{\citenamefont {Ismail-Beigi}(2013)}]{Beigi13}%
  \BibitemOpen
  \bibfield  {author} {\bibinfo {author} {\bibfnamefont {S.}~\bibnamefont
  {Ismail-Beigi}},\ }\href
  {https://volga.eng.yale.edu/sites/default/files/files/general-lorentzian-integrals.pdf}
  {\bibfield  {journal} {\bibinfo  {journal} {Yale notes}\ } (\bibinfo {year}
  {2013})}\BibitemShut {NoStop}%
\bibitem [{\citenamefont {Erdman}\ \emph {et~al.}(2023)\citenamefont {Erdman},
  \citenamefont {Rolandi}, \citenamefont {Abiuso}, \citenamefont
  {Perarnau-Llobet},\ and\ \citenamefont {No\'e}}]{Erdman2023Pareto}%
  \BibitemOpen
  \bibfield  {author} {\bibinfo {author} {\bibfnamefont {P.~A.}\ \bibnamefont
  {Erdman}}, \bibinfo {author} {\bibfnamefont {A.}~\bibnamefont {Rolandi}},
  \bibinfo {author} {\bibfnamefont {P.}~\bibnamefont {Abiuso}}, \bibinfo
  {author} {\bibfnamefont {M.}~\bibnamefont {Perarnau-Llobet}}, \ and\ \bibinfo
  {author} {\bibfnamefont {F.}~\bibnamefont {No\'e}},\ }\href {\doibase
  10.1103/PhysRevResearch.5.L022017} {\bibfield  {journal} {\bibinfo  {journal}
  {Phys. Rev. Res.}\ }\textbf {\bibinfo {volume} {5}},\ \bibinfo {pages}
  {L022017} (\bibinfo {year} {2023})}\BibitemShut {NoStop}%
\end{thebibliography}%

\newpage
\widetext
\appendix
\tableofcontents

\section{Solving the exact dynamics}\label{sm:solution_rl}
\subsection{Solving the Heisenberg equations}
We consider a single fermionic mode coupled to a fermionic bath. Without loss of generality we can set the chemical potential to $0$ and the ground state of the two-level system to $0$.
The Hamiltonians of the system and bath are
\begin{align}
	\hat H_S(t) &=  \varepsilon(t) \hat a^\dagger \hat a~, \\
	\hat H_B &= \sum_{k=1}^n \omega_k \hat b_k^\dagger \hat b_k~,
\end{align}
where $\hat a^\dagger$ is the creation operator of the two-level system and $\hat b_k^\dagger$ is the creation operator of a bath mode with frequency $\omega_k$. These ladder operators follow the canonical anticommutation relations. For the interaction between system and bath, the Hamiltonian is
\begin{equation}
	\hat H_{\rm int}(t) = g(t) \hat V = g(t) \sum_{k=1}^n \lambda_k \hat a^\dagger \hat b_k + \lambda_k^*  \hat b_k^\dagger \hat a ~,
\end{equation}
where the $\lambda_k$ are the interaction weights.\\

Will will consider that $\varepsilon(t)$ and $g(t)$ are the control parameters to then perform the erasure of information in the single mode of the system. We now proceed to solve the dynamics of the system and bath in the Heisenberg picture.\\
For an operator $\hat A$ in the Schrödinger picture, we denote by $\hat A_H(t)$ the corresponding operator in the the Heisenberg picture. The evolution of $\hat A_H(t)$ is defined by the Heisenberg equation of motion:
\begin{equation}
	\frac{d}{dt}\hat A_H(t) = i\left[\hat H(t),\hat A_H(t)\right]~,
\end{equation}
where $\hat H(t) = \hat H_S(t) + \hat H_{\rm int}(t) + \hat H_B$ and $\hbar = 1$. Applying this equation to the ladder operators $\hat a_H(t)$ and $\hat b_{k,H}(t)$ we find the following system of $n+1$ equations:
\begin{align}
	\label{eom_a_0} \frac{d}{dt}\hat a_H(t) &= - i \varepsilon(t)\hat a_H(t) - i g(t)\sum_k \lambda_k \hat b_{k,H}(t)~, \\
	\label{eom_b} \frac{d}{dt}\hat b_{k,H}(t) &=  - i \omega_k\hat b_{k,H}(t) -  i g(t) \lambda_k^* \hat a_H(t)~.
\end{align}

By defining $\hat u_k(t) = e^{i\omega_k t} \hat b_{k,H}(t)$ we can see that \eref{eom_b} becomes
$$ e^{-i\omega_k t}\frac{d}{dt}\hat u_k(t) = -  i g(t) \lambda_k^* \hat a_H(t)~, $$
which is solved by $\hat u_k(t) = \hat u_k(0) - i\lambda_k^*\int_0^tds ~ e^{i\omega_k s} g(s)\hat a_H(s)$. We therefore find
\begin{equation}\label{b_sol_0}
	\hat b_{k,H}(t) = e^{-i\omega_k t}\hat b_{k} - i\lambda_k^*\int_0^tds ~ g(s)\hat a_H(s)e^{i\omega_k (s-t)}~,
\end{equation}
where we used that $\hat u_k(0) = \hat b_{k,H}(0) = \hat b_{k}$. Therefore 
\begin{equation}\label{sum_b}
	\sum_k \lambda_k\hat b_{k,H}(t) = \hat \xi(t) - i\int_0^tds ~  \chi(s-t) g(s)\hat a_H(s)~,
\end{equation}
where we defined the noise operator $\hat\xi(t) = \sum_k e^{-i\omega_k t}\lambda_k\hat b_{k}$ and $\chi(t) = \sum_k e^{i\omega_k t} |\lambda_k|^2$. We can notice that $\chi(t)$ is the (symmetrized) noise correlation function: 
$$\left\langle\left\{\hat \xi^{\dagger}(t),\hat \xi(0)\right\}\right\rangle = \sum_{j,k}e^{i\omega_k t}\lambda_k^* \lambda_j\left\langle\left\{\hat b_k^{\dagger},\hat b_j \right\}\right\rangle = \sum_{k}e^{i\omega_k t}|\lambda_k|^2 = \chi(t) ~,$$
and its Fourier transform is the (unit-less) spectral density of the bath $\mathfrak{J}(\omega) = 2\pi\sum_k |\lambda_k|^2\delta(\omega-\omega_k)$
By inserting \eref{sum_b} into equation \eref{eom_a_0} we find an equation of motion for $\hat a_H(t)$:
\begin{equation}\label{eom_a}
	\frac{d}{dt}\hat a_H(t) = - i \varepsilon(t)\hat a_H(t) - i g(t)\hat\xi(t) - g(t)\int_0^tds~\chi(s-t)g(s)\hat a_H(s)~.
\end{equation}
In order to solve \eref{eom_a} we will need to explicitly take the continuum limit so that our bath indeed becomes a bath. We can take its spectral density to be either a Lorentzian $\mathfrak{J}(\omega) = \frac{\Lambda^2}{\Lambda^2 + \omega^2}$ or a pass-band $\mathfrak{J}(\omega) = \Theta(\Lambda-|\omega|)$ ($\Theta$ is the Heaviside step function). We will need to assume that we are working in the wide-band approximation ($\Lambda \rightarrow \infty$). More practically, we are assuming that the bath interaction is the same over the energies we are spanning with the system. This limit allows us to say that the noise correlation function is negligible for time differences larger than zero:
$$\lim_{\Lambda\rightarrow\infty}\lim_{n\rightarrow\infty}\chi(t) = \delta(t)~.$$
In this limit \eref{eom_a} becomes considerably simpler:
\begin{equation}\label{eom_a_wide}
	\frac{d}{dt}\hat a_H(t) = - \left(i \varepsilon(t) +\frac{1}{2}g(t)^2\right)\hat a_H(t) - i g(t)\hat\xi(t)~.
\end{equation}
Similarly to how we solved \eref{eom_b}, we define $\hat u(t) = \exp\left[\int_0^t z(s)ds\right] \hat a_H(t)$ for $z(t):= i\varepsilon(t) + \frac{1}{2}g(t)^2$. Now we have 
$$e^{-\int_0^t z(s)ds}\frac{d}{dt}\hat u(t) = - i g(t)\hat\xi(t)~, $$
which is solved by $\hat u(t) = \hat u(0) - i\int_0^tds~ g(s)\exp\left[\int_{0}^s z(r)dr\right]\hat\xi(s)$. We therefore find the solution of the evolution of the ladder operator of the distinguished mode:
\begin{equation}\label{a_sol}
	\hat a_H(t) = G(t,0)\hat a - i \int_0^t ds~g(s)G(t,s)\hat\xi(s)
\end{equation}
where we defined the propagator $G(t,s) = \exp\left[-\int_s^t z(r)dr\right]$~. And from \eref{b_sol_0} we find the solution for the bath modes:
\begin{equation}\label{b_sol}
	\hat b_{k,H}(t) =  e^{-i\omega_k t}\hat b_{k} - i\lambda_k^*\int_0^tds ~ g(s)G(s,0)\hat a e^{i\omega_k (s-t)} - \lambda_k^*\int_0^tds\int_0^s dr~g(s)g(r)G(s,r)\hat\xi(r)e^{i\omega_k (s-t)}~.
\end{equation}

\subsection{Relevant observables}
Since we are performing an erasure, we will assume the system starts in a factorized state and that the bath starts in a thermal state at inverse temperature $\beta$ state with respect to its Hamiltonian: 
$$\hat\rho(0) = \hat\rho_S(0)\otimes\frac{e^{-\beta\hat H_B}}{Z_B}~,\qquad Z_B = \Tr[e^{-\beta \hat H_B}]~.$$

We are interested in computing the occupation probability of the excited level $p(t) = \langle \hat a^{\dagger}\hat a\rangle$ and the system-bath interaction potential $v(t) = \langle\hat V\rangle$. From \eref{a_sol} we have
\begin{equation}\label{p_gen_0}
	p(t) = \Tr\!\left[\hat\rho(0)\hat a_H^{\dagger}(t)\hat a_H(t)\right] = \left|G(t,0)\right|^2 p(0)  +  \int_0^t dsdr~g(s)g(r)G^*(t,s) G(t,r)\Tr\!\left[\frac{e^{-\beta H_B}}{Z_B} \hat\xi^{\dagger}(s)\hat\xi(r)\right]~,
\end{equation}
where we used the CAR to get $\Tr[\hat a^{\dagger}\hat \xi(s)] = 0$ and drop the cross terms. Further using the CAR we simplify the remaining trace in the integral:
\begin{equation} \label{eval_tr_bath_1}
	\Tr\!\left[\frac{e^{-\beta H_B}}{Z_B} \hat\xi^{\dagger}(s)\hat\xi(r)\right] =  \sum_k e^{i\omega_k(s-r)} |\lambda_k|^2 \Tr\!\left[\frac{e^{-\beta H_B}}{Z_B} \hat b_k^{\dagger}\hat b_k\right] = \sum_k e^{i\omega_k(r-s)} |\lambda_k|^2 f_\beta(\omega_k)~,
\end{equation}
where $f_\beta(\omega) = (1+e^{\beta \omega})^{-1}$ is the Fermi-Dirac distribution. We can apply the continuum limit to \eref{eval_tr_bath_1} by using the equality $2\pi\sum_k|\lambda_k|^2 h(\omega_k) = \int d\omega~ \mathfrak{J}(\omega)h(\omega) $, which holds for any function $h$ by definition of $\mathfrak{J}$. We can then apply the wideband limit by using $\lim_{\Lambda\rightarrow\infty}\lim_{n\rightarrow\infty}\mathfrak{J}(\omega) = 1$. We find
\begin{equation} \label{eval_tr_bath_2}
	\Tr\!\left[\frac{e^{-\beta H_B}}{Z_B} \hat\xi^{\dagger}(s)\hat\xi(r)\right]  =\int_{-\infty}^\infty \frac{d\omega}{2\pi} e^{i\omega(s-r)} \mathfrak{J}(\omega) f_\beta(\omega) \xrightarrow{\Lambda\to\infty} \int_{-\infty}^\infty \frac{d\omega}{2\pi} e^{i\omega(s-r)} f_\beta(\omega)~.
\end{equation}
Applying \eref{eval_tr_bath_2} to \eref{p_gen_0} we get
\begin{equation}\label{p_gen}
	p(t) = \left|G(t,0)\right|^2 p(0)  + \frac{1}{2\pi}\int_{-\infty}^{\infty}d\omega~  f_\beta(\omega) \int_0^t ds\int_0^t dr~g(s)g(r)G^*(t,s) G(t,r)e^{i\omega(s-r)}~.
\end{equation}
Using \eref{sum_b}, the CAR and \eref{a_sol}, we have
\begin{align*}\label{v_gen_0}
	v(t) =&\; \Tr\!\left[\hat\rho(0) \hat a^\dagger_H(t) \left(\sum_k\lambda_k \hat b_H(t)\right)\right] + h.c.~ \\
	  = &\;\Tr\!\left[\hat\rho(0) \hat a^\dagger_H(t) \hat \xi(t)\right] - i\int_0^tds ~ \chi(s-t) g(s)\Tr\!\left[\hat\rho(0) \hat a^\dagger_H(t)\hat a_H(s)\right] + h.c. ~,\\
	 \xrightarrow{\Lambda\to \infty}&\; \Tr\!\left[\hat\rho(0) \hat a^\dagger_H(t) \hat \xi(t)\right] - i \frac{g(t)}{2} \Tr\!\left[\hat\rho(0) \hat a^\dagger_H(t)\hat a_H(t)\right] + h.c.~, \\
	 = &\;\Tr\!\left[\hat\rho(0) \hat a^\dagger_H(t) \hat \xi(t)\right] + h.c. ~, \\
	 = &\; i \int_0^t ds~g(s)G^*(t,s)\Tr\!\left[\frac{e^{-\beta H_B}}{Z_B}\hat\xi^{\dagger}(s) \hat \xi(t)\right] + h.c. ~,\\
	 = &\; \frac{i}{2\pi} \int_{-\infty}^\infty d\omega~f_\beta(\omega) \int_0^t ds~g(s)G^*(t,s)e^{i\omega(s-t)} + h.c. ~.\\
 \end{align*}
So we have
\begin{equation}\label{v_gen}
v(t) = \frac{1}{\pi}\Im \int_{-\infty}^\infty d\omega~f_\beta(\omega) \int_0^t ds~g(s)G(t,s)e^{i\omega(t-s)}~.
\end{equation}
\subsection{Proof of requirement 1}
We will now proceed to prove that, in absence of driving, $p(t)$ and $v(t)$ thermalize. We do so in two steps, we first simplify the expressions of \eref{p_gen} and \eref{v_gen} for $\eps(t) = \eps$ and $g(t) = g$ and compute the infinite time limit. Then we compute the thermal expectation value of the corresponding observables and prove that the obtained expressions are the same.
\subsubsection{Infinite time limit in absence of driving}
By assuming that the driving parameters are kept constant the propagator becomes 
\begin{equation}\label{frozenH_prop}
    G(t,s) = e^{-(t-s)(\frac{1}{2}g^2 + i\eps)}~.
\end{equation}
This allows us to compute the time integrals in \eref{p_gen} and \eref{v_gen}:
\begin{align}
    \label{p_frozen}
    p(t) &= p(0)e^{-g^2 t} + \frac{g^2}{2\pi} \int_{-\infty}^{\infty}d\omega~ f_\beta(\omega) \frac{1-2e^{-g^2t/2}\cos([\omega-\eps]t) + e^{-g^2 t}}{g^4/4 + (\omega-\eps)^2}~,\\
    \label{v_frozen}
    v(t) &= \frac{g}{\pi} \int_{-\infty}^{\infty}d\omega~ f_\beta(\omega) \frac{(\omega-\eps)\left[ 1 - e^{-g^2 t/2}\cos([\omega-\eps]t) \right]- \frac{1}{2}g^2 e^{-g^2 t/2}\sin([\omega-\eps]t)  }{g^4/4 + (\omega-\eps)^2}~.
\end{align}
As a side-note, it is interesting to note that the frequency integral of \eref{p_frozen} can be solved to give the following expression for the occupation probability
\begin{multline}
    p(t) = 1 + p(0)e^{-g^2 t} + \frac{\sinh(\beta\eps)}{\cosh(\beta\eps)+ \cos(\frac{\beta g^2}{2})} + \left(1+e^{-g^2 t}\right)\left[\frac{1}{2}-\frac{1}{\pi}\Im \psi^{(0)}\!\!\left(\frac{1}{2}+ \frac{\beta}{2\pi}\left(\frac{g^2}{2}+ i\eps \right)\right)\right] \\
    + \frac{e^{-g^2 t}}{\pi}\Im B\!\left(e^{2\pi t/\beta};\frac{1}{2}+ \frac{\beta}{2\pi}\left(\frac{g^2}{2}- i\eps \right),0\right) + \frac{1}{\pi}\Im B\!\left(e^{2\pi t/\beta};\frac{1}{2} - \frac{\beta}{2\pi}\left(\frac{g^2}{2}- i\eps \right),0\right)~,
\end{multline}
where $\psi^{(0)}(z) = \frac{d}{dz}\ln\Gamma(z)$ is the digamma function (defined as the logarithmic derivative of the Gamma function) and $B(x;a,b) = \int_0^xds~ s^{a-1}(1-s)^{b-1}$ is the incomplete beta function. This expression is useful for numerical implementations as it is faster to compute than the integral of \eref{p_frozen}.

By taking the limit $t\rightarrow\infty$ in \eref{p_frozen} and \eref{v_frozen} we find
\begin{align}
    \label{p_thm_0}
    \lim_{t\rightarrow\infty} p(t) &=\int_{-\infty}^{\infty} \frac{d\omega}{\pi} f_\beta(\omega) \frac{g^2/2}{g^4/4 + (\omega-\eps)^2}~,\\
    \label{v_thm_0}
    \lim_{t\rightarrow\infty} v(t) &= g \int_{-\infty}^{\infty} \frac{d\omega}{\pi} f_\beta(\omega) \frac{(\omega-\eps) }{g^4/4 + (\omega-\eps)^2}~.
\end{align}
Here we can notice that if we take the Laplace transform of the propagator we obtain
\begin{equation}
    \tilde G(z) := \int_0^\infty dt~ G(t,0)e^{-zt} = \frac{1}{z+i\eps+g^2/2}~,
\end{equation}
which allows us to rewrite \eref{p_thm_0} and \eref{v_thm_0} as
\begin{align}
    \label{p_thm}
    \lim_{t\rightarrow\infty} p(t) &=\int_{-\infty}^{\infty} \frac{d\omega}{\pi} f_\beta(\omega) \Re\!\left[\tilde G(-i\omega)\right]~,\\
    \label{v_thm}
    \lim_{t\rightarrow\infty} v(t) &= g \int_{-\infty}^{\infty} \frac{d\omega}{\pi} f_\beta(\omega) \Im\!\left[\tilde G(-i\omega)\right]~.
\end{align}

\subsubsection{Thermal expectation value}
We  now compute the expectation value of $\hat a^\dagger \hat a$ and $\hat V$ when the state is a Gibbs state
$$ \hat \omega_{\beta} := \frac{e^{-\beta \hat H}}{Z} = \frac{\exp\!\left[-\beta\eps \hat a^\dagger \hat a -\beta g \hat V -\beta \hat H_B\right]}{Z}~, \qquad Z = \Tr[e^{-\beta\hat H}]~.$$
Therefore we want to find $p_{th} := \Tr[\hat \omega_\beta \hat a^\dagger \hat a]$ and $v_{th} := \Tr[\hat \omega_\beta \hat V]$. Using the fact that the total Hamiltonian is quadratic, we can diagonalize it to rewrite it in the following way
\begin{equation}
    \hat H  = \sum_k \eps_k \hat c_k^\dagger\hat c_k~,
\end{equation}
where $\eps_k$ are eigen-energies and $\hat c_k$ are fermionic ladder operators that follow the CAR: $\{\hat c_j^\dagger,\hat c_k\} = \delta_{jk}\mathbb{1}$, $\{\hat c_j,\hat c_k\} = 0$. They are related to the original ones in the following way
\begin{align}
    \hat a &= \sum_k \braket{0|\hat a|k}\hat c_k~, \\
    \hat b_j &= \sum_k \braket{0|\hat b_j|k}\hat c_k~,
\end{align} 
where $\ket{k} = \hat c_k^\dagger \ket{0} $ are $1$-particle eigenstates of the Hamiltonian with eigenvalue $\eps_k$. Inserting this in the expression for the thermal expectation of the probability of occupation we find
\begin{equation}
    p_{th} = \frac{1}{Z}\sum_{jk} \braket{j|\hat a^\dagger|0}\!\braket{0|\hat a|k}\Tr\!\left[e^{-\beta \hat H}\hat c_j^\dagger \hat c_k\right] 
           = \sum_{k} \left|\braket{k|\hat a^\dagger|0}\right|^2 \frac{\Tr\!\left[e^{-\beta\eps_k\hat c_k^\dagger \hat c_k}\hat c_k^\dagger \hat c_k\right]}{\Tr\!\left[e^{-\beta\eps_k \hat c_k^\dagger \hat c_k}\right]}
           = \sum_{k} \left|\braket{k|\hat a^\dagger|0}\right|^2 f_\beta(\eps_k)~.
\end{equation} 
From \eref{a_sol} it is easy to see that we can write the propagator in the following way
\begin{equation}
    G(t,0) = \braket{0|\hat a_H(t) \hat a^\dagger|0}
           = \braket{0|\hat U^\dagger(t)\hat a \hat U(t)\hat a^\dagger|0}
           = \sum_k e^{-i \eps_k t}\braket{0|\hat U^\dagger(t)\hat a|k}\!\braket{k|\hat a^\dagger|0}
           = \sum_k e^{-i \eps_k t}\left|\braket{k|\hat a^\dagger|0}\right|^2 ~,
\end{equation}
where we used the fact that the vacuum state does not evolve $\hat U(t)\ket{0} = \ket{0}$ and that since we are performing no driving we have $\hat U(t) = e^{-it\sum_k \eps_k \hat c^\dagger_k \hat c_k}$. Note that the sum needs only to be over $1$-particle states as there is a scalar product with the $1$-particle state $\hat a^\dagger\ket{0}$. By now defining $\varphi(\omega) := \sum_k \left|\braket{k|\hat a^\dagger|0}\right|^2 \delta(\omega-\eps_k) $ we can identify
\begin{align}
    \label{eq:G_ft_phi}
    G(t,0) &= \int_{-\infty}^\infty d\omega~ \varphi(\omega) e^{-i\omega t} ~,\\
    p_{th} &= \int_{-\infty}^\infty d\omega~ f_\beta(\omega) \varphi(\omega)~.
\end{align}
Considering \eref{p_thm} it is clear that if $\varphi(\omega) = \frac{1}{\pi}\Re\!\left[\tilde G(-i\omega)\right]$ then we have proven $p_{th} = \lim_{t\rightarrow\infty}p(t)$. Therefore we compute the Laplace transform of $G(t,0)$ using \eref{eq:G_ft_phi}
\begin{align*}
    \tilde G(-i\omega) &= \int_0^\infty dt~ G(t,0)e^{i\omega t} ~,\\
                &= \int_0^\infty dt\int_{-\infty}^\infty d\omega'~ e^{i(\omega-\omega')t} \varphi(\omega')~,\\
                &= \int_{-\infty}^\infty d\omega'~\varphi(\omega')\int_{-\infty}^\infty dt~ \Theta(t)e^{i(\omega-\omega')t}~,\\
                &= \pi \varphi(\omega) + i P.\int_{-\infty}^\infty d\omega'~\frac{\varphi(\omega')}{\omega-\omega'}~,
\end{align*}
where $P.$ denotes the Cauchy principal value, $\Theta(t)$ is the Heaviside step function and we used that its Fourier transform is (in a distributional sense) $\int dt~ e^{ist}\Theta(t) = \pi \delta(s) + P. \frac{i}{s}$. Since $\varphi(\omega)$ is by definition a real function we can see that $P.\int_{-\infty}^\infty d\omega'~\frac{\varphi(\omega')}{\omega-\omega'}$ is a real number. Therefore we can conclude $\varphi(\omega) = \frac{1}{\pi}\Re\!\left[\tilde G(-i\omega)\right]$. Which concludes the proof of the thermalization of $p(t)$.

To prove the thermalization of $v(t)$ we proceed in a similar fashion. We start by computing $v_{th}$
\begin{equation}
    v_{th} = \frac{1}{Z}\sum_{jk}\lambda_j\braket{k|\hat a^\dagger|0}\! \braket{0|\hat b_j|k}\Tr\!\left[e^{-\beta \hat H}\hat c_k^\dagger \hat c_k\right] + h.c. = \sum_{jk}f_\beta(\eps_k)\left(\lambda_j\braket{k|\hat a^\dagger|0}\! \braket{0|\hat b_j|k}+ \lambda_j^*\braket{k|\hat b_j^\dagger|0}\! \braket{0|\hat a|k}\right)~.
\end{equation}
To proceed we have to define the following cross-propagators
\begin{align}
\lambda_j K_j(t) &:= \braket{0|\hat a_H(t) \hat b^\dagger_j |0} = \sum_k e^{-i\eps_k t}\braket{k|\hat b_j^\dagger|0}\! \braket{0|\hat a|k} = \int_{-\infty}^\infty d\omega ~\psi_j(\omega)e^{-i\omega t}~,\\
\lambda_j^* H_j(t) &:= \braket{0|\hat b_{j,H}(t) \hat a^\dagger |0} = \sum_k e^{-i\eps_k t}\braket{k|\hat a^\dagger|0}\! \braket{0|\hat b_j|k} = \int_{-\infty}^\infty d\omega ~\psi_j^*(\omega)e^{-i\omega t}~,
\end{align}
where we defined $\psi_j(\omega) = \sum_k \braket{k|\hat b_j^\dagger|0}\! \braket{0|\hat a|k} \delta(\omega-\eps_k)$. By further defining $\psi_0(\omega): = \sum_k \lambda_k^*\psi_k(\omega)$ and $\psi(\omega) = \psi_0(\omega)+\psi_0^*(\omega)$ we can see that
\begin{align}
    K(t) &:= \sum_j |\lambda_j|^2\left(K_j(t) + H_j(t) \right) = \int_{-\infty}^\infty d\omega ~\psi (\omega)e^{-i\omega t}~, \\
    \label{v_thm_proof}
    v_{th} &= \int_{-\infty}^\infty d\omega~ f_\beta(\omega) \psi(\omega)~.
\end{align}
Therefore, similarly to the case of $G(t,0)$, we have
\begin{equation*}
    \tilde K(-i\omega) = \pi \psi(\omega) + i P.\int_{-\infty}^\infty d\omega' \frac{\psi(\omega')}{\omega-\omega'}~,
\end{equation*}
and in particular $\pi \psi(\omega) = \Re\left[\tilde K(-i\omega)\right]$ (since $\psi(\omega)$ is real by definition). Hence, by \eref{v_thm} and \eref{v_thm_proof}, the last step to prove that $v(t)$ thermalizes is to check that $\Re\left[\tilde K(-i\omega)\right] = g\Im\left[\tilde G(-i\omega)\right]$. To do so we start by computing the components of $K(t)$: from \eref{a_sol} we can see that
\begin{equation}
\begin{split}
    K_j(t) &= -i g \int_0^t ds~G(t,s) e^{-i\omega_j s}~,\\
           &= -i g \int_0^t ds~ e^{-(t-s)(\frac{1}{2}g^2 +i \eps)} e^{-i\omega_j s} ~,\\
           &= -i g e^{-t(\frac{1}{2}g^2 +i \eps)}  \int_0^t ds~ e^{s(\frac{1}{2}g^2 +i (\eps-\omega_j))} ~,\\
           &= -ig\frac{e^{-i \omega_j t} - e^{-(\frac{1}{2}g^2 +i \eps)t} }{\frac{1}{2}g^2 +i (\eps-\omega_j)} = -ig\frac{e^{-i \omega_j t} - G(t,0)}{\frac{1}{2}g^2 +i (\eps-\omega_j)}~;
\end{split}
\end{equation}
and from \eref{b_sol}
\begin{equation}
    H_j(t) = - ig\int_0^tds ~ G(s,0) e^{i\omega_j (s-t)}
           = - ig e^{-i\omega_j t}\int_0^tds ~ e^{-(\frac{1}{2}g^2+i(\eps-\omega_j))s}
           = - ig \frac{ e^{-i\omega_j t} - e^{-(\frac{1}{2}g^2+i\eps)t} }{\frac{1}{2}g^2+i(\eps-\omega_j)} = K_j(t)~.
\end{equation}
Since the time time dependence is contained in the exponentials, it is straightforward to compute the Laplace transform
\begin{multline}
    \tilde K_j(z) = \tilde H_j(z) = \frac{- i g}{\frac{1}{2}g^2 +i (\eps-\omega_j)}\int_0^\infty dt \left[e^{-i \omega_j t} - G(t,0)\right]e^{-zt} \\
    = \frac{- i g}{\frac{1}{2}g^2 +i (\eps-\omega_j)}\left[\frac{1}{z+i\omega_j}-  \frac{1}{z+i\eps+g^2/2}\right] =  \frac{- i g}{z+i \omega_j}\tilde G(z)~.
\end{multline}
Therefore we find
\begin{equation}
    \tilde K(-i\omega) = \frac{g}{\pi} \tilde G(-i\omega) P.\int_{-\infty}^\infty  d\omega' \frac{1}{\omega-\omega'} = -ig \tilde G(-i\omega) ~,
\end{equation}
which allows us to conclude $\psi(\omega) = g\Im\!\left[\tilde G(-i\omega)\right]$. This concludes the proof of the thermalization of $v(t)$.

\section{Slow driving expansion}\label{sm:sd}
\subsection{Deriving the thermodynamic metric}
We are interested in performing an erasure protocol and minimizing the work cost of performing it. The erasure protocol is one where $\varepsilon(0) = 0$, $\varepsilon(\tau) \gg \beta^{-1}$ and $g(0) = g(\tau) = 0$, for $\tau$ the total time of the protocol. The work cost of a protocol where we control $\varepsilon$ and $g$ is
\begin{equation}\label{work_0}
	W = \int_0^\tau dt~ \Tr\!\left[\hat\rho(t) \frac{d}{dt}\hat{H}(t)\right] = \int_0^\tau dt~ \dot\varepsilon(t) p(t) + \dot g(t) v(t)~.
\end{equation}
To get a correction to Landauer's bound for finite time protocols, and to work with more tractable expressions, we  expand \eref{work_0} in the long times limit up to first order. To do that we first need to make some notation changes. First we  make the time parameter in $\varepsilon$ and $g$ dimensionless, so that the protocol starts at ``time'' input parameter $0$ and ends at ``time'' input parameter $1$. So we have the following mappings: $t\rightarrow \tau t$, $\int dt \rightarrow \tau \int dt$ and $\frac{d}{dt} \rightarrow \tau^{-1}\frac{d}{dt}$. Second we need to ``extract'' the evolution timescale of the system in order to make the slow driving expansion. From \eref{p_gen}, \eref{v_gen} and the definition of the propagator (or more clearly form \eref{p_frozen} and \eref{v_frozen}) it is quite clear that the relaxation timescale of the system, at any point of the evolution, is of the order $(g(t)^2)^{-1}$. Hence we are going take the average of the square of the coupling as normalizing factor, we therefore define (in normalized time) $\Gamma := \int_0^1 dt~ g(t)^2$. We now define a normalized version of our control parameters:
\begin{equation}\label{norm_params}
	\epsilon(t) := \frac{1}{\Gamma}\varepsilon(t)~,\qquad \gamma(t) := \frac{1}{2\Gamma}g(t)^2~.
\end{equation}
We can therefore write the expression for work cost in this new convention
\begin{equation}\label{norm_work}
	W =  \int_0^1 dt~ \dot\epsilon(t)\Gamma p(t) + \dot \gamma(t) \sqrt{\frac{\Gamma}{2\gamma(t)}}v(t)~.
\end{equation}
We can also rewrite the propagator
\begin{equation}\label{norm_prop}
	G(t,s) = \exp\left[-\tau\Gamma\int_s^t dr~ \gamma(r)+ i\epsilon(r)\right]~,
\end{equation}
and the expectation values of the observables
\begin{align}
	\label{norm_p} p(t) &= \left|G(t,0)\right|^2 p_0  + \frac{1}{\pi}\int_{-\infty}^{\infty}d\omega~  f_\beta(\omega\Gamma) \left|\tau\Gamma\! \int_0^t ds~ \gamma(s)^{\frac{1}{2}} G(t,s)e^{i\tau \omega\Gamma (t-s)}\right|^2~,\\
	\label{norm_v} v(t) &= \frac{\tau\Gamma\sqrt{2\Gamma}}{\pi}\Im \int_{-\infty}^\infty d\omega~f_\beta(\omega\Gamma)  \int_0^t ds~ \gamma(s)^{\frac{1}{2}}G(t,s)e^{i\tau\Gamma\omega(t-s)}~,
\end{align}
where we were able to insert a phase in the time integral of \eref{norm_p} because of the absolute value and rescaled $\omega$ by $\Gamma$. We can see that we have to expand in $1/\tau\Gamma$ the same integral for both \eref{norm_p} and \eref{norm_v}. To do that we  do partial integration. First, we can notice that
$$ G_\omega(t,s) := G(t,s)e^{i\tau\Gamma\omega(t-s)} = \exp\left[-\tau\Gamma\int_s^t dr~ \gamma(r)+ i(\epsilon(r)-\omega)\right]~. $$
Furthermore we have
$$\frac{d}{ds} G_\omega(t,s) =  \tau\Gamma\left[\gamma(s)+ i(\epsilon(s)-\omega)\right]G_\omega(t,s)~. $$
Therefore we can write
\begin{multline}\label{partial_integration_1}
	\tau\Gamma\int_0^t ds~\gamma(s)^{\frac{1}{2}}G_\omega(t,s) = \int_0^t ds~ \frac{\gamma(s)^{\frac{1}{2}}}{\gamma(s)+ i(\epsilon(s)-\omega)} \frac{d}{ds} G_\omega(t,s) ~,\\
	= \left. \frac{\gamma(s)^{\frac{1}{2}}}{\gamma(s)+ i(\epsilon(s)-\omega)} G_\omega(t,s)\right|_{s=0}^t - \int_0^t ds~G_\omega(t,s) \frac{d}{ds}\frac{\gamma(s)^{\frac{1}{2}}}{\gamma(s)+ i(\epsilon(s)-\omega)}~,
\end{multline}
where we can evaluate the first part as $\left.\frac{\gamma(s)^{\frac{1}{2}}}{\gamma(s)+ i(\epsilon(s)-\omega)} G_\omega(t,s)\right|_{s=0}^t =   \frac{\gamma(t)^{\frac{1}{2}}}{\gamma(t)+ i(\epsilon(t)-\omega)} + \mathcal{O}(e^{-\tau\Gamma})$. We absorbed the $G_\omega(t,0)$ term in $\mathcal{O}(e^{-\tau\Gamma})$. For the second term we evaluate the derivative and continue integrating by parts
\begin{multline}\label{partial_integration_2}
	\int_0^t ds~ \frac{\frac{\dot\gamma(s)}{2}(\gamma(s)+ i(\epsilon(s)-\omega)) - \gamma(s)(\dot \gamma(s)+i\dot\epsilon(s))}{\gamma(s)^{\frac{1}{2}}(\gamma(s)+ i(\epsilon(s)-\omega))^2} G_\omega(t,s) \\ = 
	\frac{1}{\tau\Gamma}\left. \frac{\frac{\dot\gamma(s)}{2}(\gamma(s)+ i(\epsilon(s)-\omega)) - \gamma(s)(\dot \gamma(s)+i\dot\epsilon(s))}{\gamma(s)^{\frac{1}{2}}(\gamma(s)+ i(\epsilon(s)-\omega))^3} G_\omega(t,s)\right|_{s=0}^t 
	\\- \frac{1}{\tau\Gamma}\int_0^t ds~ G_\omega(t,s)\frac{d}{ds}\frac{\frac{\dot\gamma(s)}{2}(\gamma(s)+ i(\epsilon(s)-\omega)) - \gamma(s)(\dot \gamma(s)+i\dot\epsilon(s))}{\gamma(s)^{\frac{1}{2}}(\gamma(s)+ i(\epsilon(s)-\omega))^3}~.
\end{multline}
Similarly as in \eref{partial_integration_1}, we  keep only the evaluation at $s=t$ for the first term because the evaluation at $s=0$ is of order $\mathcal{O}(e^{-\tau\Gamma})$. Whereas the remaining integral will also have to be evaluated by parts, and in doing so we will obtain another power of $1/\tau\Gamma$. But since we are only interested in the first order correction and the integral will only yield terms of order $\mathcal{O}(1/\tau^2\Gamma^2)$ we don't need to compute it. By combining \eref{partial_integration_1} and \eref{partial_integration_2} we finally find
\begin{equation}\label{sd_exp}
	\tau\Gamma\!\!\int_0^t ds~\gamma(s)^{\frac{1}{2}}G_\omega(t,s) = \frac{\gamma(t)^{\frac{1}{2}}}{\gamma(t)+ i(\epsilon(t)-\omega)} - \frac{1}{\tau\Gamma} \frac{\frac{\dot\gamma(t)}{2}(-\gamma(t)+ i(\epsilon(t)-\omega)) - i\gamma(t)\dot\epsilon(t)}{\gamma(t)^{\frac{1}{2}}(\gamma(t)+ i(\epsilon(t)-\omega))^3} + \mathcal{O}(\frac{1}{\tau^2\Gamma^2})~.
\end{equation}
The absolute value squared of \eref{sd_exp} is
\begin{multline}\label{abs2_sd_exp}
	\left|\tau\Gamma\!\!\int_0^t ds~\gamma(s)^{\frac{1}{2}}G_\omega(t,s)\right|^2 =  \frac{\gamma(t)}{\gamma(t)^2 + (\epsilon(t)-\omega)^2} \\
	+ \frac{1}{\tau\Gamma} \frac{4\dot\epsilon(t)\gamma(t)^2(\epsilon(t)-\omega) + \dot\gamma(t)\gamma(t)\left(\gamma(t)^2-3(\epsilon(t)-\omega)^2\right)}{(\gamma(t)^2 + (\epsilon(t)-\omega)^2)^3} + \mathcal{O}(\frac{1}{\tau^2\Gamma^2})~.
\end{multline}
Combining this with \eref{norm_p} we find the expansion of $p(t)$ in the slow driving regime
\begin{multline}\label{p_SD}
	p(t) =  \frac{1}{\pi}\int_{-\infty}^{\infty}d\omega~  f_\beta(\omega\Gamma) \frac{\gamma(t)}{\gamma(t)^2 + (\epsilon(t)-\omega)^2} \\
	+ \frac{1}{\tau\Gamma} \frac{1}{\pi}\int_{-\infty}^{\infty}d\omega~  f_\beta(\omega\Gamma) \frac{4\dot\epsilon(t)\gamma(t)^2(\epsilon(t)-\omega) + \dot\gamma(t)\gamma(t)\left(\gamma(t)^2-3(\epsilon(t)-\omega)^2\right)}{(\gamma(t)^2 + (\epsilon(t)-\omega)^2)^3} + \mathcal{O}(\frac{1}{\tau^2\Gamma^2})~.
\end{multline}
Whereas the imaginary part of \eref{sd_exp} is
\begin{multline}\label{im_sd_exp}
	\frac{\tau\Gamma}{\gamma(t)^{\frac{1}{2}}} \Im\!\!\int_0^t ds~\gamma(s)^{\frac{1}{2}}G_\omega(t,s) = - \frac{\epsilon(t)-\omega}{\gamma(t)^2 + (\epsilon(t)-\omega)^2} \\- \frac{1}{\tau\Gamma} \frac{2\dot\gamma(t)(\epsilon(t)-\omega)\left(\gamma(t)^2-(\epsilon(t)-\omega)^2\right) - \dot\epsilon(t)\gamma(t)\left(\gamma(t)^2-3(\epsilon(t)-\omega)^2\right)}{(\gamma(t)^2 + (\epsilon(t)-\omega)^2)^3} + \mathcal{O}(\frac{1}{\tau^2\Gamma^2})~.
\end{multline}
Therefore the slow driving expansion of $v(t)$ is
\begin{multline}\label{v_SD}
	\frac{1}{\sqrt{2\Gamma\gamma(t)}} v(t) = -\frac{1}{\pi}\int_{-\infty}^{\infty}d\omega~  f_\beta(\omega\Gamma) \frac{\epsilon(t)-\omega}{\gamma(t)^2 + (\epsilon(t)-\omega)^2} \\
	- \frac{1}{\tau\Gamma}\frac{1}{\pi}\int_{-\infty}^{\infty}d\omega~  f_\beta(\omega\Gamma) \frac{2\dot\gamma(t)(\epsilon(t)-\omega)\left(\gamma(t)^2-(\epsilon(t)-\omega)^2\right) - \dot\epsilon(t)\gamma(t)\left(\gamma(t)^2-3(\epsilon(t)-\omega)^2\right)}{(\gamma(t)^2 + (\epsilon(t)-\omega)^2)^3} \\ + \mathcal{O}(\frac{1}{\tau^2\Gamma^2})~.
\end{multline}
Therefore we can see that we can rewrite the work cost of the protocol as
\begin{equation}\label{W_slow}
	W = W^{(0)} + \frac{1}{\tau\Gamma}W^{(1)} + \mathcal{O}(\frac{1}{\tau^2\Gamma^2})~.
\end{equation}
The leading order term is
\begin{multline}\label{W_th}
	W^{(0)} = \frac{1}{\pi}\int_{-\infty}^{\infty}d\omega~f_\beta(\omega)\int_0^1dt~ \frac{\dot\epsilon(t)\gamma(t) - \dot\gamma(t)(\epsilon(t)-\omega/\Gamma)}{\gamma(t)^2 + (\epsilon(t)-\omega/\Gamma)^2} \\= \frac{1}{\pi}\int_{-\infty}^{\infty}d\omega~f_\beta(\omega)\left(\arctan\frac{\gamma(0)}{\epsilon(0)-\omega/\Gamma}-\arctan\frac{\gamma(1)}{\epsilon(1)-\omega/\Gamma}\right)~,
\end{multline}
where we re-scaled $\omega$ by $\Gamma$. Here we were able to perform the integral independently of the function describing the control parameters. Therefore $W^{(0)}$ only depends on their initial and final value. More importantly, we can identify the instantaneous thermal expectation values of $p(t)$ and $v(t)$ (from \eref{p_thm_0} and \eref{v_thm_0}) in the time integral of $W^{(0)}$. This implies very directly that $W^{(0)} = \Delta F$.\\
We can notice that we can write $W^{(1)}$ as
\begin{equation}\label{W_length}
	W^{(1)} = \int_0^1 dt~ \dot{\vec\lambda}_t^T m(\vec\lambda_t)\dot{\vec\lambda}_t~,
\end{equation}
with $\vec\lambda_t = (\epsilon(t),\gamma(t))^T$ and the metric
\begin{equation}\label{metric_0}
	m(\vec\lambda) = \frac{1}{\pi}\int_{-\infty}^{\infty}d\omega~f_\beta(\omega)m_\omega(\epsilon-\omega/\Gamma, \gamma)~,
\end{equation}
for
\begin{equation}\label{metric_omega}
	m_\omega(\epsilon,\gamma) = \frac{1}{\left(\gamma^2 + \epsilon^2\right)^3}
	\begin{pmatrix}
		4\epsilon\gamma^2 &
		\gamma(\gamma^2-3\epsilon^2) \\
		\gamma(\gamma^2-3\epsilon^2) &
		2\epsilon(\epsilon^2-\gamma^2)
	\end{pmatrix}~.
\end{equation}
Since the leading order is independent of the path taken in parameter space, minimizing the work cost of erasure only implies minimizing $W^{(1)}$, i.e., the entropy production $k_B T \Sigma$. As we see from \eref{W_length} it is equivalent to finding the shortest path in a metric space described by the metric $m(\vec\lambda)$. The length of this shortest path is known as thermodynamic length. In the main text \eref{eq:metric} and \eref{eq:metric_omega} represent the metric when the problem is rewritten in terms of the unit-full parameters.

\subsection{Weak coupling limit}\label{section:WC}
Previous works on optimization of finite-time Landauer erasure have focused on the Markovian regime~\cite{Zulkowski2014,Proesmans2020,Proesmans2020II,boyd2022shortcuts,Lee2022,Diana2013,scandi19,Zhen2021,VanVu2022,Zhen2022,Ma2022}, corresponding to the weak coupling limit. We analyze this regime in this section. 
First we assume that the coupling remains unchanged during the protocol, which means $\Gamma = g^2$ and $\gamma = 1$. We start by rewriting $p(t)$ from \eref{p_SD} in a more convenient manner under this first assumption:
\begin{equation}\label{p_weak_0}
	p(t) =  \frac{1}{2\pi}\int_{-\infty}^{\infty}d\omega~  f_\beta(\omega) \frac{\Gamma}{\Gamma^2/4 + (\varepsilon(t)-\omega)^2}
	+ \frac{1}{\tau\Gamma} \frac{1}{2\pi}\int_{-\infty}^{\infty}d\omega~  f_\beta(\omega) \frac{\dot\varepsilon(t)(\varepsilon(t)-\omega)\Gamma^3}{(\Gamma^2/4 + (\varepsilon(t)-\omega)^2)^3} + \mathcal{O}(\frac{1}{\tau^2\Gamma^2})~,
\end{equation}
where we used \eref{norm_params} to go back to unit-full parameters and re-scaled $\omega$ by $\Gamma$. Integrating by parts the second integral we get
\begin{equation}\label{p_weak_1}
	p(t) =  \frac{1}{2\pi}\int_{-\infty}^{\infty}d\omega~  f_\beta(\omega) \frac{\Gamma}{\Gamma^2/4 + (\varepsilon(t)-\omega)^2}
	+ \frac{1}{\tau\Gamma} \frac{\beta\dot\varepsilon(t)}{8\pi}\int_{-\infty}^{\infty}d\omega~  f_\beta(\omega)(1-f_\beta(\omega)) \frac{\Gamma^3}{(\Gamma^2/4 + (\varepsilon(t)-\omega)^2)^2} ~,
\end{equation}
where we used that $\frac{d}{d\omega}f_\beta(\omega) = -\beta f_\beta(\omega)(1-f_\beta(\omega))$, $\frac{d}{d\omega}\frac{1}{(\Gamma^2/4 + (\varepsilon(t)-\omega)^2)^2} = \frac{4(\varepsilon(t)-\omega)}{(\Gamma^2/4 + (\varepsilon(t)-\omega)^2)^3}$ and dropped the $\mathcal{O}(1/\tau^2\Gamma^2)$ to make the notation lighter. We now take the weak coupling limit, but we do so while keeping the slow driving assumption: $\tau\Gamma \gg 1$. Using the results of~\cite{Beigi13} we get the following 
\begin{align*}
\lim_{\Gamma\to 0} \frac{1}{\pi}\frac{\Gamma/2}{\Gamma^2/4 + (\varepsilon(t)-\omega)^2} &= \delta(\varepsilon(t)-\omega)~, \\
	\lim_{\Gamma\to 0}\frac{1}{\pi} \frac{\Gamma^3/8}{(\Gamma^2/4 + (\varepsilon(t)-\omega)^2)^2} &= \frac{1}{2}\delta(\varepsilon(t)-\omega)~,
\end{align*}
where the equalities are meant in a distributional sense. Therefore we find that the occupation probability in the weak coupling limit is
\begin{equation}\label{p_weak}
	p(t) = f_\beta(\varepsilon(t)) + \frac{1}{\tau \Gamma} \beta\dot\varepsilon(t) f_\beta(\varepsilon(t))[1-f_\beta(\varepsilon(t))]~.
\end{equation}
This result coincides with applying a slow driving expansion to a simple exponential relaxation model with characteristic time $\Gamma$ ($\dot p = -\tau \Gamma [p-f_\beta(\epsilon(t))]$). Computing the work cost yields
\begin{equation}\label{W_weak}
	 W = \Delta F + \frac{1}{\tau \Gamma} \beta \int_0^1dt~\dot\varepsilon(t)^2 f_\beta(\varepsilon(t))[1-f_\beta(\varepsilon(t))]~,
\end{equation}
with $\Delta F = \beta^{-1}\ln \frac{1+e^{-\beta\varepsilon(0)}}{1+e^{-\beta\varepsilon(1)}}$. We will now minimize the work cost of the erasure protocol, similar optimizations have been done before in~\cite{scandi19,Erdman2023Pareto,Esposito10}. From variational calculus we know that the extremal function of the integral in \eref{W_weak} will keep the integrand constant. 
So we can solve the variational problem as follows:
\begin{equation}
\begin{split}
	\dot\varepsilon(t) \sqrt{f_\beta(\varepsilon(t))[1-f_\beta(\varepsilon(t))]} = K_w ~, \\
	\int_{\varepsilon(0)}^{\varepsilon(t)}\frac{e^{-\beta \varepsilon/2}}{1+e^{-\beta \varepsilon}} d\varepsilon  = K_w \int_{0}^{t}dt' ~, \\
	2\arctan(e^{\beta \varepsilon(t)/2}) - \frac{\pi}{2} = \beta K_w t ~,\\
	\varepsilon(t) = 2\beta^{-1}\ln\tan\left(\beta K_w t/2 + \frac{\pi}{4}\right)~,
\end{split}
\end{equation}
with $K_w = 2\beta^{-1}(\arctan(e^{\beta \varepsilon(1)/2}) - \frac{\pi}{4}) \xrightarrow{\beta\epsilon(1)\to\infty} \frac{\pi}{2}\beta^{-1}$.
We therefore find
\begin{equation}\label{eps_weak}
	\varepsilon_{weak}(t) = 2\beta^{-1}\ln\tan\left(\frac{\pi}{4}(t+1)\right)~,
\end{equation}
and recover the result of \eref{eq:finitetimeboundweak} from the main text:
\begin{equation}\label{W1_weak_sol}
	W = k_B T\left( \ln 2 + \frac{\pi^2}{4\tau \Gamma} \right)~.
\end{equation}

\subsection{Solving the integral of the thermodynamic metric and finding the symmetry}
We will now try to find a more tractable version of the metric in \eref{metric_0}. First we notice that
\begin{equation}\label{metric_prim}
	m_\omega(\epsilon,\gamma) = -\frac{d}{d\epsilon} \frac{1}{\left(\gamma^2 + \epsilon^2\right)^2}
	\begin{pmatrix}
		\gamma^2 &
		-\epsilon\gamma\\
		-\epsilon\gamma &
		\epsilon^2
	\end{pmatrix} 
	=: -\frac{d}{d\epsilon} m_0(\epsilon,\gamma)~.
\end{equation}
We can remark that $m_0$ coincides with a metric of an angle distance in the $(\epsilon,\gamma)$ space. To solve the integral of \eref{metric_0} we will go in Fourier space. For a function $h(\epsilon)$ its Fourier transform $\tilde h(\xi)$ has the defining property
\begin{equation}\label{fourier_def}
h(\epsilon) =  \frac{1}{2\pi}\int_{-\infty}^\infty d\xi~ \tilde h(\xi) e^{i\xi \epsilon}~.
\end{equation}
Even though $f_\beta(\omega)$ is not an integrable function we can find its Fourier transform in a distributional sense
\begin{equation}\label{f_fourier}
	\tilde f_\beta(\xi) = \pi \delta(\xi) + \frac{\pi i}{\beta \sinh\left(\pi \xi/\beta\right)}~.
\end{equation}
For $m_0$ we find
\begin{equation}\label{m0_fourier}
	\tilde m_0(\xi,\gamma) = \frac{\pi}{2}\Theta(\xi) e^{-\xi \gamma}
	\begin{pmatrix}
		\gamma^{-1} + \xi &
		i\xi \\
		i\xi &
		\gamma^{-1}-\xi
	\end{pmatrix} 
	+ \frac{\pi}{2}\Theta(-\xi) e^{\xi \gamma}
	\begin{pmatrix}
		\gamma^{-1} - \xi &
		i\xi \\
		i\xi &
		\gamma^{-1}+\xi
	\end{pmatrix} 
	~,
\end{equation}
with $\tilde m_0(0,\gamma) = \frac{\pi}{2\gamma}\mathbb{1}$. Therefore we can rewrite \eref{metric_0} as 
\begin{equation}
\begin{split}
	m(\vec\lambda) &= -\frac{1}{4\pi^3}\frac{d}{d\epsilon}\int_{-\infty}^{\infty}d\omega d\xi d\xi'~\tilde f_\beta(\xi') \tilde m_0(\xi, \gamma) e^{i\omega\xi'}e^{i(\epsilon-\omega/\Gamma)\xi}~, \\
	&= -\frac{i}{4\pi^3}\int_{-\infty}^{\infty}d\omega d\xi d\xi'~\xi \tilde f_\beta(\xi') \tilde m_0(\xi, \gamma) e^{i\epsilon\xi }e^{i\omega(\xi'-\xi/\Gamma)}~, \\
	&= -\frac{i}{2\pi^2}\int_{-\infty}^{\infty} d\xi d\xi'~\xi \tilde f_\beta(\xi') \tilde m_0(\xi, \gamma) \delta(\xi'-\xi/\Gamma)e^{i\epsilon\xi }~, \\
	&= -\frac{i}{2\pi^2}\int_{-\infty}^{\infty} d\xi~\xi \tilde f_\beta(\xi/\Gamma) \tilde m_0(\xi, \gamma) e^{i\epsilon\xi }~, \\
	&= \frac{1}{2\beta\pi}\int_{-\infty}^{\infty} d\xi~\xi \frac{\tilde m_0(\xi, \gamma)}{\sinh(\frac{\pi\xi}{\beta\Gamma})} e^{i\epsilon\xi }~,
\end{split}
\end{equation}
where we used \eref{fourier_def}, \eref{f_fourier} and the fact that $\int_{-\infty}^\infty d\omega~ e^{i\omega x} = 2\pi\delta(x)$. If we now insert \eref{m0_fourier} and flip the sign in the second integral we find
\begin{equation}
\begin{split}
	m(\vec\lambda) &= \frac{1}{4\beta}\int_{0}^{\infty} d\xi~\frac{\xi e^{-\xi(\gamma-i\epsilon)} }{\sinh(\frac{\pi\xi}{\beta\Gamma})}
	\begin{pmatrix}
		\gamma^{-1} + \xi &
		i\xi \\
		i\xi &
		\gamma^{-1}-\xi
	\end{pmatrix} 
	+ \frac{1}{4\beta}\int_{-\infty}^{0} d\xi~\frac{\xi e^{\xi(\gamma + i\epsilon) } }{\sinh(\frac{\pi\xi}{\beta\Gamma})}
	\begin{pmatrix}
		\gamma^{-1} - \xi &
		i\xi \\
		i\xi &
		\gamma^{-1}+\xi
	\end{pmatrix} ~, \\
	&= \frac{1}{4\beta}\int_{0}^{\infty} d\xi~\frac{\xi e^{-\xi(\gamma-i\epsilon)} }{\sinh(\frac{\pi\xi}{\beta\Gamma})}
	\begin{pmatrix}
		\gamma^{-1} + \xi &
		i\xi \\
		i\xi &
		\gamma^{-1}-\xi
	\end{pmatrix} 
	+ \frac{1}{4\beta}\int_{0}^{\infty} d\xi~\frac{\xi e^{-\xi(\gamma + i\epsilon) } }{\sinh(\frac{\pi\xi}{\beta\Gamma})}
	\begin{pmatrix}
		\gamma^{-1} + \xi &
		-i\xi \\
		-i\xi &
		\gamma^{-1}-\xi
	\end{pmatrix} ~,\\
	& = \frac{1}{4\beta}\int_{0}^{\infty} d\xi~\frac{\xi e^{-\xi\gamma} }{\sinh(\frac{\pi\xi}{\beta\Gamma})}\left[
	e^{i\xi\epsilon}
	\begin{pmatrix}
		\gamma^{-1} + \xi &
		i\xi \\
		i\xi &
		\gamma^{-1}-\xi
	\end{pmatrix}
	+
	e^{-i\xi\epsilon}
	\begin{pmatrix}
		\gamma^{-1} + \xi &
		-i\xi \\
		-i\xi &
		\gamma^{-1}-\xi
	\end{pmatrix}
	\right] ~,\\
	& = \frac{1}{4\beta}\int_{0}^{\infty} d\xi~\frac{\xi e^{-\xi\gamma} }{\sinh(\frac{\pi\xi}{\beta\Gamma})}
	\begin{pmatrix}
		(\gamma^{-1} + \xi)(e^{i\xi\epsilon} + e^{-i\xi\epsilon}) &
		-i \xi (e^{-i\xi\epsilon} - e^{i\xi\epsilon}) \\
		-i \xi (e^{-i\xi\epsilon} - e^{i\xi\epsilon}) &
		(\gamma^{-1}-\xi)(e^{i\xi\epsilon} + e^{-i\xi\epsilon})
	\end{pmatrix} ~,\\
	& = \frac{1}{4\beta\gamma}\int_{0}^{\infty} d\xi~\frac{\xi e^{-\xi\gamma}(e^{i\xi\epsilon} + e^{-i\xi\epsilon}) }{\sinh(\frac{\pi\xi}{\beta\Gamma})} \mathbb{1}
	+ \frac{1}{4\beta}\int_{0}^{\infty} d\xi~\frac{\xi^2 e^{-\xi\gamma} }{\sinh(\frac{\pi\xi}{\beta\Gamma})}
	\begin{pmatrix}
		e^{i\xi\epsilon} + e^{-i\xi\epsilon} &
		\frac{e^{-i\xi\epsilon} - e^{i\xi\epsilon}}{i} \\
		\frac{e^{-i\xi\epsilon} - e^{i\xi\epsilon}}{i} &
		-e^{i\xi\epsilon} - e^{-i\xi\epsilon}
	\end{pmatrix} ~,\\
	& = \frac{1}{2\beta\gamma} \mathbb{1} \Re\int_{0}^{\infty} d\xi~\frac{\xi e^{-\xi(\gamma+i\epsilon)} }{\sinh(\frac{\pi\xi}{\beta\Gamma})} 
	+ \frac{1}{2\beta}
	\begin{pmatrix}
		\Re &
		\Im \\
		\Im &
		-\Re
	\end{pmatrix}
	\int_{0}^{\infty} d\xi~\frac{\xi^2 e^{-\xi(\gamma+i\epsilon)} }{\sinh(\frac{\pi\xi}{\beta\Gamma})} ~.
\end{split}
\end{equation}
We will now be able to compute these integrals in terms of poly-gamma functions. The poly-gamma function of order $m\geq 0$ is defined as $\psi^{(m)}(z) := \frac{d^{m+1}}{dz^{m+1}}\ln\Gamma(z)$. For $m>0$ and $\Re[z]>0$ they have an integral representation:
\begin{equation}\label{polygamma_def}
	\psi^{(m)}(z) = (-1)^{m+1}\int_0^\infty d\xi~ \frac{\xi^m e^{-\xi z}}{1-e^{-\xi}}~.
\end{equation}
Using the fact that $\frac{1}{\sinh(x)} = \frac{2 e^{-x}}{1-e^{-2x}}$, a change of variable and \eref{polygamma_def} we find
\begin{equation}\label{metric_unitless}
	\begin{split}
		m(\vec\lambda) & = \frac{1}{\beta\gamma} \mathbb{1} \Re\int_{0}^{\infty} d\xi~\frac{\xi e^{-\xi(\frac{\pi}{\beta\Gamma}+\gamma+i\epsilon)} }{1-e^{-2\frac{\pi\xi}{\beta\Gamma}}} 
		+ \frac{1}{\beta}
		\begin{pmatrix}
			\Re &
			\Im \\
			\Im &
			-\Re
		\end{pmatrix}
		\int_{0}^{\infty} d\xi~\frac{\xi^2 e^{-\xi(\frac{\pi}{\beta\Gamma}+\gamma+i\epsilon)} }{1-e^{-2\frac{\pi\xi}{\beta\Gamma}}} ~, \\
		& = \frac{\beta\Gamma^2}{4\pi^2\gamma} \mathbb{1} \Re\int_{0}^{\infty} d\xi~\frac{\xi e^{-\xi[\frac{1}{2}+\frac{\beta\Gamma}{2\pi}(\gamma+i\epsilon)]} }{1-e^{-\xi}} 
		+ \frac{\beta^2\Gamma^3}{8\pi^3}
		\begin{pmatrix}
			\Re &
			\Im \\
			\Im &
			-\Re
		\end{pmatrix}
		\int_{0}^{\infty} d\xi~\frac{\xi^2 e^{-\xi[\frac{1}{2}+\frac{\beta\Gamma}{2\pi}(\gamma+i\epsilon)]} }{1-e^{-\xi}} ~,\\
		& = \frac{\beta\Gamma^2}{4\pi^2\gamma} \mathbb{1} \Re\psi^{(1)}\!\left(\frac{1}{2}+\frac{\beta\Gamma}{2\pi}(\gamma+i\epsilon)\right)
		- \frac{\beta^2\Gamma^3}{8\pi^3}
		\begin{pmatrix}
			\Re &
			\Im \\
			\Im &
			-\Re
		\end{pmatrix}
		\psi^{(2)}\!\left(\frac{1}{2}+\frac{\beta\Gamma}{2\pi}(\gamma+i\epsilon)\right) ~.
	\end{split}
\end{equation}
We can notice that the metric explicitly depends on $\Gamma$ in such a way that it seems that the solution for the geodesic should depend on this scale factor. Though this dependence disappears if we re-parameterize the problem in terms of its original unit-full parameters. We start by rewriting the work as
\begin{equation}
	 W = \Delta F + W^{(1)} + \mathcal{O}(\frac{1}{\tau^2\Gamma^2})~,
\end{equation}
where we redefined $W^{(1)}$ with the unit-full parameters $\vec{\lambda}_t = (\varepsilon(t), \mu(t))^T$ (with $\mu(t):= \frac{1}{2}g(t)^2 = \Gamma \gamma(t)$):
\begin{gather}\label{W1_unitfull}
W^{(1)} = \frac{1}{\tau}\int_0^1 dt~ \dot{\vec\lambda}_t^T m(\vec\lambda_t)\dot{\vec\lambda}_t,\\
    \label{metric}
	m(\vec\lambda) = \frac{\beta}{4\pi^2\mu} \mathbb{1}\, \Re\psi^{(1)}\!\left(\frac{1}{2}+\frac{\beta}{2\pi}z\right)
	- \frac{\beta^2}{8\pi^3}
	\begin{pmatrix}
		\Re &
		\Im \\
		\Im &
		-\Re
	\end{pmatrix}
	\psi^{(2)}\!\left(\frac{1}{2}+\frac{\beta}{2\pi}z\right) ~.
\end{gather}
We remind the reader that $z = \mu + i\varepsilon$. This metric is the same as the one presented in the main text. Despite not looking very approachable, \eref{metric} is a much more tractable version of \eref{eq:metric} from the main text when it comes to numerical implementations (as the polygamma functions are computed much faster than integrals) and analytical studies of the geometric properties of thermodynamic protocols.

We can notice from \eref{W1_unitfull} and \eref{metric} is that there is a symmetry in the corrective term. If we perform the following transformation:
\begin{equation}\label{symmetry}
\begin{split}
	\varepsilon(t) &\rightarrow \lambda \varepsilon(t) ~, \\
	\mu(t) &\rightarrow \lambda \mu(t) ~,\\
	\beta &\rightarrow \lambda^{-1} \beta ~,
\end{split}
\end{equation}
for $\lambda >0$; then $W^{(1)}$ remains unchanged. This symmetry allows us to conclude that the minimal value of $W^{(1)}$ to perform Landauer erasure will be of the form $c/\tau$ where $c$ is a constant that does not depend on any physical quantity.

\section{High temperature limit}\label{sm:HT}
In order to find the minimal dissipation in the multi-variable case we need to numerically solve the equations of motion given by the exact metric, which (unsurprisingly) are very untractable analytically. But instead of solving an initial value problem (which we can always solve by numerical integration, in principle) we are trying to solve a boundary value problem. Generically, to solve a BVP numerically, the solver will try many IVPs until the wanted BVP is reached. But here we can notice that we can turn the BVP into an IVP by taking an analytical approximation of the problem around the point $(\varepsilon,\mu) = (0,0)$. \\

As we have seen in \eref{symmetry} there is an underlying symmetry in this problem, so a limit where $\varepsilon$ and $\mu$ are infinitesimal is the same as a limit where $\beta$ is infinitesimal but $\varepsilon$ and $\mu$ finite. Formally we are requiring $\beta|\mu + i\varepsilon|\ll 1$, which is a high-temperature limit. It is important to note that despite the fact that this approximation will yield some analytical results on how to optimize a protocol in the high temperature regime it will not give us a result that is relevant for Landauer erasure because to perform erasure we are assuming that we reach $\beta\varepsilon \gg 1$, which is a low temperature limit.

\begin{figure}[H]
		\centering
        \includegraphics[width=\textwidth]{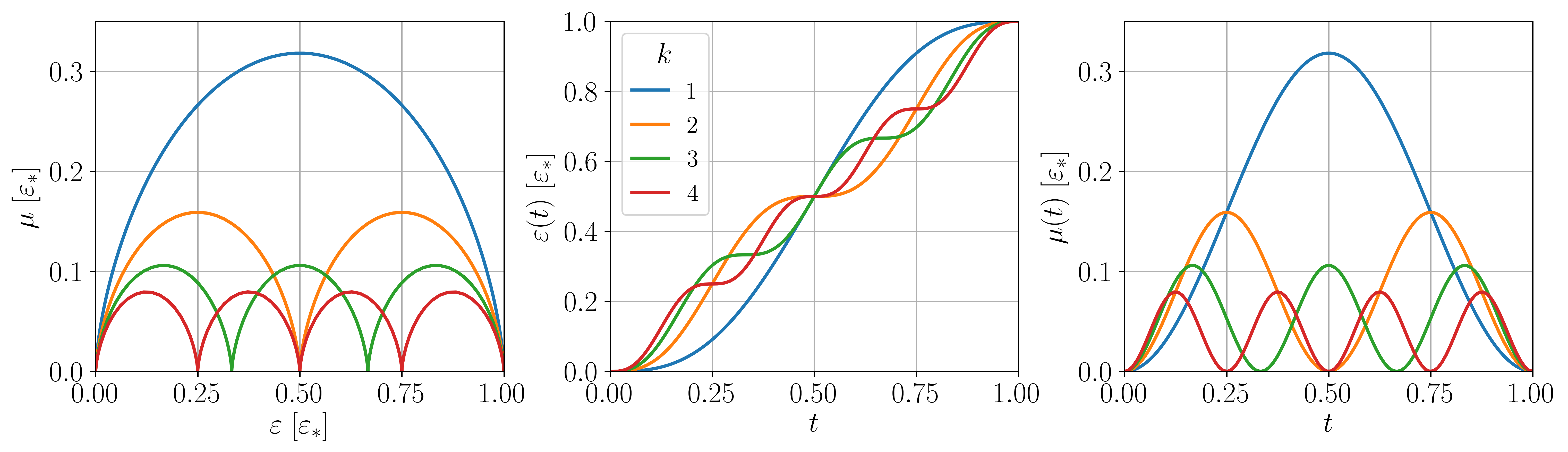}
	    \vspace{-15pt}
	    \caption{Parametrization of $\mu(t)$ and $\eps(t)$ described by \eref{geodesic_HT} for multiple values of $k$. Shown in the parameter space (left) and as a function of time (centre and right).}
	    \vspace{-10pt}
		\label{fig:HT_geo}
\end{figure}

One way to obtain an analytical result in this framework is by going back to \eref{metric_0} and apply the high-temperature expansion of the Fermi-Dirac distribution: $f_\beta(\omega) = \frac{1}{2} - \frac{1}{4}\beta\omega +\mathcal{O}(\beta^3\omega^3)$. Since $m_0(\pm \infty,\gamma) = 0$ the first term of the metric in this expansion is $0$. But from the next order we find (in unit-full parameters)
\begin{equation}\label{metric_HT}
	m_{HT}(\vec\lambda) = \frac{\beta}{4\pi}\int_{-\infty}^{\infty}d\omega~\frac{\omega-\varepsilon}{\left(\mu^2 + \omega^2\right)^3}
		\begin{pmatrix}
			4\omega\mu^2 &
			\mu(\mu^2-3\omega^2) \\
			\mu(\mu^2-3\omega^2) &
			2\omega(\omega^2-\mu^2)
		\end{pmatrix} 
	= \frac{\beta}{8\mu}\mathbb{1}~.
\end{equation}
It might not be immediate why we also require $\beta\mu \ll 1$, but it becomes clear that it is required when we want to obtain the same result by applying the same expansion on \eref{metric} (which would also allow us to get further orders). We will now compute and solve the equations of motion:
\begin{equation}\label{EOM}
	\ddot\lambda^i + \Gamma^i_{jk}\dot\lambda^j\dot\lambda^k = 0~,
\end{equation}
where we assumed the Einstein tensorial notation and $\Gamma^i_{jk}$ are the Christoffel symbols
\begin{equation}\label{Chrstf}
	\Gamma^i_{jk} :=  \frac{1}{2} m^{il}(\partial_j m_{kl} + \partial_k m_{jl} - \partial_l m_{jk})~.
\end{equation}
Here we have $m^{il} = \frac{8\mu}{\beta}\delta^{il}$ and $\partial_a m_{bc} = -\frac{\beta}{8\mu^2}\delta_{a\mu}\delta_{bc}$. We therefore find
\begin{equation}\label{Chrstf_HT_EN}
	\Gamma^i_{jk}  = -\frac{1}{2\mu}(\delta_{j\mu}\delta^{i}_k + \delta_{k\mu}\delta_{j}^i - \delta^{i\mu}\delta_{jk})~,
\end{equation}
which can be rewritten as
\begin{equation}\label{Chrstf_HT}
	\Gamma^\varepsilon = -\frac{1}{2\mu} 
	\begin{pmatrix}
		0 &
		1 \\
		1 &
		0
	\end{pmatrix} 
	\quad ,\quad
	\Gamma^\mu = \frac{1}{2\mu} 
	\begin{pmatrix}
		1 &
		0 \\
		0 &
		-1
	\end{pmatrix} ~.
\end{equation}
We get the following differential equations for $\mu$ and $\varepsilon$:
\begin{equation}\label{diff_eq_HT}
\ddot\varepsilon \mu = \dot\varepsilon \dot\mu \quad , \quad 2\mu\ddot\mu = \dot\mu^2 -\dot\varepsilon^2.
\end{equation}
From the first equation we can see that $\int d\dot\varepsilon/\dot\varepsilon = \int d\mu/\mu$, therefore $\dot\varepsilon = C \mu $ for some constant $C$ (we can already see as a sanity check that $\dot\varepsilon$ never changes sign in an optimal protocol). The equation for $\mu$ becomes
$$ 2\mu\ddot\mu = \dot\mu^2 - C^2\mu^2~,$$
when we consider that $\mu = g^2/2$ and $\ddot\mu = \dot g^2 + g\ddot g$ we can see that
$$ \ddot g = - C^2 g/4~.$$
Taking into account that the boundary conditions for $g$ are $g(0) = g(1) = 0$ we find that $g(t) = A\sin(k\pi t)$ for some constant $A$, $k\in \mathbb{N}^*$ and $C = 2k\pi$. By choosing $\varepsilon(0) = 0$ and $\varepsilon(1) = \varepsilon_*$  we have $\varepsilon(t) = k\pi A^2 \int_0^t ds \sin(k\pi s)^2$, therefore $A^2 = \frac{2\varepsilon_*}{k\pi}$. Thus the optimal protocol, portrayed in \fref{fig:HT_geo}, is
\begin{equation}\label{geodesic_HT}
	\varepsilon(t) = \varepsilon_*\left(t - \frac{\sin(2k\pi t)}{2k\pi}\right) 
	\quad , \quad 
	\mu(t) = \frac{\varepsilon_*}{k\pi}\sin(k\pi t)^2 ~.
\end{equation}

From \eref{geodesic_HT} we can compute the integrand of the dissipated work:
\begin{equation}
\begin{split}
	\dot{\vec\lambda}_t^T m_{HT}(\vec\lambda_t)\dot{\vec\lambda}_t &= \frac{\beta}{8\mu}(4k^2\pi^2\mu^2 + \dot\mu^2) \\
	& = \frac{\beta\varepsilon_*^2}{2\mu}\left(\sin(k\pi t)^4 + \sin(k\pi t)^2\cos(k\pi t)^2 \right) \\
	& = \frac{k\pi\beta\varepsilon_*}{2}~.
\end{split}
\end{equation}
We therefore find the dissipated work for the high temperature limit by inserting this in \eref{W1_unitfull} and taking $k=1$:
\begin{equation}\label{W1_HT}
	W^{(1)}_{HT} = \frac{\pi\beta\varepsilon_*}{2\tau}.
\end{equation}
We can see that in this scenario the corrective term grows extensively with the final energy $\varepsilon_*$, combining this with the fact that the exact metric goes to $0$ faster than $\mathcal{O}(|z^{-1}|)$ we can presume that most of the dissipation in the exact protocol is caused by the part of the protocol that matches with the high temperature regime.
 
\section{Low temperature limit}\label{sm:LT}
We will now study the limit of $T\rightarrow0$ we have $f_\beta(\omega)\rightarrow f_\infty(\omega) = \Theta(-\omega)$, where $\Theta$ is the Heaviside step function. Therefore the integral of \eref{metric_0} becomes (in unit-full parameters)
\begin{equation}
\begin{split}
	m_{T=0}(\vec\lambda) &= \frac{1}{\pi} \int_{-\infty}^{\infty}d\omega~\Theta(-\omega)m_\omega(\eps-\omega, \mu)~,\\
	&= \frac{1}{\pi}\int_{\eps}^{\infty}d\omega~ m_\omega(\omega, \mu)~,\\
	&= -\frac{1}{\pi}\int_{\eps}^{\infty}d\omega~ \frac{d}{d\omega} m_0(\omega, \mu) ~,\\
	&= \frac{1}{\pi}m_0(\eps,\mu)~.
\end{split}
\end{equation}
where we used \eref{metric_prim} and the fact that $m_0(+\infty,\mu) = 0$. Thus we have 
\begin{equation}
    m_{T=0}(\vec\lambda) =  
    \frac{1}{\pi}\frac{1}{\left(\mu^2 + \eps^2\right)^2}
	\begin{pmatrix}
		\mu^2 &
		-\eps\mu\\
		-\eps\mu &
		\eps^2
	\end{pmatrix} 
	~.
\end{equation}
We can compute the integrand of $W^{(1)}$ to find
\begin{equation}
    \dot{\vec\lambda}^T m_{T=0}(\vec\lambda)\dot{\vec\lambda} = \frac{1}{\pi} \frac{\dot\eps^2\mu^2 - 2\dot\eps\dot\mu\eps\mu + \dot\mu^2 \eps^2 }{\left(\mu^2 + \eps^2\right)^2} = \frac{1}{\pi} \left(\frac{\eps\dot\mu- \dot\eps\mu}{\mu^2 + \eps^2}\right)^2
\end{equation}
By defining the coordinates $\vec\lambda_{r,\phi}=(r,\phi)^T$, such that $\eps = r\cos\phi$ and $\mu = r\sin\phi$, we have $\phi = \arctan\frac{\mu}{\eps}$. Therefore 
\begin{equation}
    \dot{\vec\lambda}^T m_{T=0}(\vec\lambda)\dot{\vec\lambda} = \frac{\dot\phi^2}{\pi} ~,
\end{equation}
From which we can deduce the metric in these new coordinates
\begin{equation}\label{metric_LT}
    m_{T=0}^{(r,\phi)}(\vec\lambda_{r,\phi}) =  
    \frac{1}{\pi}
	\begin{pmatrix}
		0 &
		0\\
		0 &
		1
	\end{pmatrix}~.
\end{equation}
Crucially, we can notice that this metric is singular: here changes in the coordinate $r$ do not cause an increase in the work cost. Therefore we can parameterize geodesics at zero temperature as follows
\begin{equation}\label{geodesic_LT}
	\varepsilon(t) = r(t)\cos\!\left(\phi(0) (1-t) + \phi(1) t\right)
	\quad , \quad 
	\mu(t) = r(t)\sin\!\left(\phi(0) (1-t) + \phi(1) t\right) ~,
\end{equation}
where $r(t)$ is any function that satisfies the boundary conditions. By using \eref{W1_unitfull} we can find the dissipated work
\begin{equation}
    W_{T=0}^{(1)} = \frac{(\Delta \phi)^2}{\pi \tau}~,
\end{equation}
for $\Delta\phi = \phi(1)-\phi(0)$.

\section{Discontinuities in the protocol}\label{sm:jumps}
As is mentioned in the main text, it is well known~\cite{Schmiedl2007o} that, at weak coupling, discontinuities appear at the beginning and at the end in optimal finite-time protocols. In the main text we gave numerical evidence to the fact that these jumps disappear as one approaches the quasistatic limit. Here we make an a posteriori argument as to why these jumps should also disappear for the system we studied in the strong coupling regime.\\

We can start by immediately discarding jumps in the coupling because these lead to a diverging work cost in the wideband limit. Then, if one makes a jump in the energy at the start of the protocol from $0$ to $\eps_*$ its work cost is
\begin{equation}\label{eq:jump_start}
    W_{jump} = p(0)\eps_* = \frac{1}{2}\eps_*~,
\end{equation}
where $p(0)$ is the probability of occupation at $t=0$ and is set to be $1/2$. We can compare it to the work cost given by an optimal continuous protocol with the same boundary conditions:
\begin{equation}\label{eq:HT_start}
    W_{cont} = \Delta F + \frac{a(\beta\eps_*)\hbar}{\tau} =  k_B T \ln\!\left(\frac{2}{1+e^{-\beta\eps_*}}\right) + \frac{a(\beta\eps_*)\hbar}{\tau}~,
\end{equation}
where $a(\beta\eps_*)$ is a bounded function of $\beta\eps_*$ (c.f. figure in the main text). We note that by expanding $\Delta F$ around $\beta = 0$ we find
\begin{equation}
    \Delta F = \frac{1}{2}\eps_* - \frac{1}{8}\beta\eps_*^2 + \mathcal{O}(\beta^3)~.
\end{equation}
Therefore at $\beta > 0$ and small enough one finds that $\Delta F < W_{jump}$. But one can also note that $\lim_{\beta\rightarrow\infty}\Delta F = 0$, and since $\Delta F$ is a monotonous function of $\beta$ we conclude that $\Delta F < W_{jump}$ for any temperature and any $\eps_* > 0$. At this point it is immediate that there exists $\tau$ large enough such that $W_{cont} < W_{jump}$.\\
We finally consider jumps in the energy at the end of the protocol. These jumps would have to be performed once the coupling is very close to zero since it is at the end of the protocol. In this limit optimal protocols are found by~\cite{Esposito10}. These optimal protocols feature jumps whose magnitude is controlled by a constant of integration $K$, in particular the magnitude of these jumps is $\mathcal{O}(\sqrt{K})$. This constant is defined as follows (in terms of our notation with unit-less time)
\begin{equation}
    K = \frac{1}{\tau^2\Gamma^2}\frac{\dot p^2}{(p + \dot p/\tau\Gamma)(1 - p - \dot p/\tau\Gamma)}~.
\end{equation}
From which it is clear that $K$ goes to $0$ in the limit of $\tau\Gamma \gg 1$ and therefore the jumps disappear. This behavior is also confirmed in \fref{fig:WC_comparison} where these optimal protocols are shown.

\section{One-parameter case}\label{sm:OP}
Because of the metric we obtain in \eref{metric} it is quite clear that, in the two-parameter case, we will not be able to solve analytically the geodesics for the full problem. This even prevents us form finding an analytical expression for the distance between two points in the parameter space, as it is the length of the shortest path (for which we have no expression). But by fixing one parameter to an arbitrary value and solving for the other we can use the fact that geodesics always have a conserved quantity along their path (the integrand: $\dot{\vec\lambda}^T_t m(\vec\lambda_t)\dot{\vec\lambda}_t$) to avoid solving the geodesic equation and finding an explicit formula for the length and geodesic.
We point out the fact that the symmetry mentioned in \aref{sm:sd} does not lead to a conserved quantity because $\beta$ is a constant of the system instead of a function of time for which we are solving.

Here we will take the erasure protocol to be made of three parts, which will be optimized separately: 1. we turn on the coupling to some value $\mu_*$ while keeping the energy at zero; 2. while keeping the coupling at $\mu_*$ we increase the energy from zero to infinity; 3. we turn the coupling off.
Incidentally, this type of protocols are more realistic for an experimental realization as often setups are not able to control optimally energy an coupling at the same time. And even if the control over the coupling is only to turn it on to some value and turn it off, step 2 will remain valid. Furthermore, previous studies done at weak coupling essentially are described by this type of protocol; but they are in a regime where step 1 and 3 can be neglected. Therefore we can compare the results of this section to those of the weak coupling limit.

We start by looking at step 3, in the limit of $\beta\eps\rightarrow\infty$ we actually reach a scenario described by the $T=0$ limit. Therefore the length of this step is described by \eref{metric_LT}, for any finite value of $\mu_*$ the angle span of this step is trivially $0$. Therefore, up to first order, this step will not cause any extra dissipation, no matter how it is realized.

We notice that we can write the length of the first step as follows
\begin{equation}\label{L_step1_0}
	L_1 = \int_0^1 dt~ |\dot\mu(t)| m_{\mu\mu}(0,\mu(t))^{1/2} = \int_0^{\mu(1)} d\mu~ m_{\mu\mu}(0,\mu )^{1/2}~,
\end{equation}
where we used the fact that, since the metric is not explicitly time-dependent, the sign of $\dot\mu$ has to be always positive for this step. With \eref{metric} we find the following expression for the length
\begin{equation}\label{L_step1}
	L_1 = \frac{1}{\sqrt{2\pi}} \int_0^{\beta\mu_*/2\pi} \!\!\!\!\sqrt{\Re\!\left[ \frac{1}{x} \psi^{(1)}\!\left(\frac{1}{2}+x \right) + \psi^{(2)}\!\left(\frac{1}{2}+x\right) \right]} dx~.
\end{equation}
Now that we have an expression for $L_1$ we can recover an equation for $\mu(t)$. We can use the fact that the integrand of the time integral in \eref{L_step1_0} is constant to obtain
\begin{equation}\label{step1_mu}
	t L_1 = \frac{1}{\sqrt{2\pi}} \int_0^{\beta\mu(t)/2\pi} \!\!\!\!\sqrt{\Re\!\left[ \frac{1}{x} \psi^{(1)}\!\left(\frac{1}{2}+x \right) + \psi^{(2)}\!\left(\frac{1}{2}+x\right) \right]} dx~,
\end{equation}
which gives an implicit definition of $\mu(t)$, or rather an explicit definition of its inverse $t(\mu)$.

\begin{figure}[H]
		\centering
        \includegraphics[width=0.93\textwidth]{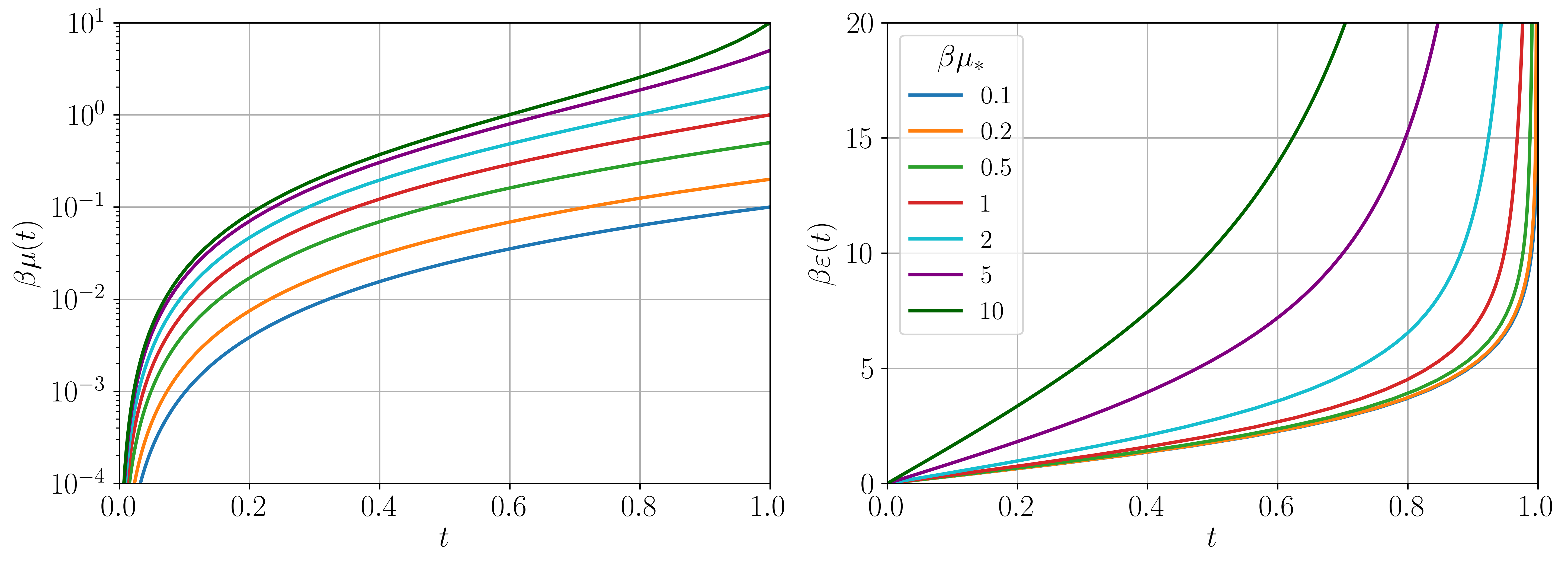}
	    \vspace{-5pt}
	    \caption{Parametrization of $\mu(t)$ and $\eps(t)$ described by \eref{step1_mu} and \eref{setp2_eps} for multiple values of $\beta\mu_*$.}
	    \vspace{-13pt}
		\label{fig:OP_geo}
\end{figure}

By following the same procedure as in step 1 we can recover the length of step 2
\begin{equation}\label{L_step2}
	L_2 = \frac{1}{\sqrt{2\pi}}\int_0^{\infty} \!\!\!\!\sqrt{\Re\!\left[ \frac{2\pi}{\beta\mu_*} \psi^{(1)}\!\left(\frac{1}{2}+\frac{\beta\mu_*}{2\pi}+iy\right)
	- \psi^{(2)}\!\left(\frac{1}{2}+\frac{\beta\mu_*}{2\pi}+iy\right) \right]} dy~,
\end{equation}
and the implicit definition of $\eps(t)$
\begin{equation}\label{setp2_eps}
	L_2 t = \frac{1}{\sqrt{2\pi}}\int_0^{\beta\eps(t)/2\pi} \!\!\!\!\sqrt{\Re\!\left[ \frac{2\pi}{\beta\mu_*} \psi^{(1)}\!\left(\frac{1}{2}+\frac{\beta\mu_*}{2\pi}+iy\right)
	- \psi^{(2)}\!\left(\frac{1}{2}+\frac{\beta\mu_*}{2\pi}+iy\right) \right]} dy~.
\end{equation}
These implicit definitions of $\eps(t)$ and $\mu(t)$ can be solved numerically, the results are shown in \fref{fig:OP_geo}.

The question remains about how to subdivide optimally the protocol times of step 1 ($\tau_1$) and step 2 ($\tau_2 = \tau-\tau_1$). The total excess work is given by $W^{(1)} = L_1^2/\tau_1 + L_2^2/\tau_2$, by taking the derivative and imposing it to be zero we find
\begin{equation}
    \tau_1 = \frac{L_1}{L_2+L_1} \tau ~.
\end{equation}
And therefore we find
\begin{equation}\label{work_one_parm}
    W^{(1)} = \frac{1}{\tau}\left(L_1 + L_2\right)^2~,
\end{equation}
which is indeed what was to be expected, as $L_1 + L_2$ is the total length of the protocol.

\begin{figure}[H]
		\centering
        \includegraphics[width=0.75\textwidth]{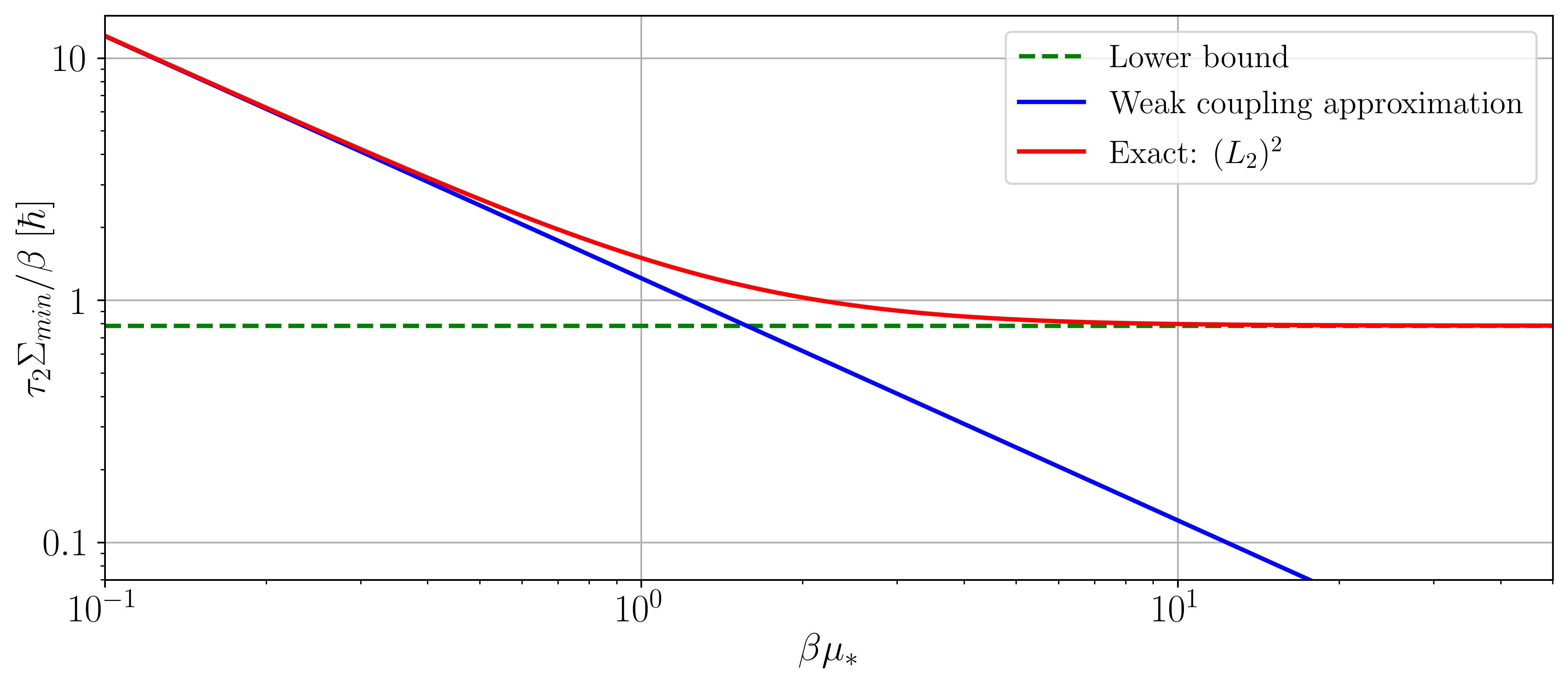}
	    \vspace{-5pt}
	    \caption{Comparison of the excess work $W^{(1)} = k_BT\Sigma$ for a slow erasure protocol at constant coupling in the exact description (\eref{L_step2}) with the weak coupling approximation (\eref{eq:finitetimeboundweak} of the main text) and the lower bound of \eref{step2_bound}.}
	    \vspace{-15pt}
		\label{fig:exact_OP}
\end{figure}

We now discuss how we can obtain an exact version of \eref{eq:finitetimeboundweak} of the main text, so that it applies also in the strong coupling regime. First we can notice that since $L_1>0$ and $L_2>0$ we have $W^{(1)}\geq L_2^2/\tau$. Then by considering that $m_{\eps\eps}(\eps,\mu_*)$ is a one-dimensional metric it has to be positive by definition. Therefore the integrand of \eref{L_step2} is always positive. Next we can consider the fact that 
\begin{equation}
    \lim_{\mu_*\rightarrow\infty} \frac{2\pi}{\beta\mu_*} \psi^{(1)}\!\left(\frac{1}{2}+\frac{\beta\mu_*}{2\pi}+iy\right)
	- \psi^{(2)}\!\left(\frac{1}{2}+\frac{\beta\mu_*}{2\pi}+iy\right) = 0~.
\end{equation}
Therefore, for all $\mu_*$ and all $\eps$
\begin{equation}
    m_{\eps\eps}(\eps,\mu_*) \geq \lim_{\mu_*\rightarrow\infty}m_{\eps\eps}(\eps,\mu_*)~,
\end{equation}
which allows us to conclude $L_2 \geq \lim_{\mu_*\rightarrow\infty} L_2$. For any finite $\mu_*$ large enough $m_{\eps\eps}(\eps,\mu_*)$ can be approximated by $[m_{T=0}]_{\eps\eps}(\eps,\mu_*)$. The angle spanned by the integral of $L_2$ is $\pi/2$. Therefore for all $\mu_*$
\begin{equation}
    \int_0^\infty d\eps~ [m_{T=0}(\eps,\mu_*)]_{\eps\eps}^{1/2} = \frac{\sqrt{\pi}}{2}~,
\end{equation}
and since the approximation becomes exact in the limit $\mu_*\rightarrow\infty$ we have $\lim_{\mu_*\rightarrow\infty} L_2 = \sqrt{\pi}/2$. Therefore, 
\begin{equation}\label{step2_bound}
    W^{(1)} \geq \frac{\pi}{4\tau}~.
\end{equation}
In \fref{fig:exact_OP} we show the minimal value of $W^{(1)}$ for step 2 as a function of $\beta\mu_*$, and we compare it to \eref{eq:finitetimeboundweak} of the main text and to \eref{step2_bound}. We can see how, when the coupling becomes small, the exact curve agrees with \eref{eq:finitetimeboundweak}.

\section{Numerical solution to the general case}\label{sm:num}
\begin{figure}[H]
		\centering
        \includegraphics[width=\textwidth]{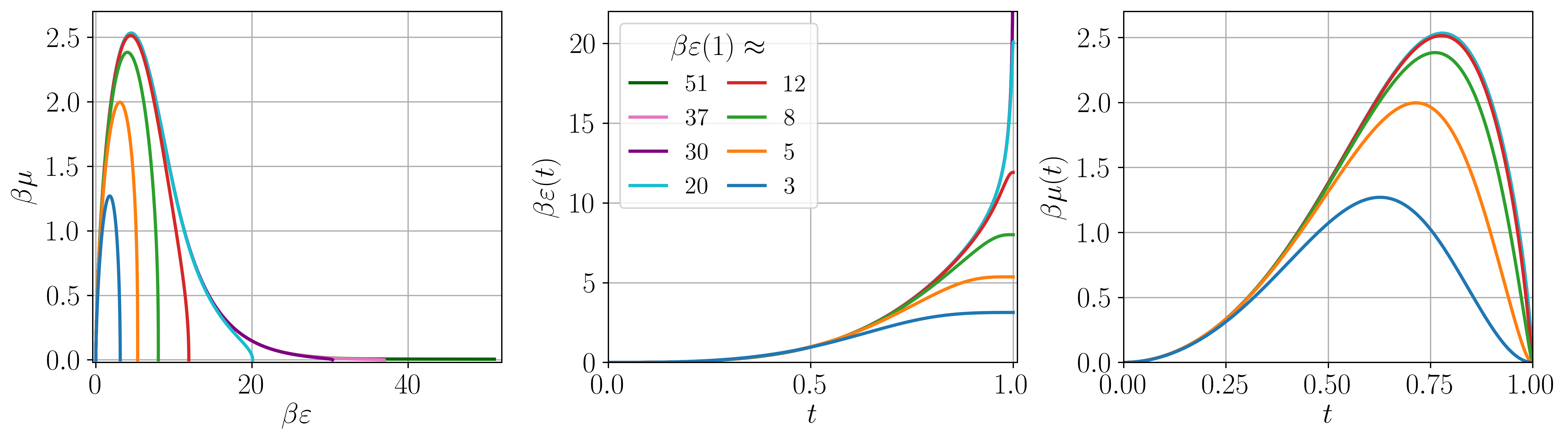}
	    \vspace{-15pt}
	    \caption{A series of optimal protocols depicted for multiple values of $\beta\eps(1)$. They all start with zero energy and coupling and end with finite energy and zero coupling. In the limit of large $\beta\eps(1)$ they can be considered as erasure protocols. Shown in the parameter space (left) and as a function of time (centre and right).}
	    \vspace{-10pt}
		\label{fig:geo}
\end{figure}
We now discuss how the numerical problem of finding the optimal erasure protocol was approached. Having found the metric \eref{metric}, all we have to do to find the optimal erasure protocol is to solve the geodesic equations
\begin{equation}
\label{num:EOM}
\begin{split}
	0 &= \ddot\eps + \Gamma^\eps_{\eps\eps}\dot\eps^2\dot + 2\Gamma^\eps_{\eps\mu}\dot\eps\dot\mu + \Gamma^\mu_{\mu\mu}\dot\mu^2 ~,\\
	0 &= \ddot\eps + \Gamma^\mu_{\eps\eps}\dot\eps^2\dot + 2\Gamma^\mu_{\eps\mu}\dot\eps\dot\mu + \Gamma^\mu_{\mu\mu}\dot\mu^2~;
\end{split}
\end{equation}
with the Christoffel symbols defined as in \eref{Chrstf}. Though the differential equations we get are quite untractable and cannot be solved analytically, we won't even write them here as they are \emph{very} long and will not bring any insight. Therefore we will solve them numerically, and indeed \eref{num:EOM} is quite practical for numerical integration since the second derivative of the parameters can be easily isolated. The boundary conditions we impose for the erasure protocol are $\{\eps(0) = \mu(0) = \mu(1) = 0~,~\eps(1)\gg k_B T\}$. Indeed we cannot impose $\beta\eps(1) = \infty$ because we are performing numerics, therefore we set it to be an arbitrarily large value.

But at this point we can notice that close to $t=0$, by continuity, we are satisfying the conditions for the high temperature approximation. And at $t=0$ the approximation becomes exact. Therefore the initial conditions of an optimal erasure protocol in the general case must match with the initial conditions of the protocols that we previously studied in the high-temperature regime. From a numerical perspective it is much more preferable to solve an initial value problem instead of a boundary value problem. Therefore we used the numerical solver \emph{DOP853}, implemented in the \emph{scipy} library in python, to solve \eref{num:EOM} with the initial conditions given by \eref{geodesic_HT}.

To be precise, we cannot start the integration from $t=0$ as the metric is formally divergent at $(\eps,\mu) = (0,0)$, therefore we evaluate \eref{geodesic_HT} at an infinitesimal time and integrate from there. The specific value we choose for $\eps_*$ sets the value of $\eps(1)$ that is reached in a monotonous way. When $\eps_*$ is chosen small the protocol closes matches with those of \eref{geodesic_HT} (as long as $\eps(1)$ is also small). Then for larger values of $\eps_*$ we get more interesting behavior, as is shown in \fref{fig:geo}.

\begin{figure}[H]
		\centering
        \includegraphics[width=\textwidth]{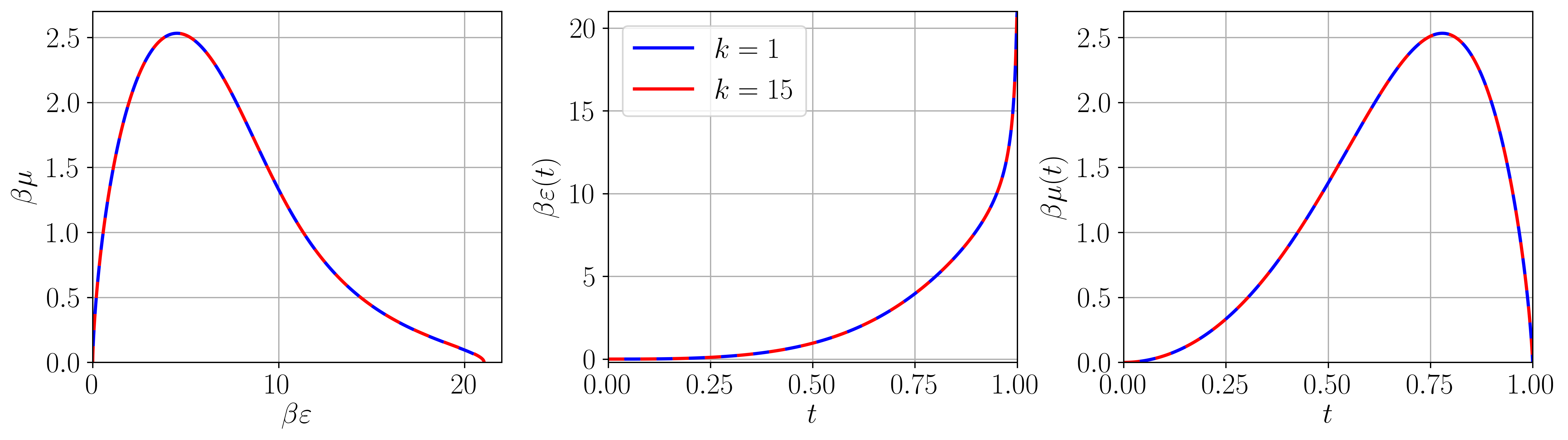}
	    \vspace{-15pt}
	    \caption{Comparing two optimal erasure protocol with $\beta\eps(1) \approx 21$ for two different values of $k$. The same is found for other values of $k$.}
	    \vspace{-10pt}
		\label{fig:k_comp}
\end{figure}

When one changes the value of $k$ in \eref{geodesic_HT} the value of $\eps(1)$ that is reached is different. But, as is shown in \fref{fig:k_comp}, by numerically searching values of $\eps_*$ such that the same $\eps(1)$ is reached for different values of $k$ we find that the protocols end up being the same.

\begin{figure}[H]
		\centering
        \includegraphics[width=\textwidth]{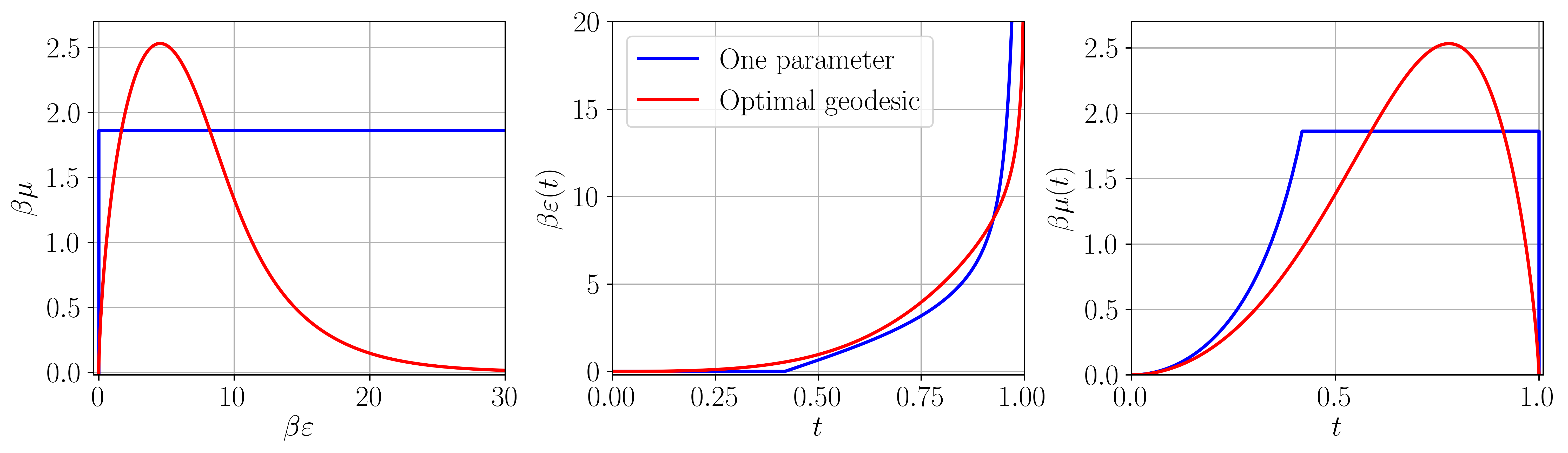}
	    \vspace{-15pt}
	    \caption{Comparing an optimal erasure protocol ($\beta\eps(1)\approx 50$) to an optimized ($\beta\mu_* \approx 1.863$) erasure protocol where we change only one parameter at a time.}
	    \vspace{-10pt}
		\label{fig:geo_vs_op}
\end{figure}

Finally we thought it might be interesting to compare the best one-parameter protocol to the geodesic erasure protocol we find numerically. And we can see from \fref{fig:geo_vs_op} that, despite seeming very different in the path taken in the parameter space, when we look at the functions of time they are actually quite similar. The apparent difference happens because the part of the protocol for $\beta\eps \gg 1$ is done very quickly since the metric is vanishing in that region.

\end{document}